\documentclass[10pt,prd,nofootinbib,preprint]{revtex4}
\usepackage[a4paper, total={7in, 10.5in}]{geometry}
\usepackage[utf8]{inputenc}
\usepackage{bm}
\usepackage{color}
\usepackage{float}
\usepackage{booktabs}
\usepackage{multirow}
\usepackage{amsmath}
\usepackage{graphicx}
\usepackage{esint}
\usepackage{epsfig}
\usepackage{caption}
\usepackage{subcaption}
\usepackage{hyperref}
\usepackage{cancel}
\makeatletter

\DeclareFontEncoding{LGR}{}{}

\ProvideTextCommand{\~}{LGR}[1]{\char126#1}

\newcommand{\lyxmathsym}[1]{\ifmmode\begingroup\def\b@ld{bold}
  \text{\ifx\math@version\b@ld\bfseries\fi#1}\endgroup\else#1\fi}

\providecommand{\tabularnewline}{\\}

\@ifundefined{textcolor}{}
{%
 \definecolor{BLACK}{gray}{0}
 \definecolor{WHITE}{gray}{1}
 \definecolor{RED}{rgb}{1,0,0}
 \definecolor{GREEN}{rgb}{0,1,0}
 \definecolor{BLUE}{rgb}{0,0,1}
 \definecolor{CYAN}{cmyk}{1,0,0,0}
 \definecolor{MAGENTA}{cmyk}{0,1,0,0}
 \definecolor{YELLOW}{cmyk}{0,0,1,0}
}

\usepackage{xcolor}\usepackage{epsfig}\usepackage{slashed}\usepackage{appendix}
\usepackage{tabularx}
\usepackage{pdfpages}
\usepackage{epstopdf}
\usepackage{multirow}
\usepackage{siunitx}
\usepackage{soul}

\usepackage[T1]{fontenc}
\usepackage{array}
\usepackage{booktabs}
\usepackage{amstext}
\usepackage{babel}
\makeatother

\begin{document}
\begin{center}
{\bf\Large\boldmath
Addressing the $R_{\tau/{\mu,e}}\left(D^{(*)}\right)$ puzzle through New Physics four-fermion operators and their impact on
$\Lambda_{b}\rightarrow\Lambda_{c}\tau\bar{\nu}_{\tau}$ decay
}\\[5mm]
\par\end{center}
\begin{center}
\setlength{\baselineskip}{0.2in} {Muhammad Arslan$^{a,}$\footnote{arslan.hep@gmail.com (corresponding author)},
Ishtiaq Ahmed$^{b,}$\footnote{ishtiaqmusab@gmail.com}, Muhammad Jamil Aslam $^{a,}$\footnote{jamil@qau.edu.pk}, Saba Shafaq$^{c,}$\footnote{saba.shafaq@iiu.edu.pk}  and Tahira Yasmeen$^{c,}$\footnote{tahira709@gmail.com}}\\[5mm] $^{a}$~\textit{Department of Physics, Quaid-i-Azam
University, Islamabad 45320, Pakistan.}\\
 $^{b}$~\textit{National Center for Physics, Islamabad 44000, Pakistan.}\\
 $^{c}$~\textit{Department of Physics, International Islamic University,
Islamabad 44000, Pakistan.}\\[5mm] 
\par\end{center}

\begin{abstract}
The Lepton Flavor Universality ratio $R_{\tau/{\mu,e}}\left(D^{(*)}\right)$ poses a notable challenge to the Standard Model (SM), as results of the B-factory experiments, BaBar, Belle, and the LHCb show $3.31\sigma$ deviations from their theoretical predictions. Utilizing the latest HFLAV averages and incorporating the branching ratio constraints $60\%$, $30\%$ and $10\%$ from the lifetime of the $B_c$ meson, we determine the values of the Wilson coefficients (WCs) for different New Physics (NP) four-fermion operator with specific Lorentz structures. Our analysis finds that the parametric region allowed for the WC scenario $\left(C_{S_{L}},C_{S_{R}}\right)$ emerged as the most probable, yielding the maximum pull from the SM, and strongly influenced by the constraints of the branching ratio. Furthermore, we identify three degenerate solutions involving $C_{V_{L}}$, $C_{V_{L}}^{\prime}$, $C_{V_{L}}^{\prime\prime}$, and $C_{S_{R}}^{\prime\prime}$ as the second most probable NP scenarios. We then studied the influence of these NP operators on various physical observables in $\Lambda_{b}\rightarrow\Lambda_{c}\tau\bar{\nu}_{\tau}$ decay by using the Lattice QCD form factors. Our results highlighted $C_{S_{L}}^{\prime\prime}$, $C_{S_{R}}$, $C_{T}$, $\left(C_{S_{L}},C_{S_{R}}\right)$, $\left(C_{S_{R}},C_{T}\right)$, and the three degenerate scenarios involving $\left(C_{S_{L}},C_{T}\right)$, $\left(C_{S_{L}}^{\prime},C_{T}^{\prime}\right)$ and $\left(C_{S_{L}}^{\prime\prime},C_{T}^{\prime\prime}\right)$ as strong indicators of NP. The correlation of different physical observables shows a direct correlation between $d\Gamma/dq^{2}$ and $P_{L}^{\tau}$ for WC $\left(C_{S_{L}},C_{S_{R}}\right)$; and between $A_{FB}$ and $P_{L}^{\Lambda_{c}}$ for three degenerate WCs involving $\left(C_{S_{L}},C_{T}\right)$. We hope that the measurements of these observables on some ongoing and future experiments will help us to scrutinize these constraints on the various NP couplings. \\[5mm]

\end{abstract}

\maketitle

\section{Introduction}\label{sec1}

The Standard Model (SM) formulated by Glashow \cite{Glashow:1961tr}, Weinberg \cite{Weinberg:1967tq} and Salam \cite{Salam:1968rm} in the 1960s is well tested experimentally. So far, we have not found direct evidence of new particles in the Large Hadron Collider (LHC) Run 3 data, taken at an energy of $13.6$ TeV and $39.7$ fb$^{-1}$ of integrated luminosity. These null results set the mass scale of New Physics (NP) in most of the cases beyond 1 TeV, with the possibility that the LHC energy is still not sufficient to make any direct NP discovery. However, there are several observations that are inconsistent with SM, such as neutrino oscillations and some anomalies in flavor physics, which require the existence of NP \cite{Arbey:2021gdg, Gonzalez-Garcia:2022pbf}. In these circumstances, the useful probe for the NP is to look for the low-energy physics observables, \textit{e.g.}, the violation of (approximate) symmetries of the SM, like the lepton flavor universality (LFU), which is only broken in the SM Lagrangian by the small Yukawa couplings. This has been explored in rare decays that occur through flavor-changing neutral-current (FCNC) transitions $b\to s\ell^{+}\ell^{-}$. These transitions are loop suppressed in the SM, offering a fertile ground to look for the possible NP; therefore, the corresponding exclusive decays $B^{\pm}\to K^{(*)\pm}\ell^{+}\ell^{-}$,
$B^{0}\to K^{0}\ell^{+}\ell^{-}$ and $B_{s}^{0}\to\phi\ell^{+}\ell^{-}$,
where the decays $\ell=\mu,\;e$ were experimentally searched extensively \cite{LHCb:2014cxe,LHCb:2014auh,LHCb:2015wdu,LHCb:2015svh,LHCb:2016ykl,LHCb:2021zwz,LHCb:2021xxq}. Specifically, the experimental measurements of LFU violation (LFUV) in $B\to K^{\left(*\right)}\ell^{+}\ell^{-}$ \cite{LHCb:2022vje, LHCb:2022qnv, Smith:2024xgo, CMS:2024syx}, where the dependence on the elements of the Cabibbo-Kobayashi-Maskawa matrix (CKM) and the uncertainties in the form factors (FFs) almost cancel, are rigorously explored in various NP models; see \textit{e.g.}, \cite{Celis:2017doq,Buttazzo:2017ixm,Aebischer:2019mlg,Alasfar:2020mne,Isidori:2021tzd,Ciuchini:2022wbq}, and these are found to be in the range of SM \cite{Hiller:2003js, Bordone:2016gaq, Mishra:2020orb, Isidori:2020acz,Bernlochner:2021vlv,Fischer:2021sqw,London:2021lfn,Crivellin:2021sff,Crivellin:2022qcj}.

However, the window of NP is still open in semileptonic decays governed by the flavor-changing-charged-current (FCCC) transitions $b\rightarrow c\ell\nu_{\ell}$ ($\ell=e,\mu,\tau$). Particularly, the LFU ratio in $B\to D^{(*)}$ decays, \textit{i.e.}, $R_{\tau/\mu,e}\left(D^{(*)}\right)\equiv\mathcal{B}\left(B\rightarrow D^{(*)}\tau\overline{\nu}_{\tau}\right)/\mathcal{B}\left(B\rightarrow D^{(*)}\ell\bar{\nu}_{\ell}\right)$, where $\ell = e,\mu$, measured experimentally by BABAR \cite{BaBar:2012obs,BaBar:2013mob}, Belle \cite{Belle:2015qfa,Belle:2019rba,Belle:leptonphoton}
and LHCb \cite{LHCb:2015gmp,LHCb:2017smo,LHCb:2017rln,LHCb:2023zxo,LHCb:2023uiv}
have marked compelling deviations from their SM predictions. Recently, the
 Heavy Flavor Averaging Group (HFLAV), took averages of almost ten years data of all these experiments and showed $3.31\sigma$ combined deviation from the SM results  \cite{MILC:2015uhg, Na:2015kha, Aoki:2016frl, Fajfer:2012vx, Bigi:2016mdz, Bernlochner:2017jka, Bigi:2017jbd, Jaiswal:2017rve, Gambino:2019sif, Bordone:2019vic, Martinelli:2021onb} with the correlation of $-0.39$ between the $R_{\tau/{\mu, e}}\left(D\right)$ and $R_{\tau/{\mu, e}}\left(D^{*}\right)$ \cite{HFLAV:2024link}. The corresponding HFLAV results and the SM predictions are:
 \begin{eqnarray}
 R_{\tau/\mu,e}\left(D\right) &=& 0.344\pm0.026\;, \quad\quad R_{\tau/\mu,e}\left(D^*\right) = 0.285\pm 0.012\;, \label{HFLAV-RDDs}\\
R^{\text{SM}}_{\tau/\mu,e}\left(D\right)&=& 0.298\pm 0.004\;,\quad\quad R^{\text{SM}}_{\tau/\mu,e}\left(D^*\right) = 0.254\pm 0.005\;. \label{SM-RDDs} 
\end{eqnarray}

The LHCb collaboration measured $R_{\tau/\mu}^{LHCb}\left(J/\psi\right) = 0.71\pm 0.17\pm 0.18$ \cite{LHCb:2017vlu}. Recently, CMS provided preliminary results using muonic $\tau$ tagging methods is $R_{\tau/\mu}^{CMS2023}\left(J/\psi\right) = 0.17^{+0.18}_{- 0.17}\left(stat\right)^{+0.21}_{-0.22}\left(syst\right)^{+0.19}_{-0.18}\left(theory\right)$ \cite{RJSi:CMS2023}, hadronic $\tau$ tagging methods is; $R_{\tau/\mu}^{CMS2024}\left(J/\psi\right) = 1.04^{+0.50}_{- 0.44}$ \cite{RJSi:CMS2024}. The observed value for the $R_{\tau/\ell}\left(\Lambda_{c}\right)$ by LHCb collaboration is $R_{\tau/\ell}^{LHCb}\left(\Lambda_{c}\right) =  0.242 \pm 0.026\pm 0.040\pm 0.059$ \cite{LHCb:2022piu}, normalizing with the SM prediction of $\Gamma\left(
\Lambda_{b}\rightarrow\Lambda_{c}\mu\bar{\nu}\right)$ improves the accuracy and slightly uplifts the central value, $R_{\tau/\ell}\left(\Lambda_{c}\right) =  \left|0.041/V_{cb}\right|^{2}\left(0.271\pm0.069\right)$ \cite{Bernlochner:2022hyz}. Here, the first and second uncertainties are statistical and systematic, respectively. The third uncertainty in the case of $R_{\tau/\ell}\left(\Lambda_{c}\right)$ corresponds to the external branching fraction measurements. So the naive average of $ R_{\tau/\mu}\left(J/\psi\right)$ and corresponding result for $R_{\tau/\ell}\left(\Lambda_{c}\right)$ are
\begin{equation}
 R_{\tau/\mu}\left(J/\psi\right) = 0.61\pm 0.18\;,\quad\quad  R_{\tau/\ell}\left(\Lambda_{c}\right) =  0.271 \pm 0.072\;,\label{Exp-RJpsiLC}
\end{equation}
These experimental results differ from the corresponding SM predictions
\begin{eqnarray}
 R^{\text{SM}}_{\tau/\mu}\left(J/\psi\right) &=&   0.258 \pm 0.038,\; \text{\cite{Watanabe:2017mip, Harrison:2020nrv} }\label{SN-RJPsi}\\
R^{\text{SM}}_{\tau/\ell}\left(\Lambda_c\right) &=& 0.324\pm0.004,\; \text{\cite{Detmold:2015aaa, Bernlochner:2018kxh}} \label{Exp-Lambdac}
\end{eqnarray}
by $1.8\sigma$. The only shortcoming of decays involving $B_c$ meson is the uncertainty in the measurement of its lifetime. Owning to this, the corresponding leptonic decay $B_c \to \tau \nu_\tau$ is not measured yet \cite{Celis:2016azn, Alonso:2016oyd}; an upper limit of
$60\%, 30\%$, and $10\%$ on its branching ratio is imposed in the literature \cite{Gershtein:1994jw,Bigi:1995fs,Beneke:1996xe,Chang:2000ac,Kiselev:2000pp,Akeroyd:2017mhr}.

In addition to these deviations in the LFU measurements, the polarization observables associated with the longitudinal polarization asymmetry of $\tau^{-}$ $\left( P_{\tau}\left(D^{*}\right)\right)$ and the longitudinal polarization of $D^{*-}$ $\left(F_{L}\left(D^{*}\right)\right)$, in $B\to D^{*}\tau\nu_{\tau}$ serve as a tool to probe NP in these decays. With regard to this, the Belle Collaboration reported the results of these observables $P_{\tau}\left(D^{*}\right) = -0.38\pm0.51_{-0.16}^{+0.21}$ \cite{Belle:2017ilt,Belle:2016dyj} and $F_{L}\left(D^{*}\right) =  0.60 \pm 0.08\pm 0.04$ \cite{Belle:2019ewo}. These results were inconsistent with the SM predictions of $P^{\text{SM}}_{\tau}\left(D^{*}\right) = -0.497\pm 0.007$ and $F^{\text{SM}}_{L}\left(D^{*}\right) =  0.464\pm0.003$ by $1.5\sigma$ \cite{Alok:2016qyh,Iguro:2020cpg}. Similarly, the LHCb Collaboration has recently reported the preliminary results of $F_{L}\left(D^{*}\right) = 0.43\pm 0.06\pm 0.03$ by combining the LHCb Run1 dataset and part of the Run2 data \cite{LHCb:2023ssl, Chen:2024zot} and in the complete integrated range $q^2$. Naively combining the results of Belle and LHCb gives \cite{Iguro:2024hyk}:
\begin{equation}
    F_{L}\left(D^{*}\right) =  0.49\pm0.05\;, \label{FLDs}
\end{equation}
which is within $1\sigma$ with the SM.

These deviations in the LFUV ratios are perplexing and have triggered much theoretical interest. For example, a number of NP studies using dimension-six operators confining to left-handed (LH) neutrinos in $b\rightarrow c\tau\overline{\nu}_{\tau}$ transitions were made \cite{Blanke:2018yud,Blanke:2019qrx,Huang:2018nnq,Alok:2019uqc, Sahoo:2019hbu,Shi:2019gxi,Bardhan:2019ljo, Fedele:2022iib, Asadi:2019xrc,Murgui:2019czp,Mandal:2020htr,Cheung:2020sbq,Colangelo:2020vhu,Arslan:2023wgk, Yasmeen:2024cki}, whereas by considering the right-handed (RH) neutrinos and/or the RH quark currents in the model independent weak effective Hamiltonian (WEH), were analyzed in a number of studies; see, \textit{e.g.} \cite{Greljo:2018ogz,Azatov:2018kzb,Heeck:2018ntp,Babu:2018vrl,He:2017bft,Gomez:2019xfw,Alguero:2020ukk,Dutta:2013qaa,Dutta:2017xmj,Dutta:2017wpq,Dutta:2018jxz}. In Ref. \cite{Freytsis:2015qca}, the WEH is extended considering all possible four-fermion operators that can contribute to $B\to D^{(*)}\tau\bar{\nu}$ decays. Particularly; for the LH neutrinos, the vectors $O_{V_L},\; O_{V_R}$, scalars $O_{S_L},\; O_{S_R}$ and the tensor $O_T$ operators, along with their primed and double-primed partners which are the product of quark-lepton bilinears, i.e., $\left(\bar{l}\Gamma_Dq\right)$, where $\Gamma_D$ correspond to different Dirac structures. By analyzing the LFUV ratio in these decays, several models with leptoquark mediators that are minimally flavor violating in the quark sector and are minimally flavor violating or $\tau$ aligned in the lepton sector were identified. Later, a refit was made after including all available data on $b\to c \tau \bar{\nu_\tau}$ in \cite{Alok:2017qsi} and obtained constraints on the NP WCs $C_{O_i}$,  $C^\prime_{O_i}$ and  $C^{\prime\prime}_{O_i}$, where $i= V_{L,R}\;,\; S_{L, R}\;,\; T$. Interestingly, the interrelations between different operators showed four NP solutions, which can be distinguished by studying the angular asymmetries and the $D^*$ polarization fraction. In addition, LHCb and Belle have recently updated their datasets; therefore, it will be interesting to update the parametric space of the various NP WCs, and see their impact on different FCCC decays. 

With this motivation, by using the model-independent WEH with LH neutrinos and real NP WCs, we take the most up-to-date HFLAV world average Moriond 2024 \cite{HFLAV:2024link} values for $R_{\tau/\mu,e}(D)$ and $R_{\tau/{\mu,e}}(D^{*})$, and the measurements of the $F_{L}\left(D^{*}\right), P_{\tau}\left(D^{*}\right)$,$R_{\tau/\ell}\left(\Lambda_{c}\right)$, and $R_{\tau/\mu}\left(J/\psi\right)$  given above (see Eqs. \ref{Exp-RJpsiLC} - \ref{FLDs}) to re-visit the global fit analysis performed in \cite{Alok:2017qsi}. For this purpose, we perform a $\chi^{2}$ analysis after considering two sets of physical observables: In set $\mathcal{S}_1$, we choose $R_{\tau/\mu,e}(D), R_{\tau/{\mu,e}}(D^{*}), F_{L}\left(D^{*}\right)$ and $ P_{\tau}\left(D^{*}\right)$, whereas; in $\mathcal{S}_2$, we also add $R_{\tau/\ell}\left(\Lambda_{c}\right)$ in the list to scrutinize the parametric space of the NP WCs.
For the set $\mathcal{S}_1$, we find that the NP scalar WCs $\left(C_{S_{L}}, C_{S_{R}}\right)$ are prominent compared to the other WCs and have a strong dependence on the constraints arising due to the branching ratio of $B_c \to \tau \nu$. With expecting $p$-values to be $\sim 50\%$ for the true solution, we observe a less favorable alignment with the observed anomalies for the set $\mathcal{S}_2$.

In the next step, we examine the phenomenological impact of the parameter space defined by the set $\mathcal{S}_1$ on various physical observables related to the decay process $\Lambda_{b} \rightarrow \Lambda_{c} \tau \bar{\nu}_{\tau}$. The study of $\Lambda_{b}$ baryon decays, which exhibit spin-1/2 characteristics, complements the information obtained from $B$ meson decays in the quark-level transition $b \rightarrow c \tau \overline{\nu}_{\tau}$. Since baryonic decays involve different kinematic properties and form factors compared to their mesonic counterparts, they provide additional insight to the nature of the $b \rightarrow c \tau \overline{\nu}_{\tau}$ transition. However, in contrast to mesonic decays, where the form factors are well-established through experimental data, the form factors for $\Lambda_{b}\rightarrow\Lambda_{c}$ are still experimentally undetermined. This uncertainty makes theoretical approaches such as the Lattice-QCD essential for their calculations. On the theoretical side, the decay $\Lambda_{b}\rightarrow\Lambda_{c}\tau\bar{\nu}_{\tau}$ has been explored within the SM and various NP scenarios in a number of studies; see \textit{e.g.}, Ref. \cite{Gutsche:2015rrt,Gutsche:2015mxa,Shivashankara:2015cta,Dutta:2015ueb,Faustov:2016pal,Li:2016pdv,Celis:2016azn}. In this work, we have used the latest form factors calculated from Lattice- QCD \cite{Bernlochner:2018kxh} and analyzed various $\Lambda_{b}\rightarrow\Lambda_{c}\tau\bar{\nu}_{\tau}$ observables such as differential decay rate, lepton forward-backward asymmetry, $\Lambda_{c}-$longitudinal polarization fraction, $\tau-$lepton longitudinal polarization fraction, and $\Lambda_{c}-$LFU ratio. Finally, we compare the results of these observables with their existing experimental values, where available.   

The benchmarks for the current study are as follows:
\begin{itemize}
    \item Our analysis incorporates updated HFLAV Moriond 2024 data for $R_{\tau/{\mu,e}}(D^{\left(*\right)})$. It also includes the naive average of $R_{\tau/\mu}\left(J/\psi\right)$ by incorporating measurements of LHCb and different methods of $\tau-$tagging at the CMS. This will slightly uplift the central value of $R_{\tau/\ell}\left(\Lambda_{c}\right)$ by normalizing it with the SM prediction of $\Gamma\left(
    \Lambda_{b}\rightarrow\Lambda_{c}\mu\bar{\nu}\right)$, and with naively combining the Belle and LHCb results for $F_{L}\left(D^{*}\right)$.
    \item We analyze all possible scenarios for the NP WCs.
    \item Scenarios with $B_{c} \leq 10\%$ constraints eliminate some cases; however, a larger number of NP solutions remain viable.
    \item $R_{\tau/\ell}\left(\Lambda_{c}\right)$ is explicitly evaluated after using the best fit points (BFPs) from parametric space of NP WCs constrained by the $\chi^2-$analysis in set $(\mathcal{S}_2)$ along with the sum rules linking it with $R_{\tau/{\mu,e}}(D^{\left(*\right)})$.
    \item The sum rule for $ R_{\tau/\mu}\left(J/\psi\right)$ is also updated with the predictions of its numerical values at the BFPs.
    \item The phenomenology analysis of various physical observables of $
    \Lambda_{b}\rightarrow\Lambda_{c}\tau\bar{\nu}_{\tau}$ is performed in detail for the different benchmark scenarios by using the form factors calculated in Lattice-QCD. The correlations among various phenomenological observables such as differential decay rate, lepton forward-backward asymmetry, $\Lambda_{c}-$, $\tau-$lepton longitudinal polarization fractions, and $\Lambda_{c}-$LFU ratio are also examined.
\end{itemize}

The paper is organized as follows. In Section \ref{sec2}, we begin by defining the WEH that includes the SM and NP operators. We then present the formulas for various observables: \( R_{\tau/\mu, e}(D) \), \( R_{\tau/\mu, e}(D^*) \), \( P_{\tau}(D^*) \), \( F_L(D^*) \), \( R_{\tau/\ell}(\Lambda_c) \), \( R_{\tau/\mu}(J/\psi) \),  and \( P_{\tau}(D) \) in terms of the NP Wilson coefficients (WCs). 
In Section \ref{sec3}, we analyze the most recent data to explore the parameter space for real NP WCs. We also examine how the constraints on the branching ratio of the decay \( B_c \to \tau \bar{\nu}_\tau \) influence the allowable regions for these NP WCs. Section \ref{sec5} gives the expressions for the decay distribution of $\Lambda_{b}\rightarrow\Lambda_{c}\tau\bar{\nu}_{\tau}$ in terms of the helicity amplitudes, the different observables mentioned above. A phenomenological analysis of the physical observables is performed in the same section using the Lattice QCD results for the form factors. In Section \ref{sumrule}, we derive the sum rules for $R_{\tau/{\mu}}\left(J/\Psi\right)$ and $R_{\tau/\ell}(\Lambda_{c})$ in terms of $R_{\tau/\mu,e}(D)$ and $R_{\tau/{\mu,e}}(D^{*})$ and discuss the correlation of $\Lambda_{b}\rightarrow\Lambda_{c}\tau\bar{\nu}_{\tau}$ observables.  Finally, in Section \ref{sec6}, we conclude our findings. This work is supplemented by four appendices, discussing the fitting procedure and the derivation of the above-mentioned $
\Lambda_{b}\rightarrow\Lambda_{c}\tau\nu_{\tau}$ observables in terms of helicity amplitudes.

\section{Theoretical Framework and Analytical Formulae}\label{sec2}
\subsection{Weak Effective Hamiltonian (WEH)}

We outline the dimension-6 semileptonic operators that contribute to the weak effective Hamiltonian (WEH) for the transition \(b \rightarrow c\tau\bar{\nu}\) at the tree level. By matching these operators with the SM Effective Field Theory (SMEFT), we derive the resulting relations among the WCs. 
The most general WEH of $b\rightarrow c\tau\bar{\nu}$ transition incorporating all possible Lorentz invariant structures is given as \cite{Freytsis:2015qca, Alok:2017qsi}:
\begin{equation}
H_{\text{eff}}=\frac{4G_{F}V_{cb}}{\sqrt{2}}\left\{ \left(C_{V_{L}}\right)_{SM}\mathcal{O}_{V_{L}}+\frac{\sqrt{2}}{4G_{F}V_{cb}}\frac{1}{\Lambda^{2}}\sum_{i}\left(C_{i}\mathcal{O}_{i}+C_{i}^{\prime}\mathcal{O}_{i}^{\prime}+C_{i}^{\prime\prime}\mathcal{O}_{i}^{\prime\prime}\right)\right\}. \label{weh} 
\end{equation}
Here, $G_{F}$ is the Fermi coupling constant, $V_{cb}$ is the CKM
matrix element, and $P_{R,L}=\left(1\pm\gamma_{5}\right)/2$ are the projection operators where $C_{i}\mathcal{O}_{i}$, $C_{i}^{\prime}\mathcal{O}_{i}^{\prime}$ and $C_{i}^{\prime\prime}\mathcal{O}_{i}^{\prime\prime}$ are the corresponding WCs and NP operators, respectively. 
Note that the first term corresponds to the SM contribution and its associated WC is normalized to unity; \textit{i.e.,} $\left(C_{V_{L}}\right)_{SM}=1$.
The new unprimed operators $\mathcal{O}_{i}$, where $i\equiv V_{L},\; V_{R},\; S_{L},\; S_{R},\; T$, read as \cite{Asadi:2018wea,Asadi:2018sym,Buchmuller:1985jz,Grzadkowski:2010es,Aebischer:2015fzz}: 
\begin{eqnarray}
\mathcal{O}_{V_{L}} & = & \left(\overline{c}\gamma^{\mu}P_{L}b\right)\left(\overline{\tau}\gamma_{\mu}P_{L}\nu\right),\; \quad\quad \mathcal{O}_{V_{R}}  =  \left(\overline{c}\gamma^{\mu}P_{R}b\right)\left(\overline{\tau}\gamma_{\mu}P_{L}\nu\right),\nonumber \\
\mathcal{O}_{S_{L}} & = & \left(\overline{c}P_{L}b\right)\left(\overline{\tau}P_{L}\nu\right),\; \quad\quad\quad\quad\;\; 
\mathcal{O}_{S_{R}}  = \left(\overline{c}P_{R}b\right)\left(\overline{\tau}P_{L}\nu\right),\nonumber \\
\mathcal{O}_{T} & = & \left(\overline{c}\sigma^{\mu\nu}P_{L}b\right)\left(\overline{\tau}\sigma_{\mu\nu}P_{L}\nu\right).\label{eq2}
\end{eqnarray}
The NP primed and double primed operators are the combination of quark-lepton bilinears. Their explicit forms of the primed operators are
\begin{eqnarray}
\mathcal{O}_{V_{L}}^{\prime} & = & \left(\overline{\tau}\gamma^{\mu}P_{L}b\right)\left(\overline{c}\gamma_{\mu}P_{L}\nu\right)\quad\leftrightarrow \mathcal{O}_{V_{L},}\nonumber \\
\mathcal{O}_{V_{R}}^{\prime} & = & \left(\overline{\tau}\gamma^{\mu}P_{R}b\right)\left(\overline{c}\gamma_{\mu}P_{L}\nu\right)\quad\leftrightarrow -2\mathcal{O}_{S_{R},}\nonumber \\
\mathcal{O}_{S_{L}}^{\prime} & = & \left(\overline{\tau}P_{L}b\right)\left(\overline{c}P_{L}\nu\right)\quad\leftrightarrow-\frac{1}{2}\mathcal{O}_{S_{L}}-\frac{1}{8}\mathcal{O}_{T},\nonumber \\
\mathcal{O}_{S_{R}}^{\prime} & = & \left(\overline{\tau}P_{R}b\right)\left(\overline{c}P_{L}\nu\right)\quad\leftrightarrow -\frac{1}{2}\mathcal{O}_{V_{R}},\nonumber \\
\mathcal{O}_{T}^{\prime} & = & \left(\overline{\tau}\sigma^{\mu\nu}P_{L}b\right)\left(\overline{c}\sigma_{\mu\nu}P_{L}\nu\right)\quad\leftrightarrow-6 \mathcal{O}_{S_{L}}+\frac{1}{2}\mathcal{O}_{T}.\label{eq2-1}
\end{eqnarray}
In Eq. (\ref{eq2-1}), these primed operators are related with the new unprimed operators through Fierz transformation \cite{Alok:2017qsi}. Similarly, the double-primed operators and their relations with the unprimed operators are 
\begin{eqnarray}
\mathcal{O}_{V_{L}}^{\prime\prime} & = & \left(\overline{\tau}\gamma^{\mu}P_{L}c^{c}\right)\left(\bar{b}^{c}\gamma_{\mu}P_{L}\nu\right)\quad\leftrightarrow-\mathcal{O}_{V_{R}},\nonumber \\
\mathcal{O}_{V_{R}}^{\prime\prime} & = & \left(\overline{\tau}\gamma^{\mu}P_{L}c^{c}\right)\left(\bar{b}^{c}\gamma_{\mu}P_{L}\nu\right)\quad\leftrightarrow-2\mathcal{O}_{V_{L}},\nonumber \\
\mathcal{O}_{S_{L}}^{\prime\prime} & = & \left(\overline{\tau}P_{L}c^{c}\right)\left(\bar{b}^{c}P_{L}\nu\right)\quad\leftrightarrow-\frac{1}{2}\mathcal{O}_{S_{L}}+\frac{1}{8}\mathcal{O}_{T},\nonumber \\
\mathcal{O}_{S_{R}}^{\prime\prime} & = & \left(\overline{\tau}P_{R}c^{c}\right)\left(\bar{b}^{c}P_{L}\nu\right)\quad\leftrightarrow\frac{1}{2}\mathcal{O}_{V_{L}},\nonumber \\
\mathcal{O}_{T}^{\prime\prime} & = & \left(\overline{\tau}\sigma^{\mu\nu}P_{L}c^{c}\right)\left(\bar{b}^{c}\sigma_{\mu\nu}P_{L}\nu\right)\quad\leftrightarrow-6 \mathcal{O}_{S_{L}}-\frac{1}{2}\mathcal{O}_{T}.\label{eq2-1-1}
\end{eqnarray}
These Fierz relations between operators will help us to write the NP WCs $C_{i}$, $C_{i}^{\prime}$, and $C_{i}^{\prime\prime}$ in the following linear combinations \cite{Alok:2017qsi}:
\begin{eqnarray}
C^{\text{eff}}_{V_{L}} & = & \alpha_{\Lambda}\left(C_{V_{L}}+C_{V_{L}}^{\prime}+0.5C_{S_{R}}^{\prime\prime}\right),\nonumber \\
C^{\text{eff}}_{V_{R}} & = & \alpha_{\Lambda}\left(C_{V_{R}}-0.5C_{S_{R}}^{\prime}-C_{V_{L}}^{\prime\prime}\right),\nonumber \\
C^{\text{eff}}_{S_{L}} & = & \alpha_{\Lambda}\left(C_{S_{L}}-0.5C_{S_{L}}^{\prime}-6C_{T}^{\prime}-0.5C_{S_{L}}^{\prime\prime}-6C_{T}^{\prime\prime}\right),\nonumber \\
C^{\text{eff}}_{S_{R}} & = & \alpha_{\Lambda}\left(C_{S_{R}}-2C_{V_{R}}^{\prime}-2C_{V_{R}}^{\prime\prime}\right),\nonumber \\
C^{\text{eff}}_{T} & = & \alpha_{\Lambda}\left(C_{T}-0.125C_{S_{L}}^{\prime}+0.5C_{T}^{\prime}+0.125C_{S_{L}}^{\prime\prime}-0.5C_{T}^{\prime\prime}\right),\label{eq2-2}
\end{eqnarray}
where $\alpha_{\Lambda} \equiv \left(2\sqrt{2}G_{F}V_{cb}\Lambda^{2}\right)^{-1}$, and setting the NP scale $\Lambda=2\; \text{TeV}$, we get $\alpha_{\Lambda}=0.186$.
The energy scale for  $b\to c\tau\bar{\nu}_\tau$ transitions is the $b-$quark mass, i.e., $\mu_{b} = m_{b}$ in the SM which is connected to NP scale through 
the renormalization group equations (RGEs) as
\cite{Gonzalez-Alonso:2017iyc, Blanke:2018yud}:

\arraycolsep=1.2pt\def\arraystretch{2.0}
\begin{align}
\widetilde{C}_{V_{L}}\left(m_{b}\right) & =1.12C_{V_{L}}^{\text{eff}}\left(2\text{TeV}\right),\nonumber \\
\widetilde{C}_{V_{R}}\left(m_{b}\right) & =1.07C_{V_{R}}^{\text{eff}}\left(2\text{TeV}\right),\nonumber \\
\widetilde{C}_{S_{R}}\left(m_{b}\right) & =2C_{S_{R}}^{\text{eff}}\left(2\text{TeV}\right),\nonumber \\
\left(\begin{array}{c}
\widetilde{C}_{S_{L}}\left(m_{b}\right)\\
\widetilde{C}_{T}\left(m_{b}\right)
\end{array}\right) & =\left(\begin{array}{cc}
1.91 & -0.38\\
0 & 0.89
\end{array}\right)\left(\begin{array}{c}
C_{S_{L}}^{\text{eff}}\left(2\text{TeV}\right)\\
C_{T}^{\text{eff}}\left(2\text{TeV}\right)
\end{array}\right).\label{eq3}
\end{align}
For the WEH given in Eq. (\ref{weh}), the physical observables under consideration can be expressed in terms of NP WCs at a scale $\mu_b = m_{b}$. Their explicit expressions are calculated in \cite{Watanabe:2017mip,Iguro:2018vqb,Asadi:2018sym,Gomez:2019xfw,Cardozo:2020uol,Fedele:2022iib,Mandal:2020htr,Kamali:2018bdp,Iguro:2022yzr,Mu:2019bin,Becirevic:2019tpx,Becirevic:2016hea,Alonso:2016gym,Hill:2019zja,Aebischer:2019zoe,Sakaki:2013bfa,Caprini:1997mu,Kumbhakar:2020jdz,Aloni:2018ipm,Duraisamy:2014sna,Duraisamy:2013pia,Alok:2011gv,Bhattacharya:2019olg,Bhattacharya:2020lfm}. Incorporating these NP WCs, their explicit expressions are summarized in Appendix \ref{AppendixA}.

\section{Analysis of the parametric space of NP WCs}\label{sec3}

In this section, we will explore the parametric space of NP WCs using the most recent data from HFLAV on FCCC transitions \cite{HFLAV:2024link}. To accomplish this, we adopt the fitting technique originally developed in \cite{Blanke:2018yud} (see Appendix \ref{GoF}). Our analysis include both primed and double-primed WCs, which can be either real or complex; however, for the purposes of this study, we focus exclusively on their real values. We have categorized the observables into two different sets;

\begin{itemize}
    \item Set 1 $(\mathcal{S}_1)$: includes $R_{\tau/\mu,e}(D), R_{\tau/{\mu,e}}(D^{*}), P_{\tau}\left(D^{*}\right)$ and $ F_{L}\left(D^{*}\right)$.
    \item In set 2 $(\mathcal{S}_2)$: we add $R_{\tau/{\ell}}\left(\Lambda_{c}\right)$ to $\mathcal{S}_1$.
\end{itemize}
Here, the number of observables $\left(N_{obs}\right)$ are $4$ and $5$ for $\mathcal{S}_1$ and $\mathcal{S}_2$, respectively.
We explore the parameter space under two scenarios involving the NP WCs: a one-dimensional case, where only one NP WC is switched on while others are set to zero, and a two-dimensional case, where two NP WCs are switched on.
Therefore, in the $\chi^2$ analysis;  the number of parameters $\left(N_{par}\right)$, equal to $1\left(2\right)$ for one (two)-dimension case, giving the number of degrees of freedom (dof):  $N_{dof}=N_{obs}-N_{par}$ to be $3\left(2\right)$ and $4\left(3\right)$ for $\mathcal{S}_1$ and $\mathcal{S}_2$, respectively. 
Using this set-up, we have found the numerical values of the BFPs, the  
$p-$value, $\chi_{\text{SM}}^{2}$, $\text{pull}_{\text{SM}}$ and $1\sigma$, $2\sigma$ intervals for NP WCs for the one-dimensional scenarios and listed in Table \ref{1d-table}. Furthermore, the effects of the constraints of the $B^-_{c}\to \tau^- \bar{\nu}_\tau$ branching ratio are incorporated to obtain these values.

\begin{figure}
    \centering
    \includegraphics[width=0.5\linewidth]{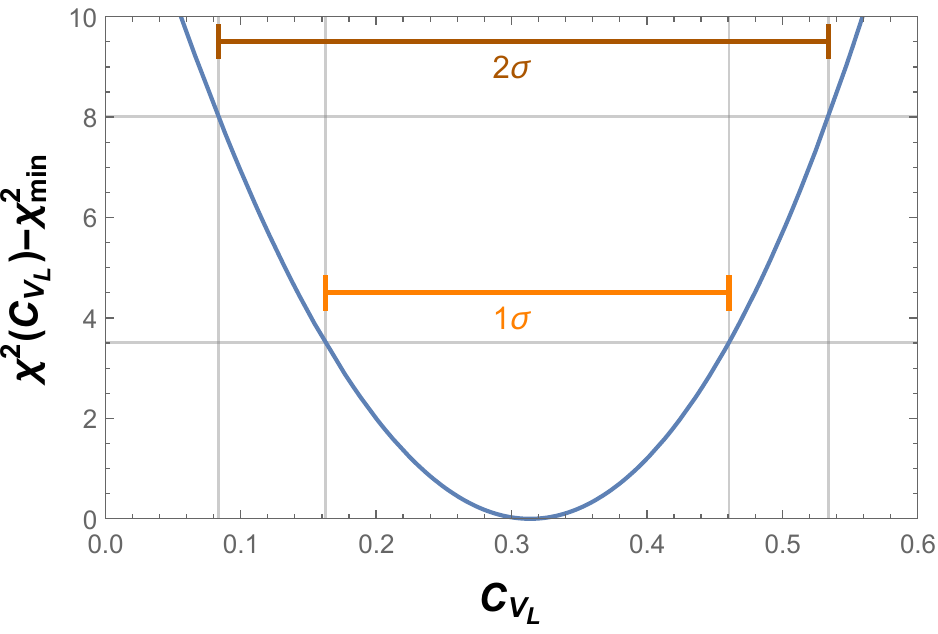}
    \caption{The plot for $\chi^2\left(C_{V_{L}}\right)-\chi^2_{min}$ vs $C_{V_{L}}$. The horizontal grid-lines at $\chi^2= 3.51\left(68\%\right)$  and $\chi^2= 8.01\left(95\%\right)$ correspond to $1\sigma$ and $2\sigma$ thresholds for $3$ dof. The vertical grid-lines show the resulting parameter ranges. As observed, the intervals are not in a $1:2$ ratio but approximately $2:3$.}
    \label{ijmpa1}
\end{figure}

In Table \ref{1d-table}, we observe that in scenario $\mathcal{S}_{1}$, the new vector-like WC $C_{V_{L}}$ has the highest $p-$value of $92\%$. By applying Fierz rearrangement, as described in Eqs. (\ref{eq2-1}, \ref{eq2-1-1}), the WCs $C_{V_{L}}^{\prime}$ and $C_{S_{R}}^{\prime\prime}$ are related to $C_{V_{L}}$. As a result, these coefficients are expected to exhibit a similar $p-$value. However, this value decreases to $62\%$ for $\mathcal{S}_{2}$, indicating the impact of the experimental measurements $R_{\tau/\ell}(\Lambda_c)$. 
Furthermore, the coefficients $C^{\prime\prime}_{S_{L}}$ also show significant effects, presenting $p-$values of nearly $54\%$ and $46\%$ in $\mathcal{S}_{1}$ and $\mathcal{S}_{2}$, respectively. It is important to note that the values of the WCs for these one-dimensional scenarios remain unaffected by the constraints imposed on the branching ratio of $B_{c}\to \tau\nu_\tau$ decay.
Fig. \ref{ijmpa1} is the plot for $\chi^2\left(C_{V_{L}}\right)-\chi2_{min}$ vs $C_{V_{L}}$. The horizontal grid lines at $\chi^2= 3.51\left(68\%\right)$  and $\chi^2= 8.01\left(95\%\right)$ correspond to the $1\sigma$ and $2\sigma$ thresholds for $3$ dof. The vertical lines show the resulting parameter ranges. As observed, the intervals are not in a $1:2$ ratio but approximately $2:3$.

\begin{figure}[H]
\centering{} \subfloat[]{\includegraphics[width=4.5cm, height=4.2cm]{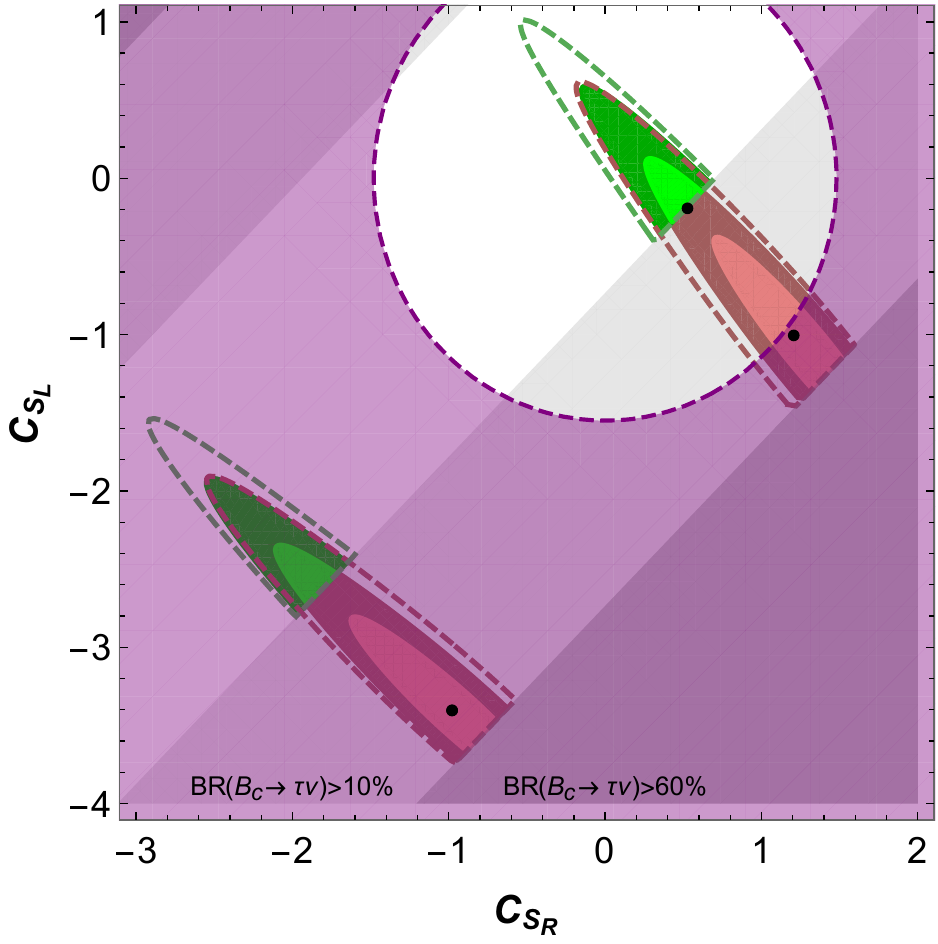}}\quad \subfloat[]{\includegraphics[width=4.5cm, height=4.2cm]{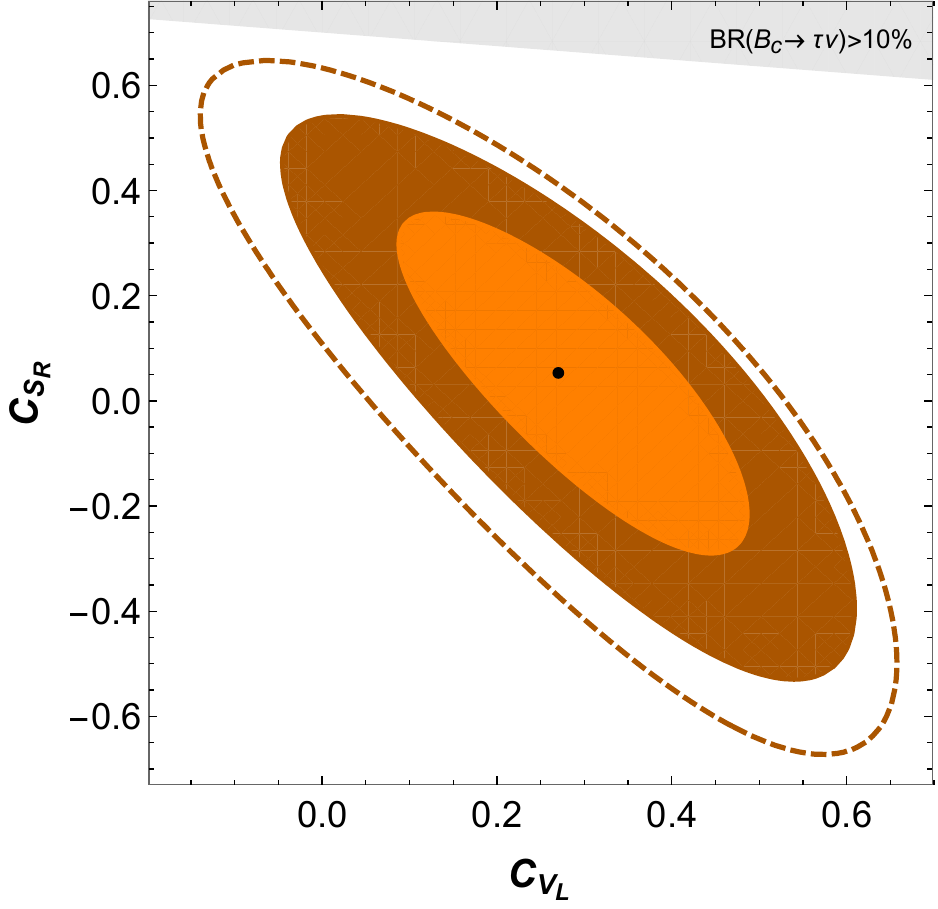}}\quad \subfloat[]{\includegraphics[width=4.5cm, height=4.2cm]{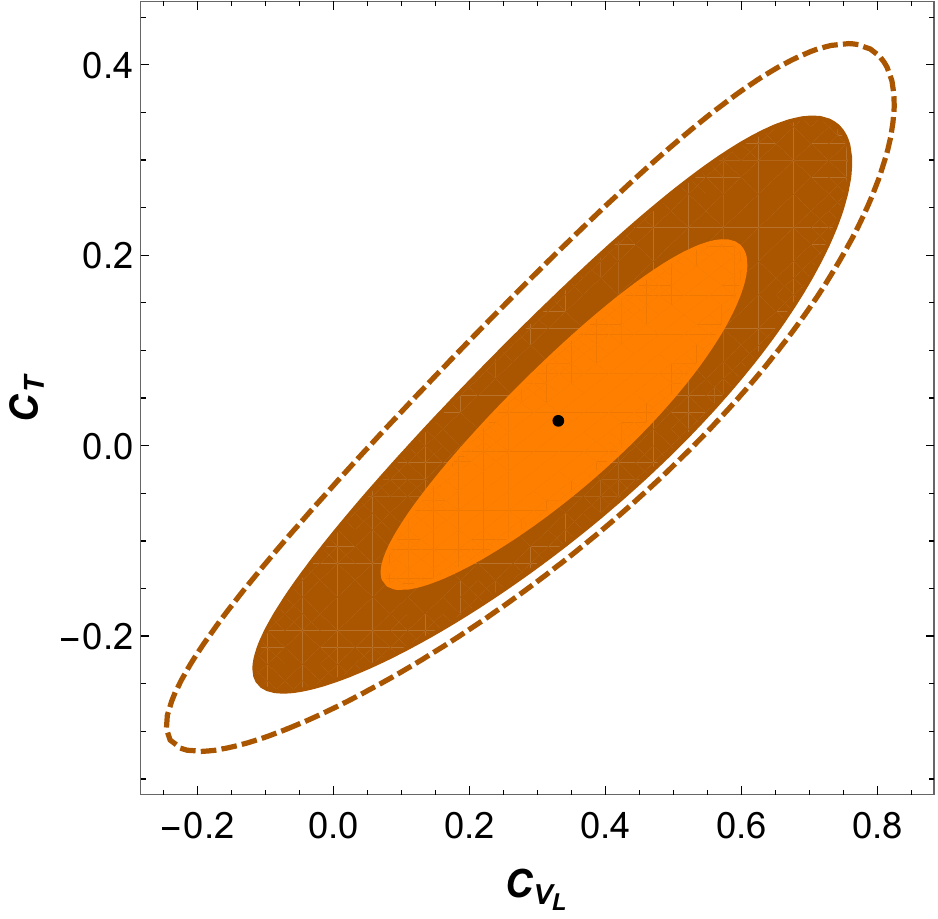}} 
\centering{} \subfloat[]{\includegraphics[width=4.5cm, height=4.2cm]{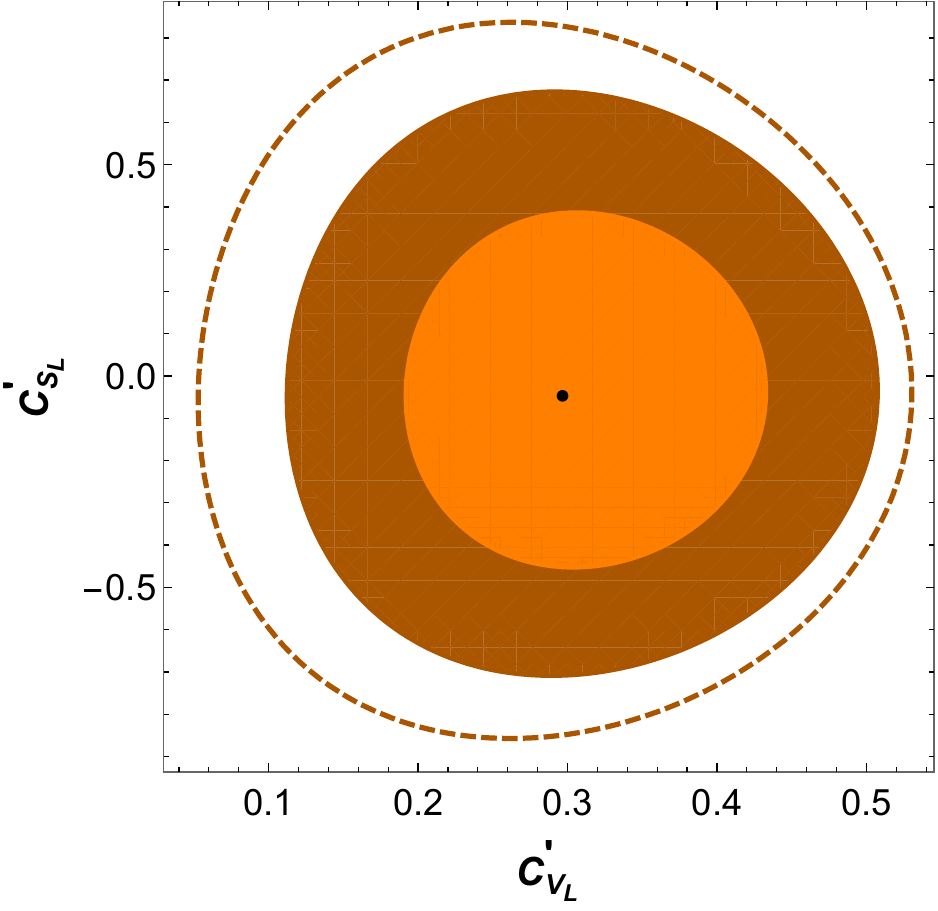}}\quad \subfloat[]{\includegraphics[width=4.5cm, height=4.2cm]{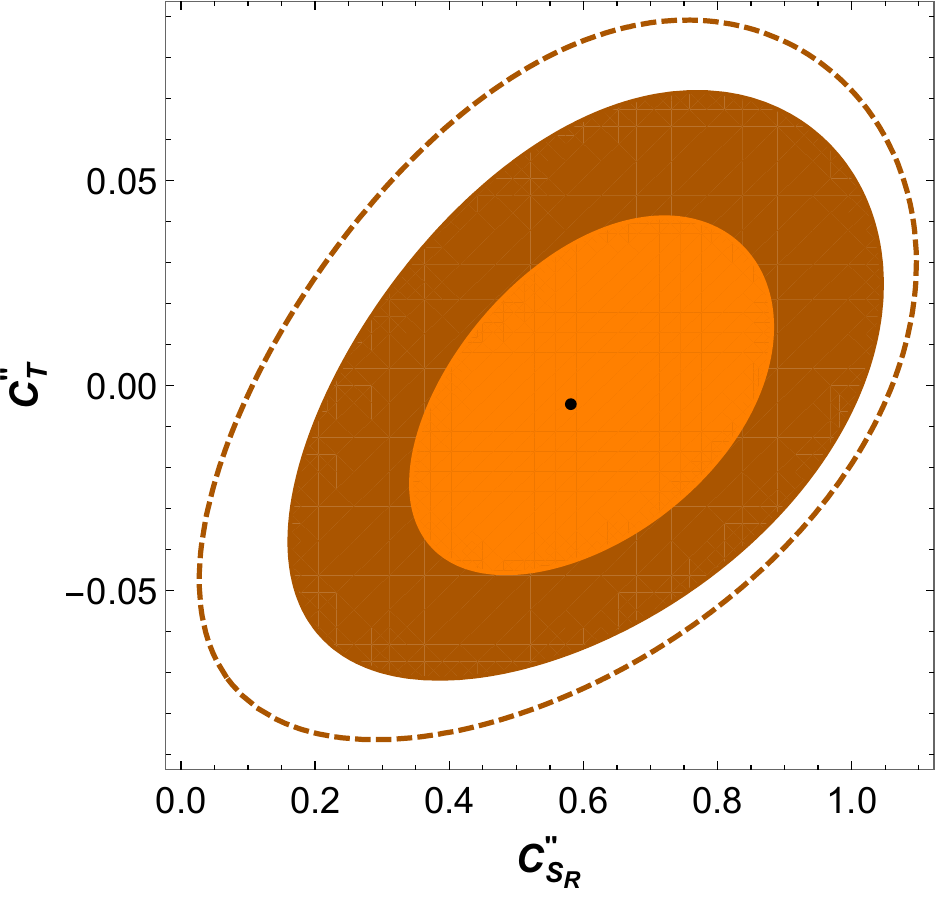}}\quad \subfloat[]{\includegraphics[width=4.5cm, height=4.2cm]{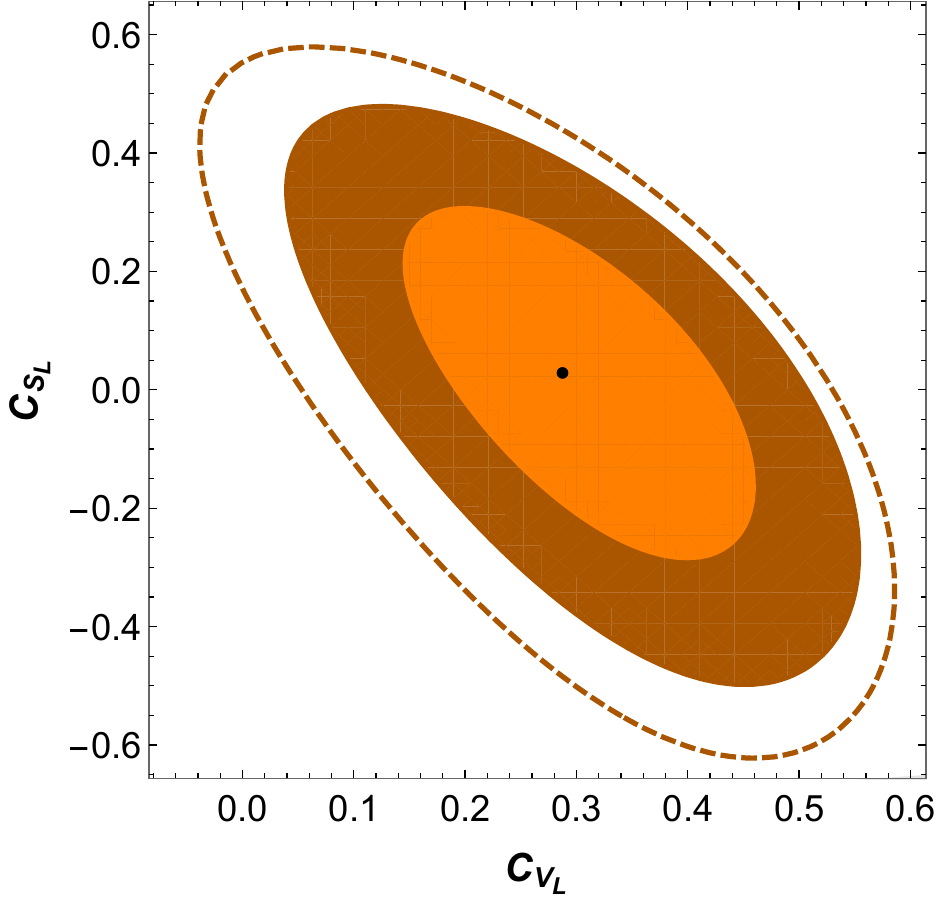}} 
\centering{} \subfloat[]{\includegraphics[width=4.5cm, height=4.2cm]{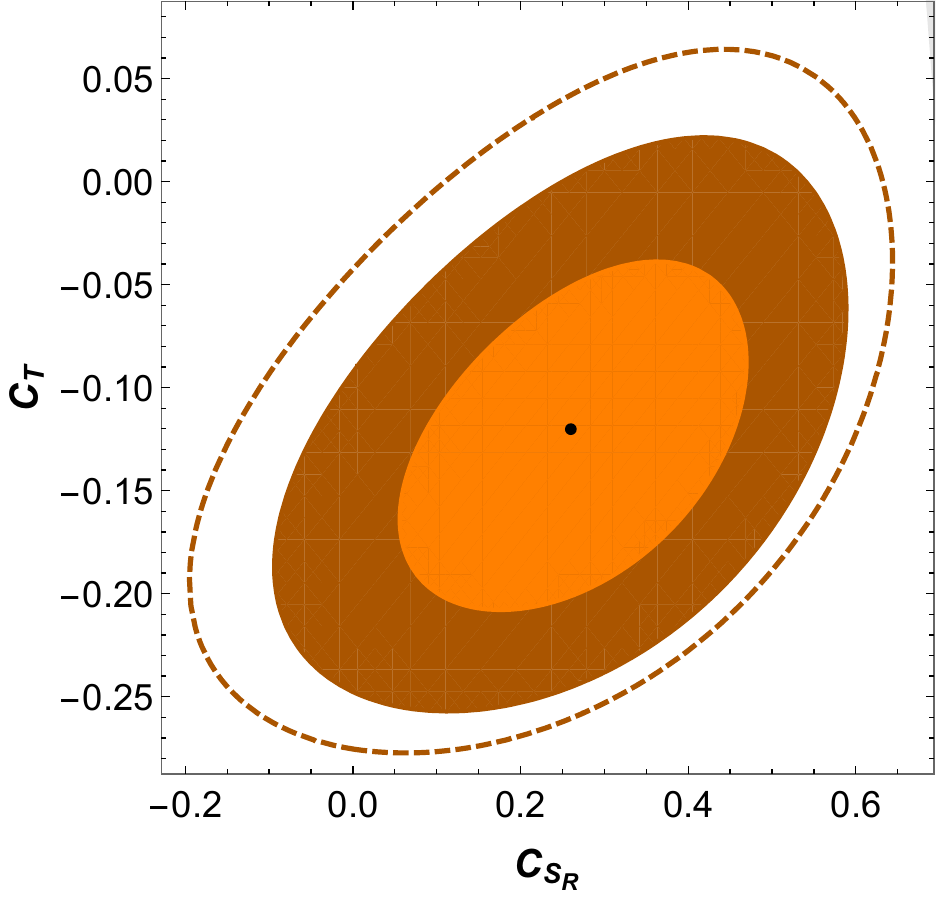}}\quad \subfloat[]{\includegraphics[width=4.5cm, height=4.2cm]{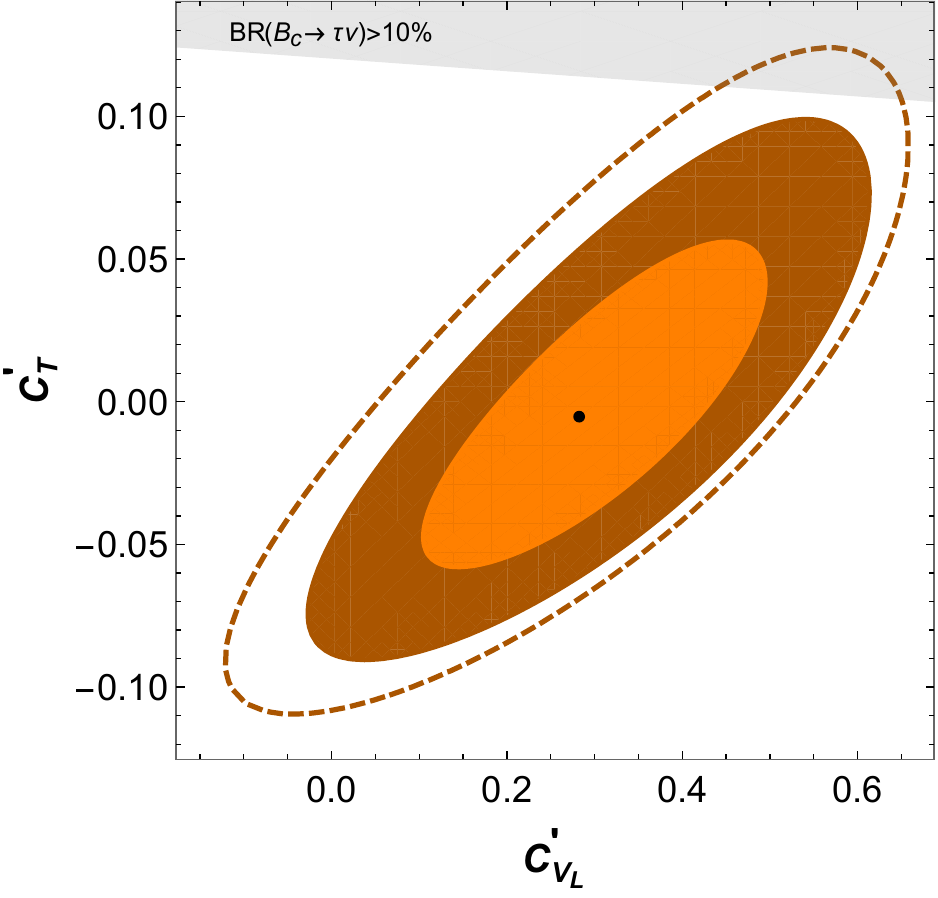}}\quad \subfloat[]{\includegraphics[width=4.5cm, height=4.2cm]{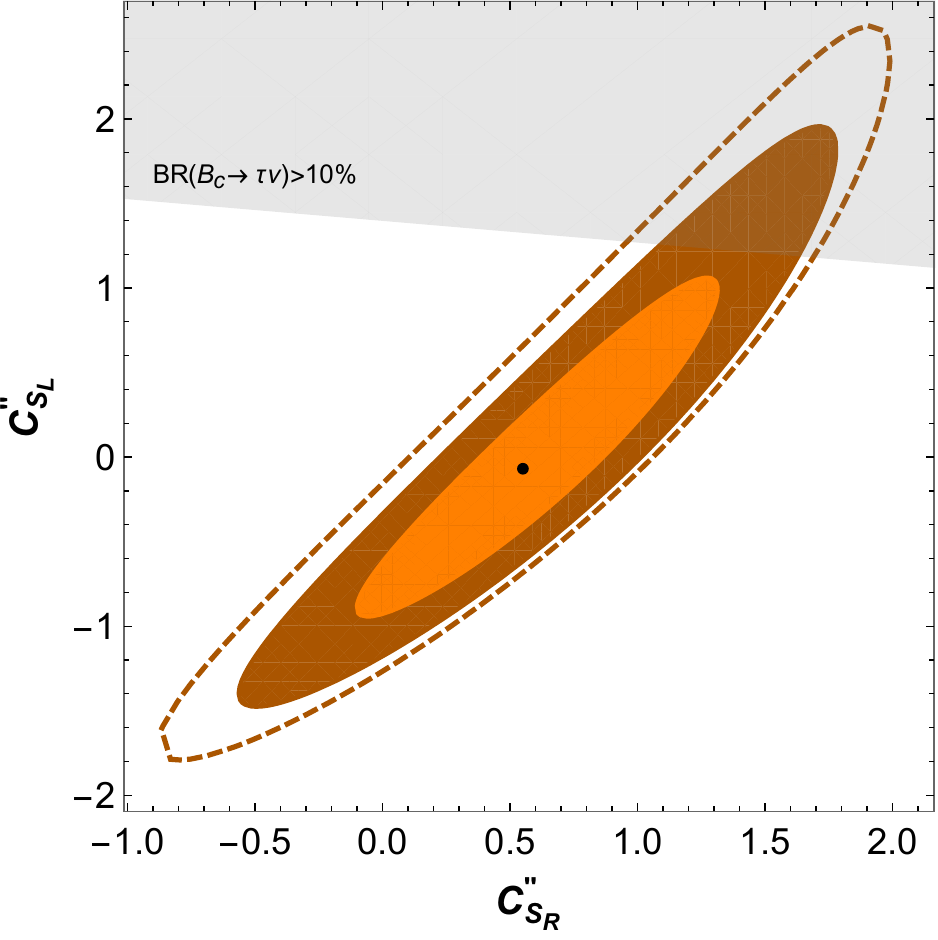}} 
\centering{} \subfloat[]{\includegraphics[width=4.5cm, height=4.2cm]{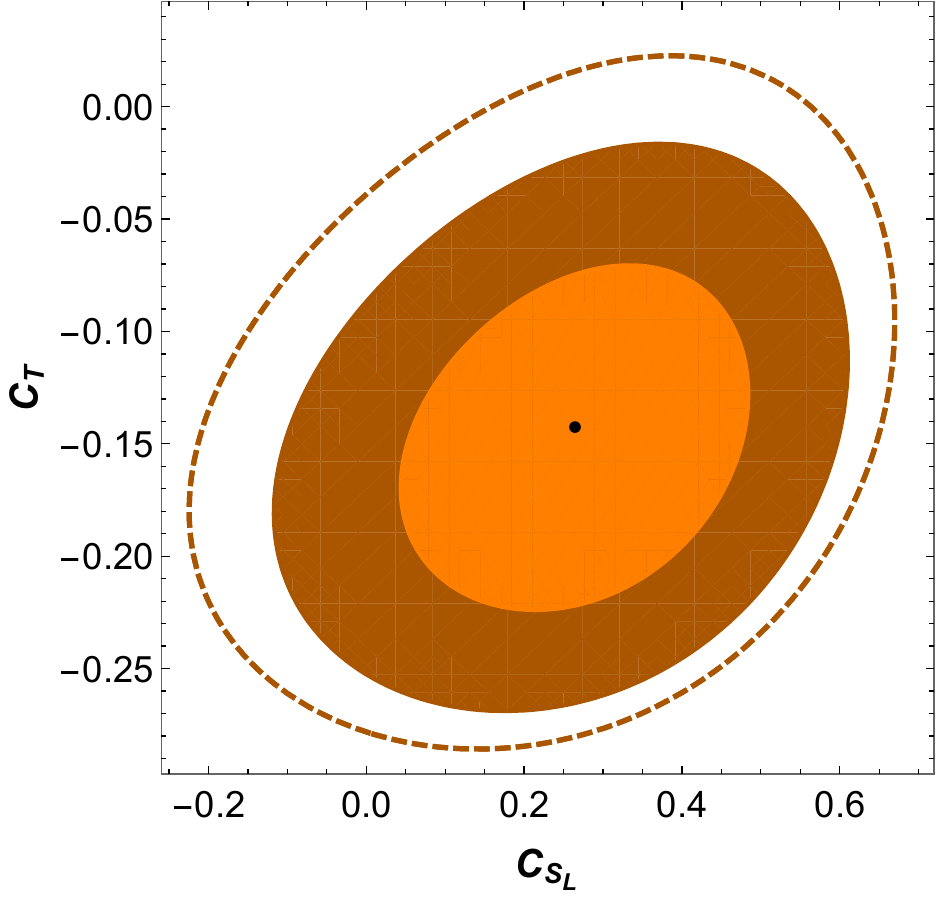}}

\caption{\label{s2d-fig} Results of the fits for NP scenarios at scale $2\;\text{TeV}$. The light and dark gray colors show the $10\%$ and $60\%$ branching ratio constraints. The light (dark) color contours represent the $1(2\sigma)$ deviations from the BFP (Black color)  for set $\mathcal{S}_{1}$and dashed contour
represents maximum of $2\sigma$ deviations for set $\mathcal{S}_{2}$. Panels (a), (b), (c), (f), (g), (j) for unprimed WCs;  (d), (h) for primed WCs; (e) and (i) for double primed WCs show the ranges of two-dimensional scenarios. Except in Fig. 1a, the orange color is not affected by either of these constraints, whereas in Figs. 1a, the red and green colors are the $60\%$ and $10\%$ constraints, respectively. Outside the dashed ellipse, the purple-shaded region represents the exclusion region due to collider bounds for the current luminosity of  $139\;\text{fb}^{-1}$.}
\end{figure}

For the two-dimensional cases, the corresponding BFPs results, $\chi_{\text{min}}^{2}$, $p-\text{value}\%$, and $\text{pull}_{\text{SM}}$ are summarized in Table \ref{2d-table-1} and their $(1 -2)\sigma$ ranges are depicted in Fig. \ref{s2d-fig} where the solid (dashed) contours represent the sets $\mathcal{S}_1$ and $\mathcal{S}_2$, respectively. The orange-colored contours represent the WCs that are not affected by $\mathcal{B}\left(B^-_c\to \tau^- \bar{\nu}_\tau\right)$ constraints. In contrast, the red and green colors show the effects of $60\%$ and $10\%$ constraints, respectively. Moreover, the $10\%$ and $60\%$ constraints of $\mathcal{B}\left(B^-_c\to \tau^- \bar{\nu}_\tau\right)$ are incorporated as light and dark gray colors, respectively. Any point falling inside the grey bands is considered to be excluded by the branching ratio constraints where one can also observed from Fig. \ref{s2d-fig} that scenario $\left(C_{S_{L}},C_{S_{R}}\right)$ is very sensitive to the these constraints. On the other hand, the scenarios $\left(C_{V_{L}}^{\prime},C_{T}^{\prime}\right)$ and $\left(C_{S_{L}}^{\prime},C_{T}^{\prime}\right)$ are dependent for $\mathcal{B}<60\%$ constraint; whereas the constraint $\mathcal{B}<30\%, 10\%$ has a negligible impact on the results. $(1, 2)\sigma$ intervals are derived
numerically from the chi-squared profiles, and do not assume symmetric
Gaussian behavior. Therefore, the scaling is not strictly $1:2$.
As for set $\mathcal{S}_{1}$ ($3$ dof), the $1\sigma$
and $2\sigma$ confidence regions correspond to $\chi{{}^2}$ thresholds
of $3.51$ and $8.01$, respectively. For set $\mathcal{S}_{2}$ ($4$
dof), the corresponding values are $4.7(1\sigma)$ and $9.69(2\sigma)$.
The apparent 2:3 ratio in the contour widths arises because the graphical
contours scale with the square root of the $\chi^2$
difference and not linearly. Although the $\chi^2$
thresholds follow $\sim1:2$ ratio for $1\sigma$ and $2\sigma$,
the contour sizes reflect $\sqrt{\Delta\chi2}$ , leading to the observed
$2:3$ scaling in the plots.
From Table \ref{2d-table-1}, we can see that the $\left(C_{S_{L}},C_{S_{R}}\right)$ scenario has the largest $p-\text{value}$ of almost $92\%$, and the maximum pull of $3.9$ among all scenarios for Set $\mathcal{S}_1$. However, when the branching ratio constraint of $(<30\%)$ is applied, then $p-\text{value}$ decreases to $\sim66\%$, which is further reduced to $\sim15\%$ by applying the branching ratio constraint of $<10\%$ . The $p-\text{value}$ of all the other scenarios of WCs (without branching constraints) is around $80\%$, except for three degenerate scenario involving $\left(C_{S_{L}},C_{T}\right)$, $\left(C_{S_{L}}^{\prime},C_{T}^{\prime}\right)$, and $\left(C_{S_{L}}^{\prime\prime},C_{T}^{\prime\prime}\right)$, which are connected by the Fierz transformation, showing somewhat moderate $p-\text{value}$ of $66\%$. The impact of branching ratio constraints on their $p-\text{value}$ is also listed in Table \ref{2d-table-1}. 

\begin{table}[H]
\centering{}%
\renewcommand{\arraystretch}{1.1}
\begin{tabular}{|c|c|c|c|c|c|c|}
\toprule 
\multicolumn{7}{c}{Set $\mathcal{S}_{1}$ ( $\chi_{\text{SM}}^{2}=15.12$, $p-\text{value}=4.46\times10^{-3}$
)}\tabularnewline
\multicolumn{7}{c}{Set $\mathcal{S}_{2}$ ( $\chi_{\text{SM}}^{2}=15.66$, $p-\text{value}=7.88\times10^{-3}$
)}\tabularnewline
\midrule
\midrule 
\hline\hline
WC & BFP & $\chi_{\text{min}}^{2}$ & $p-\text{value}$ $\%$ & $\text{pull}_{\text{SM}}$ & $1\sigma$-range & $2\sigma$-range\tabularnewline
\hline
\midrule
\midrule 
\multirow{3}{*}{$C_{V_{L}}$, $C_{V_{L}}^{'}$, $0.5C_{S_{R}}^{''}$} & $0.31$ & $0.48$ & $92.33$ & $3.93$ & $\left[0.16,0.46\right]$ & $\left[0.08,0.53\right]$\tabularnewline
 & $0.30$ & $2.23$ & $69.29$ & $3.66$ & $\left[0.12,0.47\right]$ & $\left[0.05,0.54\right]$\tabularnewline
\cmidrule{2-7} \cmidrule{3-7} \cmidrule{4-7} \cmidrule{5-7} \cmidrule{6-7} \cmidrule{7-7} 
 & $-5\%$ & $1.75$ & $-25\%$ & $-0.16$ & $15\%$ & $9\%$\tabularnewline
 \hline
\midrule 
\multirow{3}{*}{$C_{S_{L}}^{\prime\prime}$} & $-0.78$ & $2.18$ & $53.67$ & $3.60$ & $\left[-1.13,-0.39\right]$ & $\left[-1.31,-0.18\right]$\tabularnewline
 & $-0.74$ & $3.63$ & $45.87$ & $3.47$ & $\left[-1.16,-0.29\right]$ & $\left[-1.33,-0.08\right]$\tabularnewline
\cmidrule{2-7} \cmidrule{3-7} \cmidrule{4-7} \cmidrule{5-7} \cmidrule{6-7} \cmidrule{7-7} 
 & $-4\%$ & $1.45$ & $-15\%$ & $-0.13$ & $16\%$ & $10\%$\tabularnewline
 \hline
\midrule 
\multirow{3}{*}{$C_{T}$} & $-0.17$ & $4.02$ & $25.91$ & $3.33$ & $\left[-0.26,-0.08\right]$ & $\left[-0.30,-0.03\right]$\tabularnewline
 & $-0.17$ & $5.41$ & $24.74$ & $3.20$ & $\left[-0.26,-0.06\right]$ & $\left[-0.30,0\right]$\tabularnewline
\cmidrule{2-7} \cmidrule{3-7} \cmidrule{4-7} \cmidrule{5-7} \cmidrule{6-7} \cmidrule{7-7} 
 & $0\%$ & $1.39$ & $-5\%$ & $-0.13$ & $16\%$ & $10\%$\tabularnewline
 \hline
\midrule
\multirow{3}{*}{$C_{S_{R}}$} & $0.40$ & $5.08$ & $16.63$ & $3.17$ & $\left[0.17,0.61\right]$ & $\left[0.05,0.71\right]$\tabularnewline
 & $0.39$ & $6.27$ & $17.99$ & $3.06$ & $\left[0.12,0.62\right]$ & $\left[-0.01,0.72\right]$\tabularnewline
\cmidrule{2-7} \cmidrule{3-7} \cmidrule{4-7} \cmidrule{5-7} \cmidrule{6-7} \cmidrule{7-7} 
 & $-4\%$ & $1.19$ & $8\%$ & $-0.10$ & $16\%$ & $10\%$\tabularnewline
 \hline
\midrule
\multirow{3}{*}{$C_{T}^{\prime}$} &  $-0.07$ & $5.70$ & $12.70$ & $3.07$ & $\left[-0.11,-0.03\right]$ & $\left[-0.12,-0.01\right]$\tabularnewline
 & $-0.07$ & $6.83$ & $14.50$ & $2.97$ & $\left[-0.11,-0.02\right]$ & $\left[-0.12,0\right]$\tabularnewline
\cmidrule{2-7} \cmidrule{3-7} \cmidrule{4-7} \cmidrule{5-7} \cmidrule{6-7} \cmidrule{7-7} 
 & $0\%$ & $1.13$ & $14\%$ & $-0.10$ & $16\%$ & $100\%$\tabularnewline
 \hline
\midrule
\multirow{3}{*}{$C_{S_{L}}$} & $0.38$ & $8.51$ & $3.66$ & $2.57$ & $\left[0.11,0.62\right]$ & $\left[-0.04,0.74\right]$\tabularnewline
 & $0.37$ & $9.43$ & $5.12$ & $2.50$ & $\left[0.05,0.65\right]$ & $\left[-0.10,0.76\right]$\tabularnewline
\cmidrule{2-7} \cmidrule{3-7} \cmidrule{4-7} \cmidrule{5-7} \cmidrule{6-7} \cmidrule{7-7} 
 & $-3\%$ & $0.92$ & $40\%$ & $-0.07$ & $16\%$ & $10\%$\tabularnewline
 \hline
\midrule
\multirow{3}{*}{$C_{T}^{\prime\prime}$} & $-0.05$ & $11.53$ & $0.92$ & $1.89$ & $\left[-0.10,0\right]$ & $\left[-0.12,0.03\right]$\tabularnewline
 & $-0.05$ & $12.26$ & $1.55$ & $1.85$ & $\left[-0.10,0.01\right]$ & $\left[-0.12,0.04\right]$\tabularnewline
\cmidrule{2-7} \cmidrule{3-7} \cmidrule{4-7} \cmidrule{5-7} \cmidrule{6-7} \cmidrule{7-7} 
 & $0\%$ & $0.72$ & $70\%$ & $-0.05$ & $16\%$ & $11\%$\tabularnewline
 \hline
\midrule
\multirow{3}{*}{$C_{S_{L}}^{\prime}$} & $-0.06$ & $15.09$ & $0.17$ & $0.18$ & $\left[-0.68,0.61\right]$ & $\left[-0.97,0.98\right]$\tabularnewline
 & $-0.06$ & $15.63$ & $0.36$ & $0.19$ & $\left[-0.77,0.72\right]$ & $\left[-1.05,1.08\right]$\tabularnewline
\cmidrule{2-7} \cmidrule{3-7} \cmidrule{4-7} \cmidrule{5-7} \cmidrule{6-7} \cmidrule{7-7} 
 & $0\%$ & $0.54$ & $105\%$ & $0.01$ & $15\%$ & $9\%$\tabularnewline
 \hline\hline
\bottomrule
\end{tabular}
\caption{\label{1d-table} The results of the fit for real WCs which include BFPs, $\chi_{\text{min}}^{2}$,\; $p-\text{value}$ $\%$,\; $\text{pull}_{\text{SM}}$,\; $1\sigma$ and $2\sigma$-ranges of the corresponding WCs are presented here. These results are obtained with constraints on $\mathcal{B}\left(B_{c}^{-}\to\tau^{-}\bar{\nu}_{\tau}\right) <60\%,30\%$ and $<10\%$. It is important to note that these results are independent of the selection of three different limits on $\mathcal{B}\left(B_{c}^{-}\to\tau^{-}\bar{\nu}_{\tau}\right)$
for both sets of observables, i.e., $\mathcal{S}_{1}$ ($3$ dof) and $\mathcal{S}_{2}$ ($4$ dof)}. In each sub-row of WCs, the first, second, and third rows provide data for $\mathcal{S}_{1},\mathcal{S}_{2}$, and the difference for $\chi_{\text{min}}^{2}$, $\text{pull}_{\text{SM}}$ and percentage difference for BFP, $p-\text{value}$ $\%$,\; $1\sigma$ and $2\sigma$-ranges  of $\mathcal{S}_{2}$ compared to $\mathcal{S}_{1}$.
\end{table}

\begin{table}[H]
\centering{}%
\renewcommand{\arraystretch}{1.35}
\begin{tabular}{|c|c|c|c|c|c|}
\toprule 
\multicolumn{6}{c}{Set $\mathcal{S}_{1}$ ( $\chi_{\text{SM}}^{2}=16.51$, $p-\text{value}=2.41\times10^{-3}$
)}\tabularnewline
\multicolumn{6}{c}{Set $\mathcal{S}_{2}$ ( $\chi_{\text{SM}}^{2}=20.92$, $p-\text{value}=8.38\times10^{-4}$
)}\tabularnewline
\midrule
\midrule 
\hline\hline
WC & BR & BFP & $\chi_{\text{min}}^{2}$ & $p-\text{value}$ $\%$ & $\text{pull}_{\text{SM}}$\tabularnewline
\hline
\midrule
\midrule 
\multirow{9}{*}{$\left(C_{S_{L}},C_{S_{R}}\right)$} & \multirow{3}{*}{$<60\%$} & $\left(-1.07,1.28\right)$ & $0.17$ & $91.65$ & $3.87$\tabularnewline
 &  & $\left(-1.01,1.21\right)$ & $1.95$ & $37.69$ & $3.70$\tabularnewline
\cmidrule{3-6} \cmidrule{4-6} \cmidrule{5-6} \cmidrule{6-6} 
 &  & $5\%$ & $1.78$ & $-59\%$ & $-0.16$\tabularnewline
\cmidrule{2-6} \cmidrule{3-6} \cmidrule{4-6} \cmidrule{5-6} \cmidrule{6-6} 
 & \multirow{3}{*}{$<30\%$} & $\left(-0.71,0.98\right)$, & $0.84$ & $65.62$ & $3.78$\tabularnewline
 &  & $\left(-0.72,0.98\right)$ & $2.38$ & $30.38$ & $3.64$\tabularnewline
\cmidrule{3-6} \cmidrule{4-6} \cmidrule{5-6} \cmidrule{6-6} 
 &  & $1\%$ & $1.54$ & $-54\%$ & $-0.13$\tabularnewline
\cmidrule{2-6} \cmidrule{3-6} \cmidrule{4-6} \cmidrule{5-6} \cmidrule{6-6} 
 & \multirow{3}{*}{$<10\%$} & $\left(-0.18,0.53\right)$, & $3.79$ & $15.02$ & $3.37$\tabularnewline
 &  & $\left(-0.19,0.52\right)$, & $5.02$ & $8.15$ & $3.26$\tabularnewline
\cmidrule{3-6} \cmidrule{4-6} \cmidrule{5-6} \cmidrule{6-6} 
 &  & $2\%$ & $1.22$ & $-46\%$ & $-0.10$\tabularnewline
 \hline
\midrule
\multirow{3}{*}{$\left(C_{V_{L}},C_{S_{R}}\right)$} & \multirow{3}{*}{-} & $\left(0.29,0.05\right)$ & $0.43$ & $80.71$ & $3.83$\tabularnewline
 &  & $\left(0.27,0.05\right)$ & $2.17$ & $33.75$ & $3.67$\tabularnewline
\cmidrule{3-6} \cmidrule{4-6} \cmidrule{5-6} \cmidrule{6-6} 
 &  & $7\%$ & $1.74$ & $-58\%$ & $-0.16$\tabularnewline
\hline
\midrule 
\multirow{3}{*}{$\left(C_{V_{L}},C_{T}\right)$} & \multirow{3}{*}{-} & $\left(0.35,0.03\right)$ & $0.43$ & $80.68$ & $3.83$\tabularnewline
 &  & $\left(0.33,0.03\right)$ & $2.19$ & $33.46$ & $3.67$\tabularnewline
\cmidrule{3-6} \cmidrule{4-6} \cmidrule{5-6} \cmidrule{6-6} 
 &  & $-5\%$ & $1.76$ & $-59\%$ & $-0.16$\tabularnewline
 \hline
\midrule
\multirow{3}{*}{$\left(C_{V_{L}}^{\prime},C_{S_{L}}^{\prime}\right)$} & \multirow{3}{*}{-} & $\left(0.31,-0.04\right)$ & $0.46$ & $79.56$ & $3.83$\tabularnewline
 &  & $\left(0.30,-0.05\right)$ & $2.21$ & $33.18$ & $3.67$\tabularnewline
\cmidrule{3-6} \cmidrule{4-6} \cmidrule{5-6} \cmidrule{6-6} 
 &  & $5\%$ & $1.75$ & $-58\%$ & $-0.16$\tabularnewline
 \hline
 \midrule
\multirow{3}{*}{$\left(C_{S_{R}}^{\prime\prime},C_{T}^{\prime\prime}\right)$} & \multirow{3}{*}{-} & $\left(0.62,0\right)$ & $0.46$ & $79.35$ & $3.83$\tabularnewline
\cmidrule{3-6} \cmidrule{4-6} \cmidrule{5-6} \cmidrule{6-6} 
 &  & $\left(0.58,0\right)$ & $2.21$ & $33.12$ & $3.67$\tabularnewline
\cmidrule{3-6} \cmidrule{4-6} \cmidrule{5-6} \cmidrule{6-6} 
 &  & $6\%$ & $1.75$ & $-58\%$ & $-0.16$\tabularnewline
 \hline
\midrule 
\multirow{3}{*}{$\left(C_{V_{L}},C_{S_{L}}\right)$} & \multirow{3}{*}{-} & $\left(0.31,0.02\right)$ & $0.47$ & $79.22$ & $3.83$\tabularnewline
 &  & $\left(0.29,0.03\right)$ & $2.19$ & $33.46$ & $3.67$\tabularnewline
\cmidrule{3-6} \cmidrule{4-6} \cmidrule{5-6} \cmidrule{6-6} 
 &  & $6\%$ & $1.75$ & $-58\%$ & $-0.16$\tabularnewline
  \hline
\midrule
\multirow{3}{*}{$\left(C_{S_{R}},C_{T}\right)$} & \multirow{3}{*}{-} & $\left(0.27,-0.13\right)$ & $0.47$ & $79.16$ & $3.83$\tabularnewline
 &  & $\left(0.26,-0.12\right)$ & $2.13$ & $34.43$ & $3.68$\tabularnewline
\cmidrule{3-6} \cmidrule{4-6} \cmidrule{5-6} \cmidrule{6-6} 
 &  & $4\%$ & $1.67$ & $-57\%$ & $-0.15$\tabularnewline
  \hline
\midrule
\multirow{3}{*}{$\left(C_{V_{L}}^{\prime},C_{T}^{\prime}\right)$} & \multirow{3}{*}{-} & $\left(0.30,0\right)$ & $0.47$ & $79.07$ & $3.83$\tabularnewline
 &  & $\left(0.28,-0.01\right)$ & $2.22$ & $33.03$ & $3.67$\tabularnewline
\cmidrule{3-6} \cmidrule{4-6} \cmidrule{5-6} \cmidrule{6-6} 
 &  & $7\%$ & $1.75$ & $-58\%$ & $-0.16$\tabularnewline
 \hline
 \midrule
\multirow{3}{*}{$\left(C_{S_{L}}^{\prime\prime},C_{S_{R}}^{\prime\prime}\right)$} & \multirow{3}{*}{-} & $\left(-0.03,0.61\right)$ & $0.48$ & $78.74$ & $3.83$\tabularnewline
 &  & $\left(-0.06,0.55\right)$ & $2.22$ & $32.89$ & $3.67$\tabularnewline
\cmidrule{3-6} \cmidrule{4-6} \cmidrule{5-6} \cmidrule{6-6} 
 &  & $11\%$ & $1.75$ & $-58\%$ & $-0.16$\tabularnewline
 \hline
\midrule
\multirow{3}{*}{$\begin{array}{c}
\left(C_{S_{L}},C_{T}\right),\\
\left(-0.5C_{S_{L}}^{'}-6C_{T}^{'},-0.125C_{S_{L}}^{'}+0.5C_{T}^{'}\right),\\
\left(-0.5C_{S_{L}}^{''}-6C_{T}^{''},0.125C_{S_{L}}^{''}-0.5C_{T}^{''}\right)
\end{array}$} & \multirow{3}{*}{-} & $\left(0.27,-0.15\right)$ & $0.83$ & $65.99$ & $3.78$\tabularnewline
 &  & $\left(0.26,-0.14\right)$ & $2.45$ & $29.36$ & $3.63$\tabularnewline
\cmidrule{3-6} \cmidrule{4-6} \cmidrule{5-6} \cmidrule{6-6} 
 &  & $4\%$ & $1.62$ & $-56\%$ & $-0.15$\tabularnewline
\tabularnewline
 \hline\hline
\bottomrule
\end{tabular}\caption{\label{2d-table-1} The results of the two-dimensional fit for real WCs, including BFP, $\chi_{\text{min}}^{2}$,\; $p-\text{value}$ $\%$,\; $\text{pull}_{\text{SM}}$,\; $1\sigma$ and $2\sigma$-ranges of the corresponding WCs. These numbers are obtained by incorporating bounds on $\mathcal{B}\left(B_{c}^{-}\to\tau^{-}\bar{\nu}_{\tau}\right) <60\%,30\%$ and $<10\%$
for both sets of observables, i.e., $\mathcal{S}_{1}$ ($2$ dof) and $\mathcal{S}_{2}$ ($3$ dof)}. In each sub-row of WCs, the first, second, and third rows provide data for $\mathcal{S}_{1},\mathcal{S}_{2}$, and the difference for $\chi_{\text{min}}^{2}$, $\text{pull}_{\text{SM}}$ and percentage difference for BFP and $p-\text{value}$ of $\mathcal{S}_{2}$ compared to $\mathcal{S}_{1}$.
\end{table}
It is important to note from Tables \ref{1d-table} and \ref{2d-table-1}, where the BFPs and the parametric space for all NP scenarios are summarized, that
compared to $\mathcal{S}_1$ the value $\chi_{\text{min}}^{2}$ and the allowed parametric space of NP WCs are increased for $\mathcal{S}_2$, whereas the values of BFPs of NP WCs in both cases are close. Our analysis reveals that the difference in the best-fit values is no more than $10\%$; therefore, it does not significantly affect the parametric space allowed for $\mathcal{S}_1$. Consequently, we discuss the phenomenology only for the scenarios in $\mathcal{S}_1$ that have $\chi_{\text{min}}^{2} \leq 1$ and single WCs in degenerate scenarios.

In our global fits, the SM point $\left(0,0\right)$ lies outside the $1\sigma\left(68\right)\%$ and $2\sigma\left(95\right)\%$ contours mainly due to the $3\sigma$ tension in $R_{\tau/{\mu,e}}\left(D^{*}\right)$, which strongly affects the total $\chi2$ because of their precise measurements. To demonstrate the impact of $R_{\tau/{\mu,e}}\left(D^{*}\right)$ on the fit, we performed a separate fit excluding these observables. In Table III, the results of the fit which includes $P_{\tau}\left(D^{*}\right)$ and $ F_{L}\left(D^{*}\right)$ ($1$ dof) for real WCs showing BFPs, $\chi_{\text{min}}^{2}$,\; $p-\text{value}$ $\%$,\; $\text{pull}_{\text{SM}}$,\; $1\sigma$ and $2\sigma$-ranges of the corresponding WCs are presented. These results are obtained with constraints on $\mathcal{B}\left(B_{c}^{-}\to\tau^{-}\bar{\nu}_{\tau}\right) <60\%,30\%$ and $<10\%$. It is important to note that these results are independent of the selection of three different limits on $\mathcal{B}\left(B_{c}^{-}\to\tau^{-}\bar{\nu}_{\tau}\right)$
for all observables. A BFP (e.g., $C_{V_{L}} = -0.99$) shows slight tensions in multiple observables, but the corresponding pulls are small $(0.5-0.56\sigma)$, and $\chi^{2}$ values are very low $(0.05-0.32\sigma)$, leading to $p-values > 80-90\%$, which means the fit is still consistent with the SM. Furthermore, the $1\sigma$ interval widens, e.g., for $C_{V_{L}}$ in the range $[-2.03, 1.54]$ now includes the SM $\left(0,0\right)$ point. This reflects weaker constraints as the fit is made without including $R_{\tau/{\mu,e}}\left(D^{\left(*\right)}\right)$. This is consistent for all other observables in both one-dimensional and two-dimensional scenarios, which are depicted in  Tables III and IV. For two-dimensional scenarios, We use three observables, i.e. $P_{\tau}\left(D^{*}\right)$, $ F_{L}\left(D^{*}\right)$ and $R_{\tau/{\ell}}\left(\Lambda_{c}\right)$ for the 2D fit to ensure $N_{dof}=3-2=1$, as using only two observables with two parameters would give zero dof, making the fit exactly constrained and statistically meaningless.

\begin{table}[h]
\centering{}
\renewcommand{\arraystretch}{1}
\scalebox{1}{
\begin{tabular}{|c|c|c|c|c|c|c|}
\toprule 
\multicolumn{7}{c}{$\chi_{\text{SM}}^{2}=0.32$, $p-\text{value}=8.53\times10^{-1}$
}\tabularnewline
\midrule
\hline\hline 
WC & BFP & $\chi_{\text{min}}^{2}$ & $p-\text{value}$ $\%$ & $\text{pull}_{\text{SM}}$ & $1\sigma$-range & $2\sigma$-range\tabularnewline
\hline 
\hline 
$C_{V_{L}}$ & $-0.99$ & $0.05$ & $82.85$ & $0.52$ & $\left[-2.03,1.54\right]$ & $\left[-2.61,16.89\right]$\tabularnewline
\hline 
$C_{S_{L}}^{''}$ & $1.36$ & $0.02$ & $87.48$ & $0.54$ & $\left[-1.19,4.36\right]$ & $\left[-4.28,16.08\right]$\tabularnewline
\hline 
$C_{T}$ & $0.45$ & $0.07$ & $78.78$ & $0.49$ & $\left[-0.26,0.84\right]$ & $\left[-0.79,1.04\right]$\tabularnewline
\hline 
$C_{S_{R}}$ & $1.40$ & $0.$ & $96.78$ & $0.56$ & $\left[-1.18,3.87\right]$ & $\left[-24.01,6.70\right]$\tabularnewline
\hline 
$C_{T}^{'}$ & $0.17$ & $0.01$ & $91.43$ & $0.55$ & $\left[-0.15,0.44\right]$ & $\left[-0.72,0.72\right]$\tabularnewline
\hline 
$C_{S_{L}}$ & $-1.40$ & $0.$ & $96.02$ & $0.56$ & $\left[-3.73,1.24\right]$ & $\left[-6.22,19.12\right]$\tabularnewline
\hline 
$C_{T}^{''}$ & $0.40$ & $0.01$ & $91.16$ & $0.55$ & $\left[-0.28,1.61\right]$ & $\left[-1.09,24.43\right]$\tabularnewline
\hline 
$C_{S_{L}}^{'}$ & $-0.01$ & $0.32$ & $57.38$ & $0.$ & $\left[-3.79,5.51\right]$ & $\left[-5.62,13.12\right]$\tabularnewline
\hline 
\end{tabular}}
\caption{\label{table1RDs} The results of the one- dimension ($1$ dof) fit which includes $P_{\tau}\left(D^{*}\right)$ and $ F_{L}\left(D^{*}\right)$ for real WCs showing BFPs, $\chi_{\text{min}}^{2}$,\; $p-\text{value}$ $\%$,\; $\text{pull}_{\text{SM}}$,\; $1\sigma$ and $2\sigma$-ranges of the corresponding WCs are presented. These results are obtained with constraints on $\mathcal{B}\left(B_{c}^{-}\to\tau^{-}\bar{\nu}_{\tau}\right) <60\%,30\%$ and $<10\%$. It is important to note that these results are independent of the selection of three different limits on $\mathcal{B}\left(B_{c}^{-}\to\tau^{-}\bar{\nu}_{\tau}\right)$
for all observables.}s
\end{table}
\begin{table}[h]
\begin{centering}
\renewcommand{\arraystretch}{1.1}
\begin{tabular}{|c|c|c|c|c|c|}
\toprule 
\multicolumn{6}{c}{$\chi_{\text{SM}}^{2}=0.86$, $p-\text{value}=8.36\times10^{-1}$
}\tabularnewline
\midrule
\hline\hline 
WC & BR & BFP & $\chi_{\text{min}}^{2}$ & $p-\text{value}$ $\%$ & $\text{pull}_{\text{SM}}$\tabularnewline
\hline 
\hline 
$\left(C_{S_{L}},C_{S_{R}}\right)$ & $\begin{array}{c}
<60\%\\
\&30\%
\end{array}$ & $\left(-1.58,-0.43\right)$ & $0.07$ & $78.81$ & $0.89$\tabularnewline
\hline 
$\left(C_{V_{L}},C_{T}\right)$ & - & $\left(-0.04,0.36\right)$ & $0.07$ & $78.59$ & $0.88$\tabularnewline
\hline 
$\left(C_{V_{L}}^{'},C_{S_{L}}^{'}\right)$ & - & $\left(-0.48,0.32\right)$ & $0.15$ & $70.20$ & $0.84$\tabularnewline
\hline 
$\left(C_{S_{R}}^{''},C_{T}^{''}\right)$ & $\begin{array}{c}
<60\%\\
\&30\%
\end{array}$ & $\left(-0.82,0.27\right)$ & $0.$ & $99.13$ & $0.93$\tabularnewline
\hline 
$\left(C_{V_{L}},C_{S_{L}}\right)$ & $\begin{array}{c}
<60\%\\
\&30\%
\end{array}$ & $\left(-0.21,-1.14\right)$ & $0.01$ & $93.73$ & $0.92$\tabularnewline
\hline 
$\left(C_{S_{R}},C_{T}\right)$ & - & $\left(0.36,0.59\right)$ & $0.03$ & $87.23$ & $0.91$\tabularnewline
\hline 
$\left(C_{V_{L}}^{'},C_{T}^{'}\right)$ & $\begin{array}{c}
<60\%\\
\&30\%
\end{array}$ & $\left(-0.14,0.15\right)$ & $0.01$ & $90.34$ & $0.92$\tabularnewline
\hline 
$\left(C_{S_{L}}^{''},C_{S_{R}}^{''}\right)$ & - & $\left(1.30,-0.13\right)$ & $0.03$ & $87.07$ & $0.91$\tabularnewline
\hline 
$\left(C_{S_{L}},C_{T}\right)$ & - & $\left(-0.79,0.17\right)$ & $0.03$ & $86.40$ & $0.91$\tabularnewline
\hline 
\end{tabular}
\par\end{centering}
\caption{\label{table2RDs} The results of the two- dimensions ($1$ dof) fit which includes $P_{\tau}\left(D^{*}\right)$, $ WeF_{L}\left(D^{*}\right)$ and $R_{\tau/{\ell}}\left(\Lambda_{c}\right)$ for real WCs showing BFPs, $\chi_{\text{min}}^{2}$,\; $p-\text{value}$ $\%$,\; $\text{pull}_{\text{SM}}$ of the corresponding WCs are presented. These results are obtained with constraints on $\mathcal{B}\left(B_{c}^{-}\to\tau^{-}\bar{\nu}_{\tau}\right) <60\%,30\%$ and $<10\%$. Here, we have only presented the scenarios with higher branching constraints.}
\end{table}

\subsection{Observable circumstance at BFP}
In this work, we have considered one (two) dimensional scenario (where one (two)  NP WCs is non-zero at a time) to obtain the BFPs. Regarding this, it is useful to study the impact of the BFPs of all the NP WCs on the given observables and calculate their deviations from their experimental (central) values. The discrepancies between the experimental and predicted values are defined in units of $\sigma$ as follows: 
$dO_{i}=\frac{O_{i}^{\text{NP}}-O_{i}^{\text{exp.}}}{\sigma^{O_{i}^{\text{exp}}}}$. By using this definition, the results of the analysis are presented in Tables \ref{predbfp1d} and \ref{predbfp2d} for one and two-dimensional scenarios, respectively.\\
\begin{table}[h]
\centering{}
\renewcommand{\arraystretch}{1}
\scalebox{1}{
\begin{tabular}{|c|c|c|c|c|c|c|c|c|}
\hline 
WC & BFP & $R_{\tau/{\mu,e}}(D)$ & $R_{\tau/{\mu,e}}(D^*)$ & $P_{\tau}\left(D^{*}\right)$ & $F_{L}\left(D^{*}\right)$ & $R_{\tau/\ell}\left(\Lambda_{c}\right)$ & $R_{\tau/\mu}\left(J/\psi\right)$ & $P_{\tau}\left(D\right)$\tabularnewline
\hline 
\hline 
$C_{V_{L}}$ & $0.31$ & $\begin{array}{c}
0.34\\
-0.1\sigma
\end{array}$ & $\begin{array}{c}
0.29\\
+0.1\sigma
\end{array}$ & $\begin{array}{c}
-0.50\\
-0.2\sigma
\end{array}$ & $\begin{array}{c}
0.46\\
-0.6\sigma
\end{array}$ & $\begin{array}{c}
0.37\\
+1.3\sigma
\end{array}$ & $\begin{array}{c}
0.29\\
-1.7\sigma
\end{array}$ & $0.33$\tabularnewline
\hline 
$C_{S_{L}}^{''}$ & $-0.78$ & $\begin{array}{c}
0.36\\
+0.8\sigma
\end{array}$ & $\begin{array}{c}
0.27\\
-1.2\sigma
\end{array}$ & $\begin{array}{c}
-0.50\\
-0.2\sigma
\end{array}$ & $\begin{array}{c}
0.45\\
-0.8\sigma
\end{array}$ & $\begin{array}{c}
0.36\\
+1.2\sigma
\end{array}$ & $\begin{array}{c}
0.28\\
-1.7\sigma
\end{array}$ & $0.47$\tabularnewline
\hline 
$C_{T}$ & $-0.17$ & $\begin{array}{c}
0.29\\
-1.8\sigma
\end{array}$ & $\begin{array}{c}
0.29\\
+0.6\sigma
\end{array}$ & $\begin{array}{c}
-0.47\\
-0.2\sigma
\end{array}$ & $\begin{array}{c}
0.45\\
-0.8\sigma
\end{array}$ & $\begin{array}{c}
0.36\\
+1.2\sigma
\end{array}$ & $\begin{array}{c}
0.30\\
-1.7\sigma
\end{array}$ & $0.36$\tabularnewline
\hline 
$C_{S_{R}}$ & $0.40$ & $\begin{array}{c}
0.37\\
+1.1\sigma
\end{array}$ & $\begin{array}{c}
0.26\\
-2.4\sigma
\end{array}$ & $\begin{array}{c}
-0.47\\
-0.2\sigma
\end{array}$ & $\begin{array}{c}
0.47\\
-0.4\sigma
\end{array}$ & $\begin{array}{c}
0.35\\
+1.1\sigma
\end{array}$ & $\begin{array}{c}
0.26\\
-1.8\sigma
\end{array}$ & $0.46$\tabularnewline
\hline 
$C_{T}^{'}$ & $-0.07$ & $\begin{array}{c}
0.37\\
+1.0\sigma
\end{array}$ & $\begin{array}{c}
0.26\\
-2.4\sigma
\end{array}$ & $\begin{array}{c}
-0.52\\
-0.3\sigma
\end{array}$ & $\begin{array}{c}
0.45\\
-0.7\sigma
\end{array}$ & $\begin{array}{c}
0.35\\
+1.1\sigma
\end{array}$ & $\begin{array}{c}
0.26\\
-1.8\sigma
\end{array}$ & $0.47$\tabularnewline
\hline 
$C_{S_{L}}$ & $0.38$ & $\begin{array}{c}
0.36\\
+0.8\sigma
\end{array}$ & $\begin{array}{c}
0.25\\
-3.1\sigma
\end{array}$ & $\begin{array}{c}
-0.52\\
-0.3\sigma
\end{array}$ & $\begin{array}{c}
0.46\\
-0.7\sigma
\end{array}$ & $\begin{array}{c}
0.34\\
+1.0\sigma
\end{array}$ & $\begin{array}{c}
0.25\\
-1.8\sigma
\end{array}$ & $0.45$\tabularnewline
\hline 
$C_{T}^{''}$ & $-0.05$ & $\begin{array}{c}
0.35\\
+0.3\sigma
\end{array}$ & $\begin{array}{c}
0.25\\
-3.4\sigma
\end{array}$ & $\begin{array}{c}
-0.52\\
-0.3\sigma
\end{array}$ & $\begin{array}{c}
0.46\\
-0.6\sigma
\end{array}$ & $\begin{array}{c}
0.33\\
+0.9\sigma
\end{array}$ & $\begin{array}{c}
0.25\\
-1.9\sigma
\end{array}$ & $0.43$\tabularnewline
\hline 
$C_{S_{L}}^{'}$ & $-0.06$ & $\begin{array}{c}
0.30\\
-1.5\sigma
\end{array}$ & $\begin{array}{c}
0.25\\
-2.9\sigma
\end{array}$ & $\begin{array}{c}
-0.50\\
-0.2\sigma
\end{array}$ & $\begin{array}{c}
0.46\\
-0.5\sigma
\end{array}$ & $\begin{array}{c}
0.32\\
+0.7\sigma
\end{array}$ & $\begin{array}{c}
0.26\\
-1.8\sigma
\end{array}$ & $0.34$\tabularnewline
\hline 
\end{tabular}}
\caption{\label{predbfp1d}The BFPs of NP WCs in one-dimensional scenario and the values of possible observables at these points with their discrepancy from the experimental values expressed in multiples of $\sigma^{O_{i}^{exp}}$ for set $\mathcal{S}_1$. }
\end{table}

From Table \ref{predbfp1d}, we observe that the predicted and measured values of $R_{\tau/{\mu,e}}(D)$  generally differ by $\leq 1\sigma$. An exception arises for the WCs: $C_T$, $C_{S_R}$, and $C_{S_L}^\prime$. This discrepancy can be traced to the interference term in Eq. (\ref{eqn1}). Specifically, in $\widetilde{C}_T$, the contributions from both $C_T$ and $C_{S_L}^{\prime\prime}$ are positive. However, the interference terms associated with these coefficients exhibit opposing signs, leading to a small deviation. Therefore, one also expects the relatively small difference between calculated and measured values $R_{\tau/{\mu,e}}\left(D^{*}\right)$ (c.f Eq. (\ref{eqn2})) which can be seen in Table \ref{predbfp1d} where the values of $R_{\tau/{\mu,e}}\left(D^{*}\right)$ differ by $\geq 2\sigma$, except for $C_{V_{L}}$, $C_{S_{L}}^{\prime\prime}$, $C_{T}$. On the other hand, the $\tau$ polarization asymmetry, $P_{\tau}\left(D^*\right)$, deviates by only $0.3\sigma$ from its measured value for all the NP WCs while for $F_{L}\left(D^*\right)$, we can see that the agreement between the predicted and measured values lies only within $(0.4 - 0.8)\sigma$. In contrast to these observables, the values of $R_{\tau/\ell}\left(\Lambda_c\right)$ and $R_{\tau/\mu}\left(J/\psi\right)$ at BFPs can deviate more than $1\sigma$ from their experimental measurements. In the case of $R_{\tau/\ell}\left(\Lambda_c\right)$ this value lies within $(0.7 - 1.3)\sigma$, whereas, for $R_{\tau/\mu}\left(J/\psi\right)$ the deviations are by $(1.7 -1.9)\sigma$. Likewise, the predicted values of $P_{\tau}\left(D\right)$ at the best-fit value are greater than the SM value and lie within the $0.33 - 0.47$.

\begin{table}[h]
\begin{centering}
\renewcommand{\arraystretch}{1.1}
\begin{tabular}{|c|c|c|c|c|c|c|c|c|c|}
\hline 
WCs & BR & BFP & $R_{\tau/\mu,e}\left(D\right)$ & $R_{\tau/\mu,e}\left(D^{*}\right)$ & $P_{\tau}\left(D^{*}\right)$ & $F_{L}\left(D^{*}\right)$ & $R_{\tau/\ell}\left(\Lambda_{c}\right)$ & $R_{\tau/\mu}\left(J/\psi\right)$ & $P_{\tau}\left(D\right)$\tabularnewline
\hline 
\hline 
\multirow{3}{*}{$\left(C_{S_{L}},C_{S_{R}}\right)$} & $<60\%$ & $\left(-1.07,1.28\right)$ & $\begin{array}{c}
0.34\\
+0.1\sigma
\end{array}$ & $\begin{array}{c}
0.29\\
-0.1\sigma
\end{array}$ & $\begin{array}{c}
-0.33\\
+0.1\sigma
\end{array}$ & $\begin{array}{c}
0.51\\
+0.4\sigma
\end{array}$ & $\begin{array}{c}
0.37\\
+1.4\sigma
\end{array}$ & $\begin{array}{c}
0.29\\
-1.8\sigma
\end{array}$ & $0.42$\tabularnewline
\cline{2-10} \cline{3-10} \cline{4-10} \cline{5-10} \cline{6-10} \cline{7-10} \cline{8-10} \cline{9-10} \cline{10-10} 
 & $<30\%$ & $\left(-0.71,0.97\right)$ & $\begin{array}{c}
0.35\\
+0.4\sigma
\end{array}$ & $\begin{array}{c}
0.28\\
-1.0\sigma
\end{array}$ & $\begin{array}{c}
-0.38\\
+0\sigma
\end{array}$ & $\begin{array}{c}
0.50\\
+0.1\sigma
\end{array}$ & $\begin{array}{c}
0.36\\
+1.2\sigma
\end{array}$ & $\begin{array}{c}
0.28\\
-1.8\sigma
\end{array}$ & $0.43$\tabularnewline
\cline{2-10} \cline{3-10} \cline{4-10} \cline{5-10} \cline{6-10} \cline{7-10} \cline{8-10} \cline{9-10} \cline{10-10} 
 & $<10\%$ & $\left(-0.18,0.53\right)$ & $\begin{array}{c}
0.36\\
+0.8\sigma
\end{array}$ & $\begin{array}{c}
0.26\\
-2.1\sigma
\end{array}$ & $\begin{array}{c}
-0.45\\
-0.1\sigma
\end{array}$ & $\begin{array}{c}
0.48\\
-0.3\sigma
\end{array}$ & $\begin{array}{c}
0.35\\
+1.1\sigma
\end{array}$ & $\begin{array}{c}
0.27\\
-1.9\sigma
\end{array}$ & $0.45$\tabularnewline
\hline 
$\left(C_{V_{L}},C_{S_{R}}\right)$ & - & $\left(0.29,0.05\right)$ & $\begin{array}{c}
0.34\\
+0.1\sigma
\end{array}$ & $\begin{array}{c}
0.29\\
-0.1\sigma
\end{array}$ & $\begin{array}{c}
-0.49\\
-0.2\sigma
\end{array}$ & $\begin{array}{c}
0.46\\
-0.6\sigma
\end{array}$ & $\begin{array}{c}
0.37\\
+1.3\sigma
\end{array}$ & $\begin{array}{c}
0.29\\
-1.8\sigma
\end{array}$ & $0.35$\tabularnewline
\hline 
$\left(C_{V_{L}},C_{T}\right)$ & - & $\left(0.35,0.03\right)$ & $\begin{array}{c}
0.34\\
+0.1\sigma
\end{array}$ & $\begin{array}{c}
0.29\\
-0.1\sigma
\end{array}$ & $\begin{array}{c}
-0.50\\
-0.2\sigma
\end{array}$ & $\begin{array}{c}
0.46\\
-0.6\sigma
\end{array}$ & $\begin{array}{c}
0.37\\
+1.3\sigma
\end{array}$ & $\begin{array}{c}
0.29\\
-1.8\sigma
\end{array}$ & $0.33$\tabularnewline
\hline 
$\left(C_{V_{L}}^{'},C_{S_{L}}^{'}\right)$ & - & $\left(0.31,-0.04\right)$ & $\begin{array}{c}
0.34\\
+0\sigma
\end{array}$ & $\begin{array}{c}
0.29\\
+0\sigma
\end{array}$ & $\begin{array}{c}
-0.50\\
-0.2\sigma
\end{array}$ & $\begin{array}{c}
0.46\\
-0.6\sigma
\end{array}$ & $\begin{array}{c}
0.37\\
+1.3\sigma
\end{array}$ & $\begin{array}{c}
0.29\\
-1.8\sigma
\end{array}$ & $0.34$\tabularnewline
\hline 
$\left(C_{S_{R}}^{''},C_{T}^{''}\right)$ & - & $\left(0.62,0\right)$ & $\begin{array}{c}
0.34\\
+0\sigma
\end{array}$ & $\begin{array}{c}
0.29\\
+0\sigma
\end{array}$ & $\begin{array}{c}
-0.50\\
-0.2\sigma
\end{array}$ & $\begin{array}{c}
0.46\\
-0.6\sigma
\end{array}$ & $\begin{array}{c}
0.37\\
+1.3\sigma
\end{array}$ & $\begin{array}{c}
0.29\\
-1.8\sigma
\end{array}$ & $0.34$\tabularnewline
\hline 
$\left(C_{V_{L}},C_{S_{L}}\right)$ & - & $\left(0.31,0.02\right)$ & $\begin{array}{c}
0.34\\
+0\sigma
\end{array}$ & $\begin{array}{c}
0.29\\
+0\sigma
\end{array}$ & $\begin{array}{c}
-0.50\\
-0.2\sigma
\end{array}$ & $\begin{array}{c}
0.46\\
-0.6\sigma
\end{array}$ & $\begin{array}{c}
0.37\\
+1.3\sigma
\end{array}$ & $\begin{array}{c}
0.29\\
-1.8\sigma
\end{array}$ & $0.34$\tabularnewline
\hline 
$\left(C_{S_{R}},C_{T}\right)$ & - & $\left(0.27,-0.13\right)$ & $\begin{array}{c}
0.34\\
+0.1\sigma
\end{array}$ & $\begin{array}{c}
0.29\\
-0.1\sigma
\end{array}$ & $\begin{array}{c}
-0.46\\
-0.2\sigma
\end{array}$ & $\begin{array}{c}
0.46\\
-0.7\sigma
\end{array}$ & $\begin{array}{c}
0.37\\
+1.3\sigma
\end{array}$ & $\begin{array}{c}
0.29\\
-1.8\sigma
\end{array}$ & $0.45$\tabularnewline
\hline 
$\left(C_{V_{L}}^{'},C_{T}^{'}\right)$ & - & $\left(0.30,0\right)$ & $\begin{array}{c}
0.34\\
-0.2\sigma
\end{array}$ & $\begin{array}{c}
0.29\\
+0.2\sigma
\end{array}$ & $\begin{array}{c}
-0.17\\
+0.4\sigma
\end{array}$ & $\begin{array}{c}
0.57\\
+1.6\sigma
\end{array}$ & $\begin{array}{c}
0.38\\
+1.6\sigma
\end{array}$ & $\begin{array}{c}
0.29\\
-1.8\sigma
\end{array}$ & $0.24$\tabularnewline
\hline 
$\left(C_{S_{R}}^{''},C_{S_{L}}^{''}\right)$ & - & $\left(0.61,-0.03\right)$ & $\begin{array}{c}
0.34\\
-0.1\sigma
\end{array}$ & $\begin{array}{c}
0.29\\
+0.1\sigma
\end{array}$ & $\begin{array}{c}
-0.50\\
-0.2\sigma
\end{array}$ & $\begin{array}{c}
0.46\\
-0.6\sigma
\end{array}$ & $\begin{array}{c}
0.37\\
+1.3\sigma
\end{array}$ & $\begin{array}{c}
0.29\\
-1.8\sigma
\end{array}$ & $0.34$\tabularnewline
\hline 
$\left(C_{S_{L}},C_{T}\right)$ & - & $\left(0.27,-0.15\right)$ & $\begin{array}{c}
0.34\\
+0\sigma
\end{array}$ & $\begin{array}{c}
0.29\\
-0.1\sigma
\end{array}$ & $\begin{array}{c}
-0.49\\
-0.2\sigma
\end{array}$ & $\begin{array}{c}
0.45\\
-0.9\sigma
\end{array}$ & $\begin{array}{c}
0.36\\
+1.3\sigma
\end{array}$ & $\begin{array}{c}
0.29\\
-1.8\sigma
\end{array}$ & $0.45$\tabularnewline
\hline 
\end{tabular}
\par\end{centering}
\caption{\label{predbfp2d} The BFPs of NP WCs in one-dimensional scenario and the values of possible observables at these points with their discrepancy from the experimental values expressed in multiples of $\sigma^{O_{i}^{exp}}$ for set $\mathcal{S}_1$.}
\end{table}
For two dimension scenarios, the situation is improved for most of the observables; \textit{e.g.,} the difference between the predicted and measured values of $R_{\tau/{\mu,e}}\left(D, D^*\right)$ differ by $\leq 0.2\sigma$, except for WCs $\left(C_{S_{L}}, C_{S_{R}}\right)$ with branching constraint $30\%$ and $10\%$. As one can see from Table. (\ref{predbfp2d}), the $\tau$ polarization asymmetry, $P_{\tau}\left(D^*\right)$, deviates from its measured value by $\leq0.2\sigma$ for all NP WCs, except for $\left(C_{V_{L}}^{\prime},C_{T}^{\prime}\right)$, where the difference is $0.4\sigma$. Similarly, for $F_{L}\left(D^*\right)$, we can see that the agreement between the predicted and measured values is $\leq0.9\sigma$, except for $\left(C_{V_{L}}^{\prime}, C_{T}^{\prime}\right)$, which gives $1.6\sigma$ mismatch. And similar to the case of one-dimensional scenarios, the other observables such as $R_{\tau/\ell}\left(\Lambda_c\right)$ and $R_{\tau/\mu}\left(J/\psi\right)$ in two-dimensional scenarios at the BFPs can deviate by more than $1\sigma$ from their experimental measurements. In the case of $R_{\tau/\ell}\left(\Lambda_c\right)$ this value lies within $(1.1 - 1.6)\sigma$, whereas, for $R_{\tau/\mu}\left(J/\psi\right)$ the deviations are by $(1.8, 1.9)\sigma$. Likewise, the predicted values of $P_{\tau}\left(D\right)$ at the best-fit value are greater than the SM value and lie within the $0.33 - 0.45$ range, except for $\left(C_{V_{L}}^{\prime}, C_{T}^{\prime}\right)$, which predicts $0.24$ . 

\subsection{Inclusion of $C_{V_{R}}$}
It is well-established that in the SMEFT, the right-handed vector operators \( \mathcal{O}_{V_R} \) for quarks does not appear at the dimension-six level, which allows us to ignore its contribution at this order. However, in our study, we find that through Fierz transformations, these operators mix with the right-handed scalar  $\left(\mathcal{O}^{\prime}_{S_{R}}\right)$ and the new left-handed vector $\left(\mathcal{O}^{\prime \prime}_{V_{L}}\right)$ operators. Therefore, we include them in the present study. As a result, the relevant parametric space for the corresponding WCs is given in Tables. \ref{global-t1} and \ref{global-t2} for one and two-dimensional NP scenarios after incorporating the relevant WC $\left(C_{V_{R}}\right)$ of $\mathcal{O}_{V_R}$ . The analysis reaffirms the interdependence of primed and unprimed WCs established by the Fierz identity, as previously discussed in Section \ref{sec3}.

In one-dimensional case, we presented the WCs in two groups in Table \ref{global-t1}, \textit{i.e.,} one with $-2C_{V_{R}}^{'}$, $-2C_{V_{R}}^{''}$, and $C_{S_{R}}$ and the other with $C_{V_{R}}$, $-0.5C_{S_{R}}^{'}$, and $-C_{V_{L}}^{''}$. However, for the latter case, the $p$-value of $0.85$ indicates a lack of statistical significance, coupled with a low $pull_{SM}=1.85$; therefore, we can ignore this scenario.

\begin{table}[H]
\begin{centering}
\renewcommand{\arraystretch}{1.1}
\begin{tabular}{|c|c|c|c|c|c|c|}
\hline 
WC & BFP & $\chi_{min}^{2}$ & $p-value$ $\%$ & $pull_{SM}$ & $1\sigma-$Range & $2\sigma-$Range\tabularnewline
\hline 
\hline  
$C_{S_{R}}$, $-2C_{V_{R}}^{'}$, $-2C_{V_{R}}^{''}$ & $0.40$ & $5.08$ & $16.63$ & $3.17$ & $[0.18,0.60]$ & $[0.04,0.70]$\tabularnewline
\hline 
$C_{V_{R}}$, $-0.5C_{S_{R}}^{'}$, $-C_{V_{L}}^{''}$ & $-0.26$ & $11.68$ & $0.85$ & $1.85$ & $[-0.52,0]$ & $[-0.64,0.14]$\tabularnewline
\hline 
\end{tabular}
\par\end{centering}
\centering{}\caption{\label{global-t1}Results of the fit for WCs, after inclusion of WC $C_{V_{R}}$ including all available
data of observables with $BR(B_{c}\rightarrow\tau\nu)<60\%$,
and $BR(B_{c}\rightarrow\tau\nu)<30\%$. In case no constraint on
$BR(B_{c}\rightarrow\tau\nu)<10\%$ is used, the fit is valid for
all three scenarios.}
\end{table}
Table \ref{global-t2} summarizes the results obtained for two-dimensional scenarios after including $C_{V_{R}}$, which extends the list of scenarios that are presented in Table {\ref{predbfp2d} to $13$ additional NP combinations. Among them WCs $\left(C_{V_{L}}^{'},-C_{V_{R}}^{'}\right)$,
$\left(0.5C_{S_{R}}^{''},-2C_{V_{R}}^{''}\right)$ are related to $\left(C_{V_{L}},C_{S_{R}}\right)$ which we have already discussed. By imposing a cut on $\chi_{min}^{2}<1$, we can exclude the combinations $\left(C_{V_{R}}^{''}, C_{S_{L}}^{''}\right)$, $\left(C_{V_{R}}, C_{T}\right)$, and $\left(C_{V_{R}}^{'}, C_{T}^{'}\right)$  which reduces the analysis to nine additional NP scenarios. 
For these scenarios, we have found that the parametric space of $\left(C_{V_{R}}^{''}, C_{T}^{''}\right)$ and $\left(C_{V_{R}}^{'},C_{S_L}^{'}\right)$ is influenced by $10\%$ branching ratio constraint that changes the BFP too (c.f. Table \ref{global-t2}). We also found that the branching ratio constraint reduces the  $p-values$  of $\left(C_{V_{R}}^{''}, C_{T}^{''}\right)$ and $\left(C_{V_{R}}^{'},C_{S_{L}}^{'}\right)$ scenarios from $99\%$ to $24\%$ and from $94\%$ to $58\%$, respectively, when branching ratio constraint $\mathcal{B}<10\%$ is used. However, the scenario $\left(C_{S_{L}}^{'}, C_{S_{R}}^{'}\right)$ and three interdependent scenarios related through Fierz identities (degenerate), \textit{i.e.,} $\left(C_{V_{R}}, C_{S_{R}}\right)$, $\left(-0.5C_{S_{R}}^{'},-2C_{V_{R}}^{'}\right)$, and
$\left(-C_{V_{L}}^{''},-C_{V_{R}}^{''}\right)$ which have explicit dependence on $C_{V_{R}}$ are independent of branching constraints. 
To enhance understanding, the accompanying $(1-2)\sigma$ contour plots are presented in Fig. \ref{Global-s2d-fig}.

\begin{table}[H]
\begin{centering}
\renewcommand{\arraystretch}{1.1}
\begin{tabular}{|c|c|c|c|c|c|}
\hline 
WC & BR & BFP & $\chi_{min}^{2}$ & $p-value$ $\%$ & $pull_{SM}$\tabularnewline
\hline 
\hline 
\multirow{2}{*}{$\left(C_{V_{R}}^{''},C_{T}^{''}\right)$} & $\begin{array}{c}
<60\%\\
\&30\%
\end{array}$ & $\left(-0.46,0.12\right)$ & $0$ & $99.97$ & $3.89$\tabularnewline
\cline{2-6} \cline{3-6} \cline{4-6} \cline{5-6} \cline{6-6} 
 & $<10\%$ & $\left(-0.26,0.03\right)$ & $2.83$ & $24.30$ & $3.51$\tabularnewline
\hline 
$\left(C_{S_{L}}^{'},C_{S_{R}}^{'}\right)$ & - & $\left(-1.76,1.85\right)$ & $0.07$ & $96.54$ & $3.88$\tabularnewline
\hline 
$\left(C_{V_{R}},C_{S_{R}}\right)$, $\left(-0.5C_{S_{R}}^{'},-2C_{V_{R}}^{'}\right)$,
$\left(-C_{V_{L}}^{''},-C_{V_{R}}^{''}\right)$ & - & $\left(-0.30,0.45\right)$ & \multirow{1}{*}{$0.08$} & \multirow{1}{*}{$96.15$} & \multirow{1}{*}{$3.88$}\tabularnewline
\hline 
\multirow{2}{*}{$\left(C_{V_{R}}^{'},C_{S_{L}}^{'}\right)$} & $\begin{array}{c}
<60\%\\
\&30\%
\end{array}$ & $\left(-0.31,0.76\right)$ & $0.12$ & $94.22$ & $3.87$\tabularnewline
\cline{2-6} \cline{3-6} \cline{4-6} \cline{5-6} \cline{6-6} 
 & $<10\%$ & $\left(0.24,0.47\right)$ & $1.10$ & $57.74$ & $3.74$\tabularnewline
\hline 
$\left(C_{V_{L}}^{''},C_{T}^{''}\right)$ & - & $\left(0.52,-0.10\right)$ & $0.28$ & $86.96$ & $3.85$\tabularnewline
\hline 
$\left(C_{V_{R}},C_{S_{L}}\right)$ & - & $\left(-0.40,0.53\right)$ & $0.38$ & $82.56$ & $3.84$\tabularnewline
\hline 
$\left(C_{V_{L}},C_{S_{R}}\right)$, $\left(C_{V_{L}}^{'},-C_{V_{R}}^{'}\right)$,
$\left(0.5C_{S_{R}}^{''},-2C_{V_{R}}^{''}\right)$ & - & $\left(0.29,0.05\right)$ & \multirow{1}{*}{$0.43$} & \multirow{1}{*}{$80.71$} & \multirow{1}{*}{$3.83$}\tabularnewline
\hline 
$\left(C_{V_{L}},C_{V_{R}}\right)$, $\left(C_{V_{L}}^{'},-0.5C_{S_{R}}^{'}\right)$,
$\left(0.5C_{S_{R}}^{''},-C_{V_{L}}^{''}\right)$ & - & $\left(0.32,0.01\right)$ & \multirow{1}{*}{$0.47$} & \multirow{1}{*}{$79.07$} & \multirow{1}{*}{$3.83$}\tabularnewline
\hline 
$\left(C_{S_{R}}^{'},C_{T}^{'}\right)$ & - & $\left(0.59,-0.08\right)$ & $0.48$ & $78.79$ & $3.83$\tabularnewline
\hline 
$\left(C_{V_{L}}^{''},C_{S_{L}}^{''}\right)$ & - & $\left(0.16,-0.74\right)$ & $0.63$ & $72.95$ & $3.81$\tabularnewline
\hline 
$\left(C_{V_{R}}^{''},C_{S_{L}}^{''}\right)$ & - & $\left(0.16,-1.27\right)$ & $1.41$ & $49.40$ & $3.70$\tabularnewline
\hline 
$\left(C_{V_{R}},C_{T}\right)$ & - & $\left(0.36,-0.28\right)$ & $1.47$ & $47.83$ & $3.78$\tabularnewline
\hline 
\multirow{3}{*}{$\left(C_{V_{R}}^{'},C_{T}^{'}\right)$} & $<60\%$ & $\left(-0.73,0.20\right)$ & $2.20$ & $33.22$ & $3.59$\tabularnewline
\cline{2-6} \cline{3-6} \cline{4-6} \cline{5-6} \cline{6-6} 
 & $<30\%$ & $\left(-0.50,0.11\right)$ & $3.40$ & $18.23$ & $3.42$\tabularnewline
\cline{2-6} \cline{3-6} \cline{4-6} \cline{5-6} \cline{6-6} 
 & $<10\%$ & $\left(-0.38,0.20\right)$ & $5.51$ & $6.36$ & $3.10$\tabularnewline
\hline 
\end{tabular}
\par\end{centering}
\centering{}\caption{\label{global-t2} Results of the fit for WCs, after inclusion of WC $C_{V_{R}}$ for two-dimensions including
all available data of observables $\mathcal{S}_1$ with $BR(B_{c}\rightarrow\tau \nu)<60\%$,
$BR(B_{c}\rightarrow\tau \nu)<30\%$. The constraint $BR(B_{c}\rightarrow\tau \nu)<10\%$ is explicitly mentioned where it is used. The dash in branching ratio column shows that the fit is valid for
all three scenarios.}
\end{table}

\begin{figure}[H]
\centering{} \subfloat[]{\includegraphics[width=4.5cm, height=4.2cm]{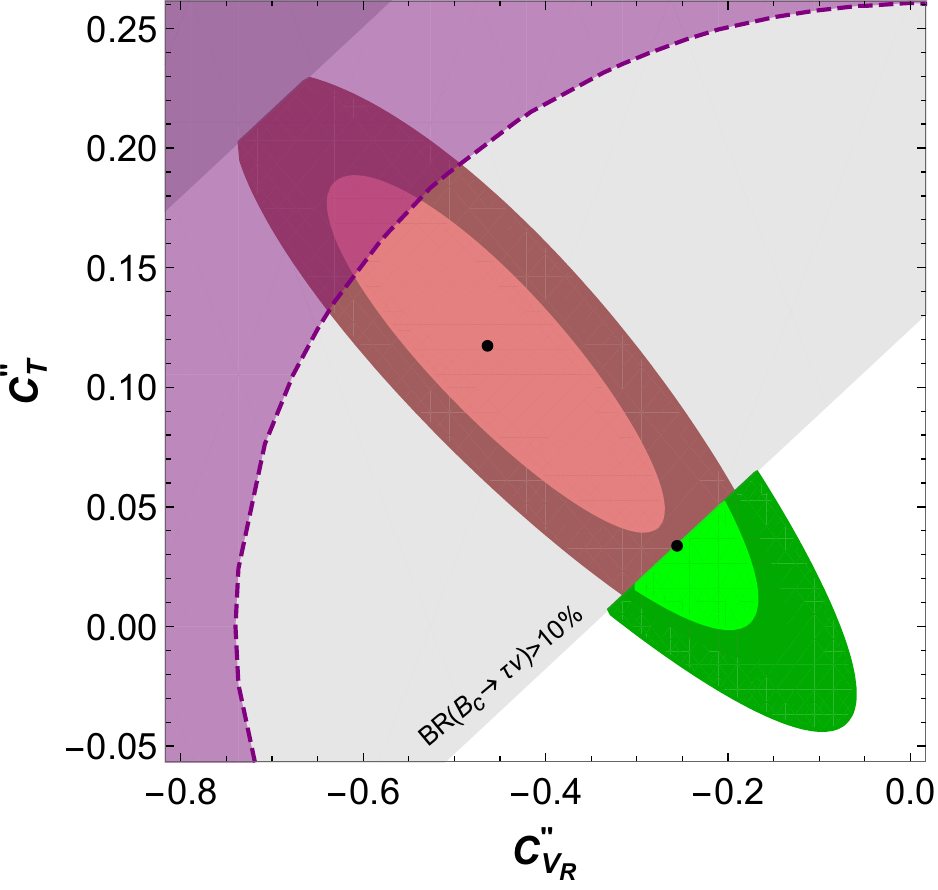}}\quad \subfloat[]{\includegraphics[width=4.5cm, height=4.2cm]{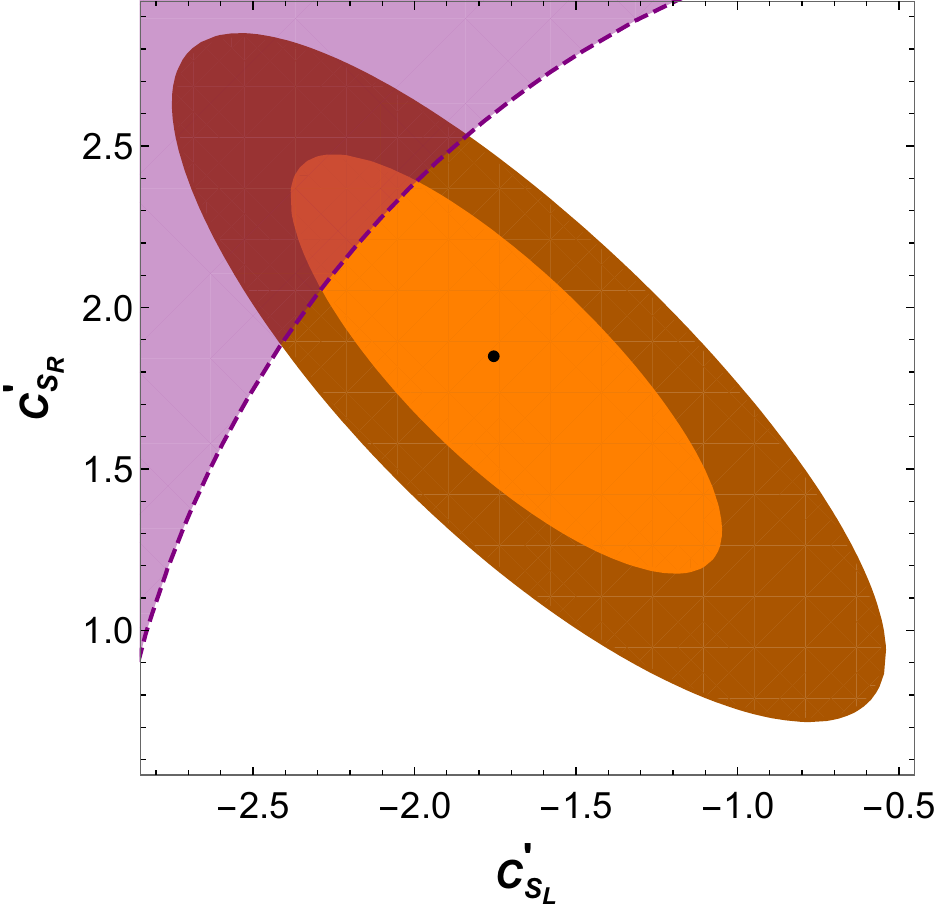}}\quad \subfloat[]{\includegraphics[width=4.5cm, height=4.2cm]{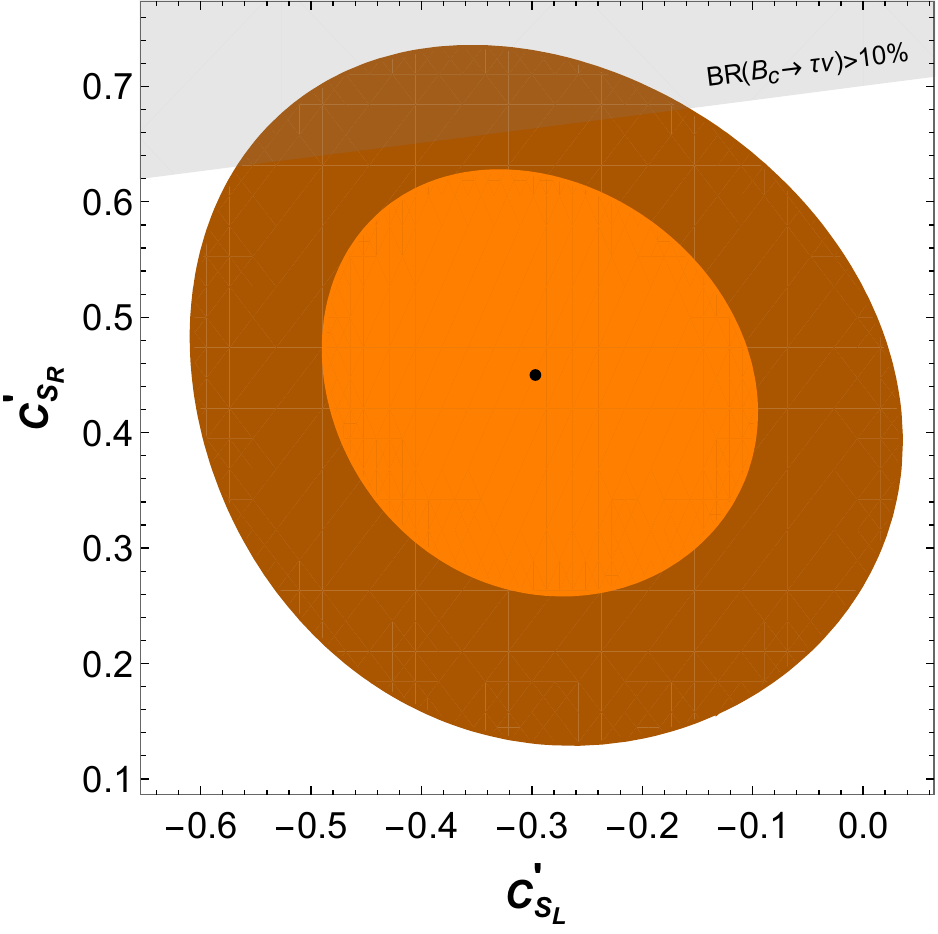}} 
\centering{} \subfloat[]{\includegraphics[width=4.5cm, height=4.2cm]{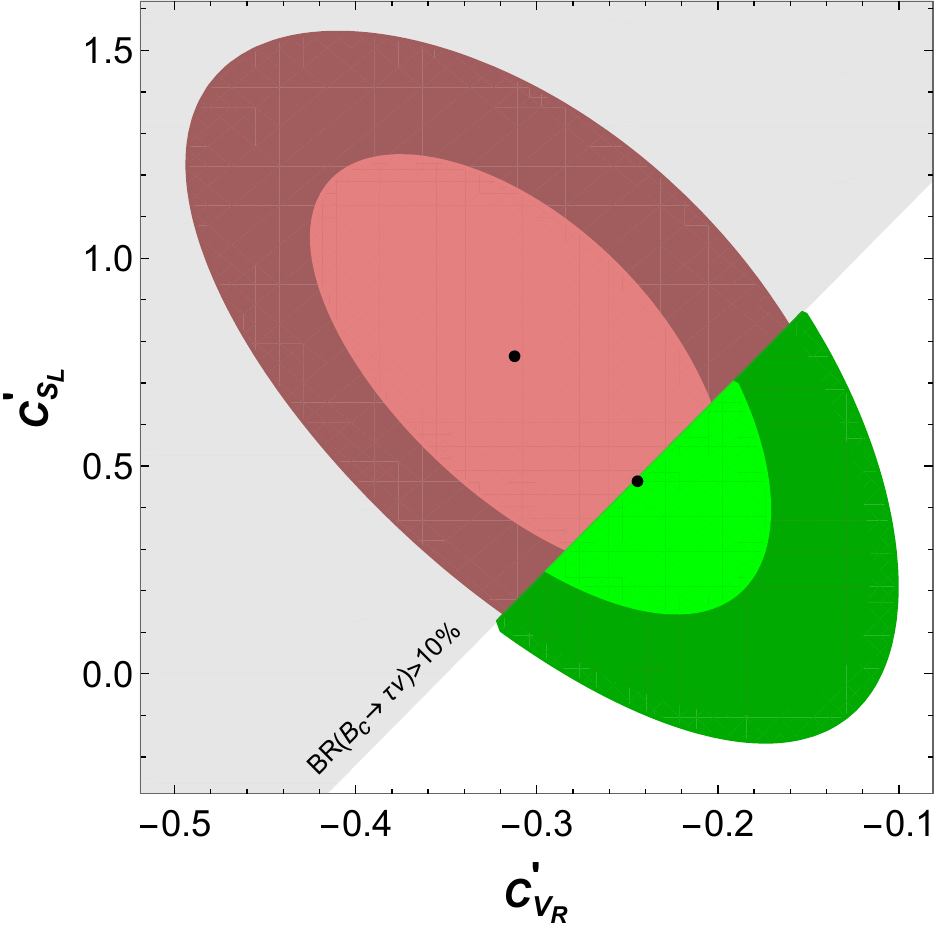}}\quad \subfloat[]{\includegraphics[width=4.5cm, height=4.2cm]{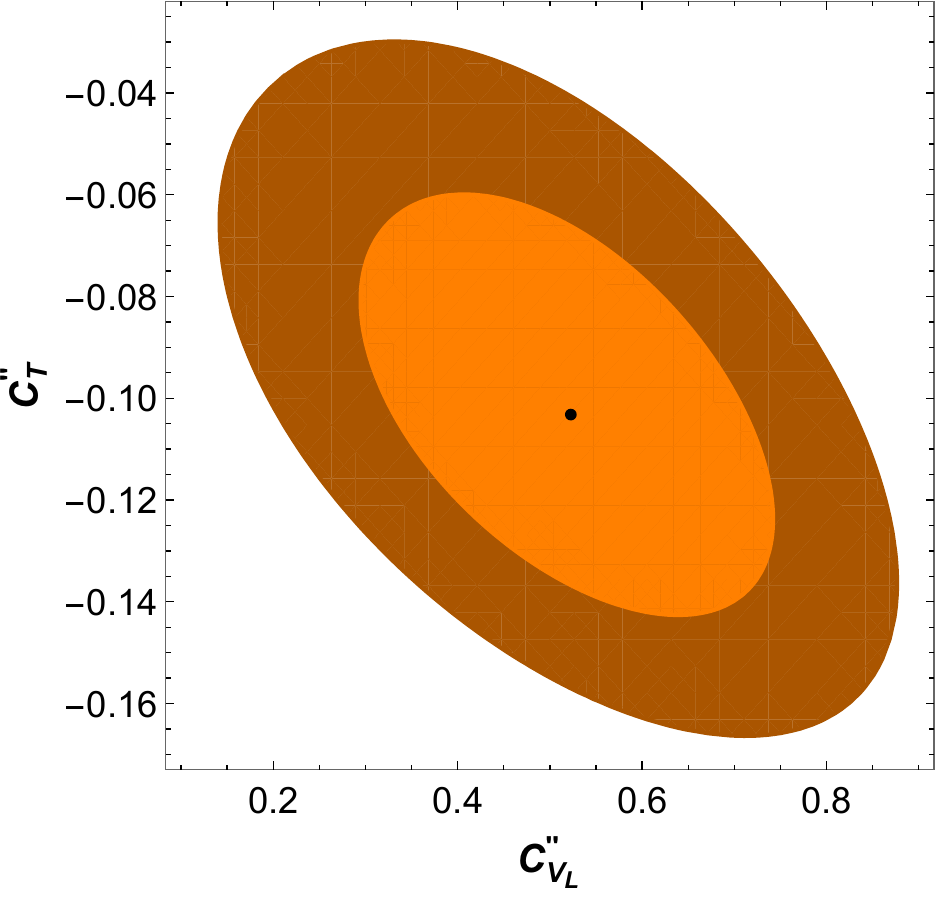}}\quad \subfloat[]{\includegraphics[width=4.5cm, height=4.2cm]{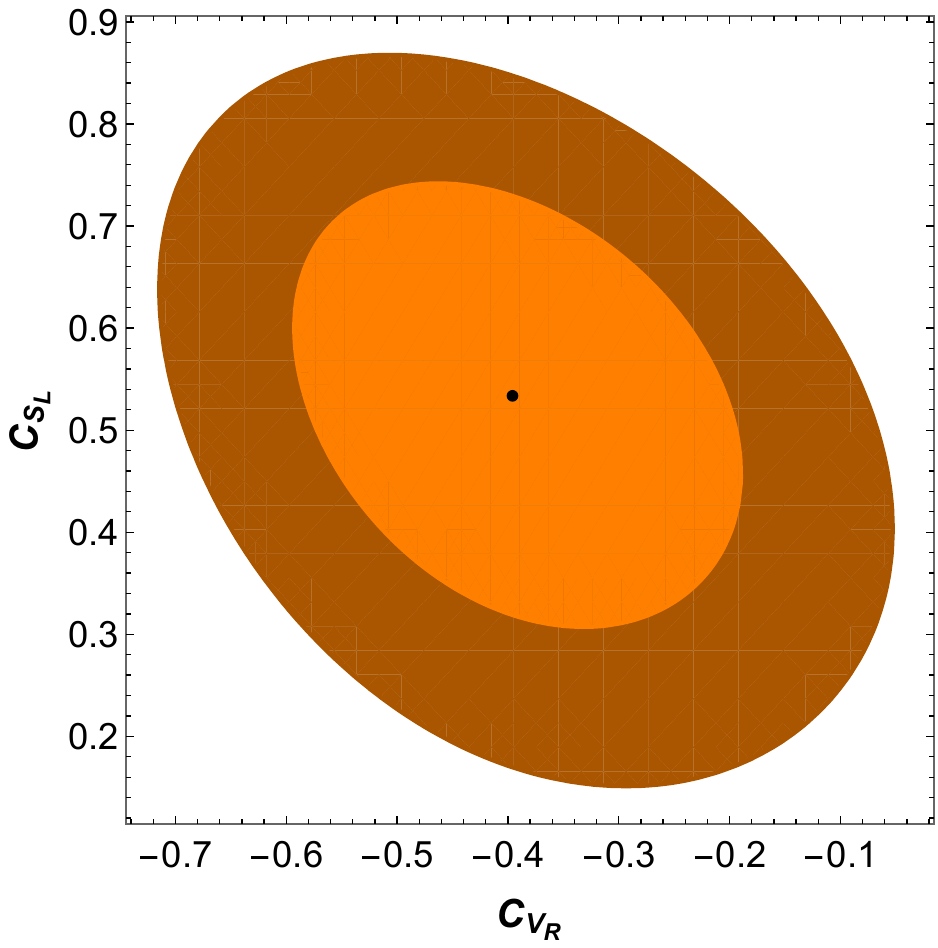}} 
\centering{} \subfloat[]{\includegraphics[width=4.5cm, height=4.2cm]{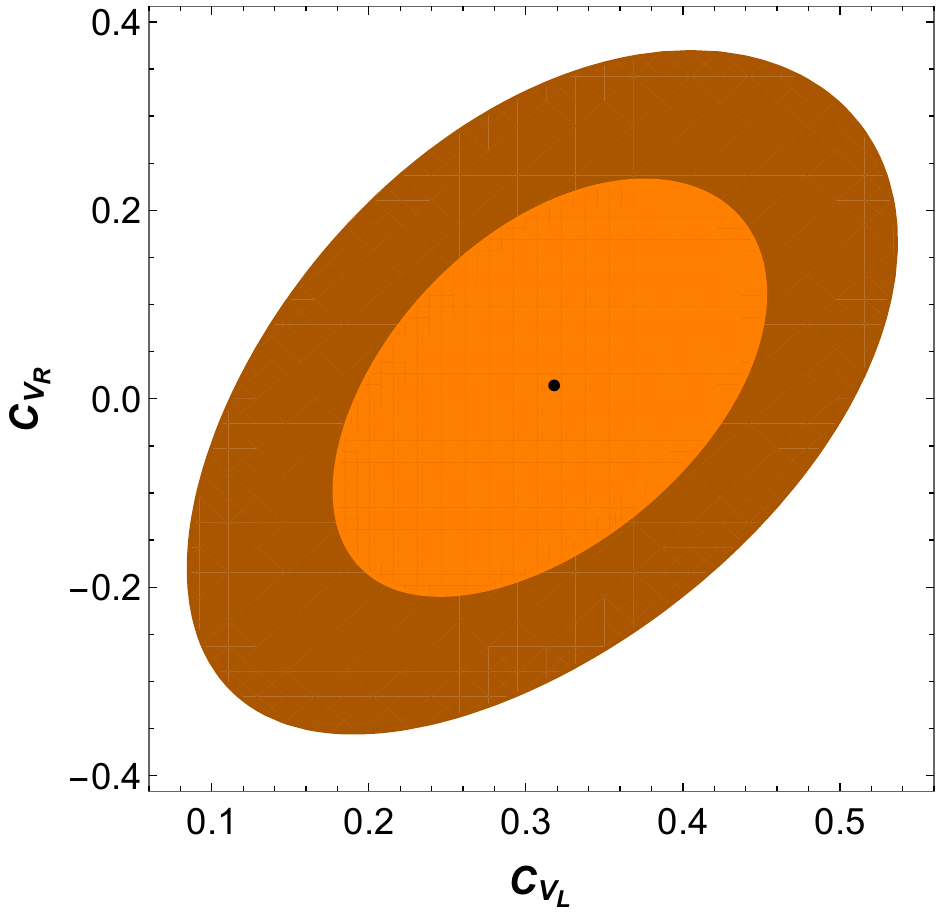}}\quad \subfloat[]{\includegraphics[width=4.5cm, height=4.2cm]{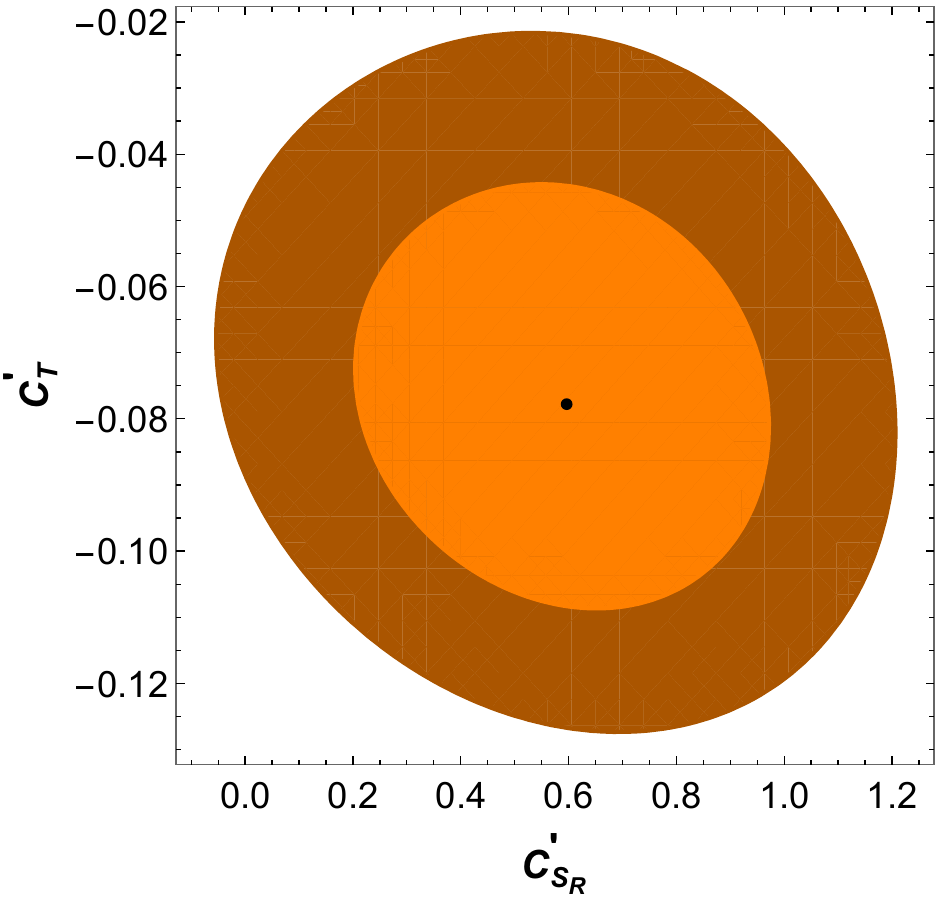}}\quad \subfloat[]{\includegraphics[width=4.5cm, height=4.2cm]{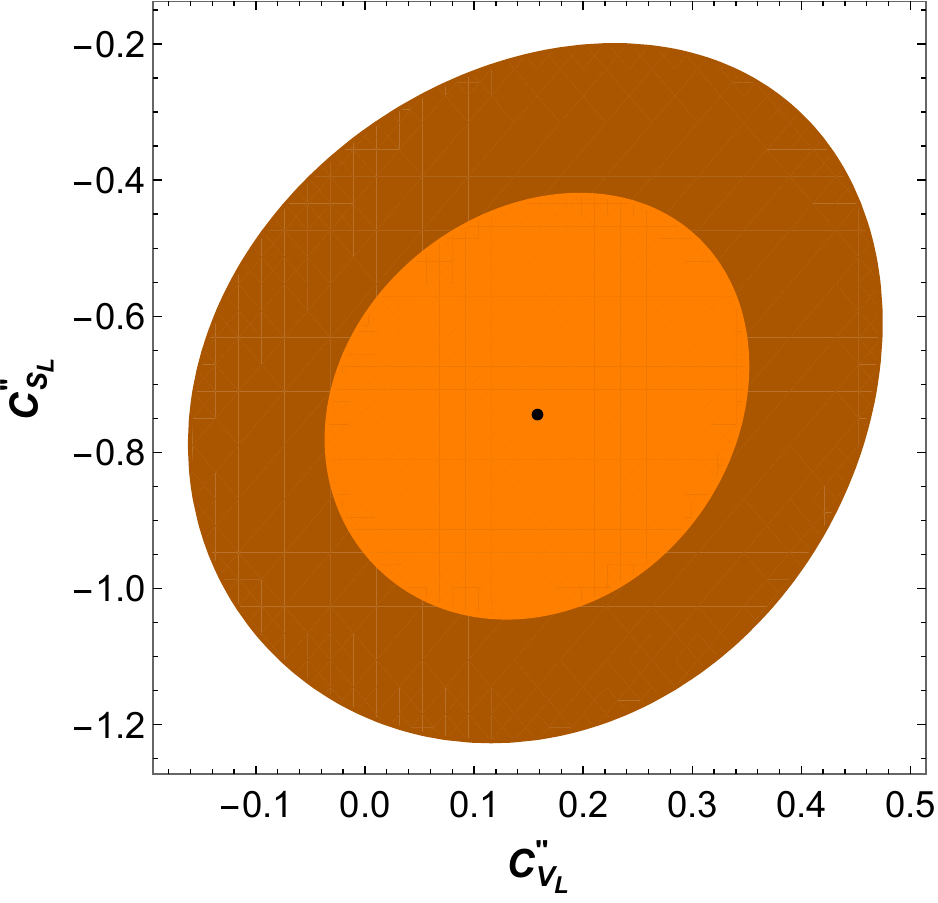}} 

\caption{\label{Global-s2d-fig}Results of the fits for NP scenarios when we include WC $C_{V_{R}}$ at scale $2\text{TeV}$. The light and dark gray colors show the $10\%$ and $60\%$ branching ratio constraints. The light (dark) color contours represent the $1(2\sigma)$ deviations from the BFP (Black color). Panels are restricted by the $\mathcal{S}_1$ and the dashed ellipse shows the measurement at $2\sigma$ level by the $\mathcal{S}_2$. Except Fig. 2(a), (b), (d), the orange color is not affected by either of these constraints, whereas in Figs. 2(a), (b), (d), the red and green colors are the $60\%$ and $10\%$ constraints, respectively. The purple-shaded region outside the dashed ellipse is excluded by the collider bounds for the current luminosity of  $139\text{fb}^{-1}$.}
\end{figure}

Using the BFPs values of the scenarios listed in Table \ref{global-t2}, calculated for $\mathcal{S}_1$, we predicted the deviation of various physical observables from their experimental measurements. We tabulated them in Table \ref{predbfp2dc}.  If we compare this with the case presented in Table \ref{predbfp1d}, it is interesting to observe that the difference between predicted and observed values reduces for $R_{\tau/{\mu,e}}\left(D\right)$, $R_{\tau/{\mu,e}}\left(D^*\right)$. Also the $\tau$ polarization asymmetry, $P_{\tau}\left(D^*\right)$, deviates by $\leq0.2\sigma$ from measurements, for the NP WCs, except for $\left(C_{V_{R}}^{\prime},C_{T}^{\prime}\right)$ with branching constraint of $10\%$, which deviates $0.4\sigma$. Similarly, for $F_{L}\left(D^*\right)$, we can see that the agreement between the predicted and measured values is $\leq1.3\sigma$ for all WCs while the values of $R_{\tau/\ell}\left(\Lambda_c\right)$ and $R_{\tau/\mu}\left(J/\psi\right)$ at the BFPs depart from their experimental measurements by $(1.1 - 1.3)\sigma$ and $(1.8, 1.9)\sigma$, respectively. Likewise, the predicted values of $P_{\tau}\left(D\right)$ at the BFP are higher than the SM results and are in the $0.33 - 0.52$ range. Therefore, exploring the models in which the right-handed vector currents are possible will be interesting. 

\begin{table}[H]
\begin{centering}
\renewcommand{\arraystretch}{1.1}
\begin{tabular}{|c|c|c|c|c|c|c|c|c|c|}
\hline 
WC & BR & BFP & $R_{\tau/\mu,e}\left(D\right)$ & $R_{\tau/\mu,e}\left(D^{*}\right)$ & $P_{\tau}\left(D^{*}\right)$ & $F_{L}\left(D^{*}\right)$ & $R_{\tau/\ell}\left(\Lambda_{c}\right)$ & $R_{\tau/\mu}\left(J/\psi\right)$ & $P_{\tau}\left(D\right)$\tabularnewline
\hline 
\hline 
\multirow{2}{*}{$\left(C_{V_{R}}^{''},C_{T}^{''}\right)$} & $\begin{array}{c}
<60\%\\
\&30\%
\end{array}$ & $\left(-0.46,0.12\right)$ & $\begin{array}{c}
0.34\\
+0\sigma
\end{array}$ & $\begin{array}{c}
0.29\\
+0\sigma
\end{array}$ & $\begin{array}{c}
-0.38\\
+0\sigma
\end{array}$ & $\begin{array}{c}
0.49\\
+0\sigma
\end{array}$ & $\begin{array}{c}
0.37\\
+1.3\sigma
\end{array}$ & $\begin{array}{c}
0.29\\
-1.8\sigma
\end{array}$ & $0.43$\tabularnewline
\cline{2-10} \cline{3-10} \cline{4-10} \cline{5-10} \cline{6-10} \cline{7-10} \cline{8-10} \cline{9-10} \cline{10-10} 
 & $<10\%$ & $\left(-0.26,0.03\right)$ & $\begin{array}{c}
0.36\\
+0.5\sigma
\end{array}$ & $\begin{array}{c}
0.27\\
-1.8\sigma
\end{array}$ & $\begin{array}{c}
-0.45\\
-0.1\sigma
\end{array}$ & $\begin{array}{c}
0.48\\
-0.3\sigma
\end{array}$ & $\begin{array}{c}
0.35\\
+1.1\sigma
\end{array}$ & $\begin{array}{c}
0.27\\
-1.9\sigma
\end{array}$ & $0.44$\tabularnewline
\hline 
$\left(C_{S_{L}}^{'},C_{S_{R}}^{'}\right)$ & - & $\left(-1.76,1.85\right)$ & $\begin{array}{c}
0.34\\
+0\sigma
\end{array}$ & $\begin{array}{c}
0.29\\
+0\sigma
\end{array}$ & $\begin{array}{c}
-0.42\\
-0.1\sigma
\end{array}$ & $\begin{array}{c}
0.48\\
-0.3\sigma
\end{array}$ & $\begin{array}{c}
0.37\\
+1.3\sigma
\end{array}$ & $\begin{array}{c}
0.29\\
-1.8\sigma
\end{array}$ & $0.57$\tabularnewline
\hline 
$\left(C_{V_{R}},C_{S_{R}}\right)$ & - & $\left(-0.30,0.45\right)$ & $\begin{array}{c}
0.34\\
+0\sigma
\end{array}$ & $\begin{array}{c}
0.29\\
+0\sigma
\end{array}$ & $\begin{array}{c}
-0.42\\
-0.1\sigma
\end{array}$ & $\begin{array}{c}
0.48\\
-0.3\sigma
\end{array}$ & $\begin{array}{c}
0.37\\
+1.3\sigma
\end{array}$ & $\begin{array}{c}
0.29\\
-1.8\sigma
\end{array}$ & $0.48$\tabularnewline
\hline 
\multirow{2}{*}{$\left(C_{V_{R}}^{'},C_{S_{L}}^{'}\right)$} & $\begin{array}{c}
<60\%\\
\&30\%
\end{array}$ & $\left(-0.31,0.76\right)$ & $\begin{array}{c}
0.34\\
+0\sigma
\end{array}$ & $\begin{array}{c}
0.29\\
+0\sigma
\end{array}$ & $\begin{array}{c}
-0.42\\
-0.1\sigma
\end{array}$ & $\begin{array}{c}
0.47\\
-0.3\sigma
\end{array}$ & $\begin{array}{c}
0.37\\
+1.3\sigma
\end{array}$ & $\begin{array}{c}
0.29\\
-1.8\sigma
\end{array}$ & $0.44$\tabularnewline
\cline{2-10} \cline{3-10} \cline{4-10} \cline{5-10} \cline{6-10} \cline{7-10} \cline{8-10} \cline{9-10} \cline{10-10} 
 & $<10\%$ & $\left(-0.24,0.47\right)$ & $\begin{array}{c}
0.34\\
+0.1\sigma
\end{array}$ & $\begin{array}{c}
0.27\\
-1\sigma
\end{array}$ & $\begin{array}{c}
-0.44\\
-0.1\sigma
\end{array}$ & $\begin{array}{c}
0.47\\
-0.4\sigma
\end{array}$ & $\begin{array}{c}
0.36\\
+1.2\sigma
\end{array}$ & $\begin{array}{c}
0.28\\
-1.8\sigma
\end{array}$ & $0.43$\tabularnewline
\hline 
$\left(C_{V_{L}}^{''},C_{T}^{''}\right)$ & - & $\left(0.52,-0.10\right)$ & $\begin{array}{c}
0.34\\
+0\sigma
\end{array}$ & $\begin{array}{c}
0.29\\
+0\sigma
\end{array}$ & $\begin{array}{c}
-0.45\\
-0.1\sigma
\end{array}$ & $\begin{array}{c}
0.46\\
-0.5\sigma
\end{array}$ & $\begin{array}{c}
0.37\\
+1.3\sigma
\end{array}$ & $\begin{array}{c}
0.29\\
-1.8\sigma
\end{array}$ & $0.52$\tabularnewline
\hline 
$\left(C_{V_{R}},C_{S_{L}}\right)$ & - & $\left(-0.40,0.53\right)$ & $\begin{array}{c}
0.34\\
+0\sigma
\end{array}$ & $\begin{array}{c}
0.29\\
+0\sigma
\end{array}$ & $\begin{array}{c}
-0.46\\
-0.2\sigma
\end{array}$ & $\begin{array}{c}
0.46\\
-0.6\sigma
\end{array}$ & $\begin{array}{c}
0.37\\
+1.3\sigma
\end{array}$ & $\begin{array}{c}
0.29\\
-1.8\sigma
\end{array}$ & $0.50$\tabularnewline
\hline 
$\left(C_{V_{L}},C_{V_{R}}\right)$ & - & $\left(0.32,0.01\right)$ & $\begin{array}{c}
0.34\\
-0.1\sigma
\end{array}$ & $\begin{array}{c}
0.29\\
+0\sigma
\end{array}$ & $\begin{array}{c}
-0.50\\
-0.2\sigma
\end{array}$ & $\begin{array}{c}
0.46\\
-0.6\sigma
\end{array}$ & $\begin{array}{c}
0.37\\
+1.3\sigma
\end{array}$ & $\begin{array}{c}
0.29\\
-1.8\sigma
\end{array}$ & $0.33$\tabularnewline
\hline 
$\left(C_{S_{R}}^{'},C_{T}^{'}\right)$ & - & $\left(0.59,-0.08\right)$ & $\begin{array}{c}
0.34\\
+0\sigma
\end{array}$ & $\begin{array}{c}
0.29\\
+0\sigma
\end{array}$ & $\begin{array}{c}
-0.47\\
-0.2\sigma
\end{array}$ & $\begin{array}{c}
0.46\\
-0.7\sigma
\end{array}$ & $\begin{array}{c}
0.37\\
+1.3\sigma
\end{array}$ & $\begin{array}{c}
0.29\\
-1.8\sigma
\end{array}$ & $0.49$\tabularnewline
\hline 
$\left(C_{V_{L}}^{''},C_{S_{L}}^{''}\right)$ & - & $\left(0.16,-0.74\right)$ & $\begin{array}{c}
0.34\\
-0.1\sigma
\end{array}$ & $\begin{array}{c}
0.29\\
+0\sigma
\end{array}$ & $\begin{array}{c}
-0.48\\
-0.2\sigma
\end{array}$ & $\begin{array}{c}
0.45\\
-0.8\sigma
\end{array}$ & $\begin{array}{c}
0.36\\
+1.3\sigma
\end{array}$ & $\begin{array}{c}
0.29\\
-1.8\sigma
\end{array}$ & $0.47$\tabularnewline
\hline 
$\left(C_{V_{R}}^{''},C_{S_{L}}^{''}\right)$ & - & $\left(0.16,-1.27\right)$ & $\begin{array}{c}
0.35\\
+0.2\sigma
\end{array}$ & $\begin{array}{c}
0.28\\
-0.4\sigma
\end{array}$ & $\begin{array}{c}
-0.52\\
-0.3\sigma
\end{array}$ & $\begin{array}{c}
0.43\\
-1.1\sigma
\end{array}$ & $\begin{array}{c}
0.36\\
+1.3\sigma
\end{array}$ & $\begin{array}{c}
0.29\\
-1.8\sigma
\end{array}$ & $0.46$\tabularnewline
\hline 
$\left(C_{V_{R}},C_{T}\right)$ & - & $\left(0.36,-0.28\right)$ & $\begin{array}{c}
0.34\\
-0.2\sigma
\end{array}$ & $\begin{array}{c}
0.29\\
+0\sigma
\end{array}$ & $\begin{array}{c}
-0.51\\
-0.2\sigma
\end{array}$ & $\begin{array}{c}
0.43\\
-1.2\sigma
\end{array}$ & $\begin{array}{c}
0.36\\
+1.3\sigma
\end{array}$ & $\begin{array}{c}
0.29\\
-1.8\sigma
\end{array}$ & $0.38$\tabularnewline
\hline 
\multirow{3}{*}{$\left(C_{V_{R}}^{'},C_{T}^{'}\right)$} & $<60\%$ & $\left(-0.73,0.20\right)$ & $\begin{array}{c}
0.36\\
+0.6\sigma
\end{array}$ & $\begin{array}{c}
0.27\\
-1.4\sigma
\end{array}$ & $\begin{array}{c}
-0.31\\
+0.1\sigma
\end{array}$ & $\begin{array}{c}
0.53\\
+0.7\sigma
\end{array}$ & $\begin{array}{c}
0.36\\
+1.3\sigma
\end{array}$ & $\begin{array}{c}
0.27\\
-1.9\sigma
\end{array}$ & $0.42$\tabularnewline
\cline{2-10} \cline{3-10} \cline{4-10} \cline{5-10} \cline{6-10} \cline{7-10} \cline{8-10} \cline{9-10} \cline{10-10} 
 & $<30\%$ & $\left(-0.50,0.11\right)$ & $\begin{array}{c}
0.36\\
+0.9\sigma
\end{array}$ & $\begin{array}{c}
0.26\\
-2\sigma
\end{array}$ & $\begin{array}{c}
-0.38\\
+0\sigma
\end{array}$ & $\begin{array}{c}
0.50\\
+0.2\sigma
\end{array}$ & $\begin{array}{c}
0.35\\
+1.2\sigma
\end{array}$ & $\begin{array}{c}
0.27\\
-1.9\sigma
\end{array}$ & $0.42$\tabularnewline
\cline{2-10} \cline{3-10} \cline{4-10} \cline{5-10} \cline{6-10} \cline{7-10} \cline{8-10} \cline{9-10} \cline{10-10} 
 & $<10\%$ & $\left(0.38,-0.20\right)$ & $\begin{array}{c}
0.37\\
+0.9\sigma
\end{array}$ & $\begin{array}{c}
0.26\\
-2.1\sigma
\end{array}$ & $\begin{array}{c}
-0.57\\
-0.4\sigma
\end{array}$ & $\begin{array}{c}
0.43\\
-1.3\sigma
\end{array}$ & $\begin{array}{c}
0.35\\
+1.1\sigma
\end{array}$ & $\begin{array}{c}
0.27\\
-1.9\sigma
\end{array}$ & $0.44$\tabularnewline
\hline 
\end{tabular}
\par\end{centering}
\centering{}\caption{\label{predbfp2dc} The BFPs and predictions at these points with
discrepancy from the experimental values expressed in multiples of
$\sigma^{O_{i}^{exp}}$ for of each WC by allowing value different
from zero and aligned with the constraint on the branching ratio,
for the set of observables $\mathcal{S}_{1}$.}
\end{table}

\subsection{Impact of collider (LHC) bounds}
The high $-p_{T}$ tails in mono$-\tau$ searches at LHC provide the severe collider restrictions on the $b\rightarrow c\tau\overline{\nu}_{\tau}$ operators \cite{Greljo:2018tzh, Faroughy:2016osc, Iguro:2018fni}, which enable us to restrict the two-dimensional NP scenarios.
Ref. \cite{Endo:2021lhi} provides the latest expected sensitivities for each single operator for one-dimensional scenario in the $\tau^{\pm}\nu$ searches. 
The current collider bounds of the NP WC with luminosity $139\text{fb}^{-1}$ based on the $\tau^{\pm}\nu$ search at $\mu=m_{b}$ and for HL-LHC $1000(3000)\text{fb}^{-1}$ are:
\begin{equation}
\left|\widetilde{C}_{V_{L}}\right|<0.30\left(0.14\right),\quad\left|\widetilde{C}_{V_{R}}\right|<0.32\left(0.15\right),\quad\left|\widetilde{C}_{S_{L,R}}\right|<0.55\left(0.25\right),\quad\left|\widetilde{C}_{T}\right|<0.15\left(0.07\right).
\end{equation}
By applying collider bounds to various combinations of the NP WCs, we have demonstrated their impact on the parametric space by the purple-shaded ellipse in Fig. \ref{s2d-fig}. Notably, the combination $\left(C_{S_{L}}, C_{S_{R}}\right)$ faces stringent restrictions, as the lower portion of the previously allowed parameter region has been excluded now. Furthermore, under the constraint of a \(60\%\) branching ratio, the bounds for the BFP are also excluded, resulting in a significant reduction in the viable parameter space. In contrast, all other scenarios remain consistent with the collider bounds. Including $C_{V_{R}}$, the combination $\left(C_{V_{R}}^{''},C_{T}^{''}\right)$ for branching $<60\%$, and $\left(C_{S_{L}}^{'},C_{S_{R}}^{'}\right)$ restrict the allowed parametric space, shown in Fig. \ref{Global-s2d-fig} (a) and (b), respectively

\section{Phenomenology of $\Lambda_{b}\rightarrow\Lambda_{c}\tau\bar{\nu}_{\tau}$ decay}\label{sec5}
The purpose of this section is to investigate the impact of the NP bounds computed in Section \ref{sec3} on different physical observables in $\Lambda_b\to \Lambda_{c}\tau\bar{\nu}_{\tau}$ decay. For this purpose, 
the differential decay rate for this process reads as \cite{Datta:2017aue}
\begin{equation}
\frac{d\Gamma}{dq^{2}d\cos\theta_{\tau}}=\frac{G_{F}^{2}\left|V_{cb}\right|^{2}}{2048\pi^{3}}\left(1-\frac{m_{\tau}^{2}}{q^{2}}\right)\frac{\sqrt{Q_{+}Q_{-}}}{m_{1}^{3}}\sum_{\lambda_{2}}\sum_{\lambda_{\tau}}\left|M_{\lambda_{2}}^{\lambda_{\tau}}\right|^{2},\label{DecayRate}
\end{equation}
where 
\begin{align*}
q & =p_{1}-p_{2},\quad Q_{\pm}=\left(m_{1}\pm m_{2}\right)^{2}-q^{2}
\end{align*}
and the helicity amplitude $M_{\lambda_{2}}^{\lambda_{\tau}}$ is
written as
\begin{equation}
M_{\lambda_{2}}^{\lambda_{\tau}}=H_{\lambda_{2}}^{SP}L^{\lambda_{\tau}}+\sum_{\lambda}\eta_{\lambda}H_{\lambda_{2},\lambda}^{VA}L_{\lambda}^{\lambda_{\tau}}+\sum_{\lambda,\lambda'}\eta_{\lambda}H_{\lambda_{2},\lambda,\lambda'}^{\left(T\right)\lambda_{1}}L_{\lambda,\lambda'}^{\lambda_{\tau}}.
\end{equation}
Here, $\left(\lambda,\lambda'\right)$ indicate the helicities of the
virtual vector boson, and $\lambda_{2}$ and $\lambda_{\tau}$ are
the helicities of the $\Lambda_{c}-$baryon and $\tau-$lepton, respectively. The scalar/pseudo-scalar-type, vector/axial-vector-type, and tensor-type hadronic helicity amplitudes are defined as:
\begin{equation}
H_{\lambda_{2}}^{SP}  =H_{\lambda_{2}}^{S}+H_{\lambda_{2}}^{P},\hspace{1.1cm}
H_{\lambda_{2}}^{S}  =\left(\widetilde{C}_{S_{L}}+\widetilde{C}_{S_{R}}\right)\left\langle \Lambda_{c}\left|\bar{c}b\right|\Lambda_{b}\right\rangle ,\hspace{1.1cm}
H_{\lambda_{2}}^{P}  =\left(-\widetilde{C}_{S_{L}}+\widetilde{C}_{S_{R}}\right)\left\langle \Lambda_{c}\left|\bar{c}b\right|\Lambda_{b}\right\rangle ,\label{eq:hscalar}
\end{equation}
\begin{eqnarray}
H_{\lambda_{2},\lambda}^{VA}  &=&H_{\lambda_{2},\lambda}^{V}-H_{\lambda_{2},\lambda}^{A},\notag \\ 
H_{\lambda_{2},\lambda}^{V} &=&\left(1+\widetilde{C}_{V_{L}}+\widetilde{C}_{V_{R}}\right)\epsilon^{*\mu}\left(\lambda\right)\left\langle \Lambda_{c}\left|\bar{c}\gamma_{\mu}b\right|\Lambda_{b}\right\rangle ,\hspace{1.2cm}
H_{\lambda_{2},\lambda}^{A} =\left(1+\widetilde{C}_{V_{L}}-\widetilde{C}_{V_{R}}\right)\epsilon^{*\mu}\left(\lambda\right)\left\langle \Lambda_{c}\left|\bar{c}\gamma_{\mu}\gamma_{5}b\right|\Lambda_{b}\right\rangle ,\label{eq:hvector}
\end{eqnarray}
and
\begin{eqnarray}
H_{\lambda_{2},\lambda,\lambda'}^{\left(T\right)\lambda_{1}} &=&H_{\lambda_{2},\lambda,\lambda'}^{\left(T1\right)\lambda_{1}}-H_{\lambda_{2},\lambda,\lambda'}^{\left(T2\right)\lambda_{1}},
\notag \\
H_{\lambda_{2},\lambda,\lambda'}^{\left(T1\right)\lambda_{1}} &=&\widetilde{C}_{T}\epsilon^{*\mu}\left(\lambda\right)\epsilon^{*\mu}\left(\lambda'\right)\left\langle \Lambda_{c}\left|\bar{c}i\sigma_{\mu\nu}b\right|\Lambda_{b}\right\rangle ,\hspace{2.2cm}
H_{\lambda_{2},\lambda,\lambda'}^{\left(T2\right)\lambda_{1}} =\widetilde{C}_{T}\epsilon^{*\mu}\left(\lambda\right)\epsilon^{*\mu}\left(\lambda'\right)\left\langle \Lambda_{c}\left|\bar{c}i\sigma_{\mu\nu}\gamma_{5}b\right|\Lambda_{b}\right\rangle .\label{eq:htensor}
\end{eqnarray}

The leptonic parts of the amplitude can be written as:
\begin{align}
L^{\lambda_{\tau}} & =\left\langle \tau\bar{\nu}_{\tau}\left|\bar{\tau}\left(1-\gamma_{5}\right)\nu_{\tau}\right|0\right\rangle ,\nonumber \\
L_{\lambda}^{\lambda_{\tau}} & =\epsilon^{*\mu}\left(\lambda\right)\left\langle \tau\bar{\nu}_{\tau}\left|\bar{\tau}\gamma_{\mu}\left(1-\gamma_{5}\right)\nu_{\tau}\right|0\right\rangle ,\nonumber \\
L_{\lambda,\lambda'}^{\lambda_{\tau}} & =-\epsilon^{*\mu}\left(\lambda\right)\epsilon^{*\mu}\left(\lambda'\right)\left\langle \tau\bar{\nu}_{\tau}\left|\bar{\tau}i\sigma_{\mu\nu}\left(1-\gamma_{5}\right)\nu_{\tau}\right|0\right\rangle ,
\end{align}
where $\epsilon^{\mu}$ defines the polarization vector of the virtual vector boson, and its different components are given in Appendix \ref{hspa}.

In this work, we make use of the helicity-based definition of the $\Lambda_{b}\rightarrow\Lambda_{c}$ form factors as introduced in \cite{Feldmann:2011xf} and then extended them to include tensor form factors from \cite{Detmold:2016pkz}. The matrix elements of the vector and axial
vector currents are expressed using six helicity form factors: $F_{+}$, $F_{\perp}$, $F_{0}$, $G_{+}$, $G_{\perp}$, and $G_{0}$; and four tensor form factor $h_{+}$, $\widetilde{h}_{+}$, $h_{\perp}$, and $\widetilde{h}_{\perp}$ . There explicit interpolation with $q^2$ is summarized in Appendix \ref{hspa}.  Using the spinors for $\Lambda_b$ and $\Lambda_c$ along with the kinematical relations given in \ref{hspa}, the scalar and pseudo-scalar hadronic helicity amplitudes are
\begin{equation}
H_{\pm1/2}^{SP}=F_{0}\left(\widetilde{C}_{S_{L}}+\widetilde{C}_{S_{R}}\right)\frac{\sqrt{Q_{+}}}{m_{b}-m_{c}}m_{-}\pm G_{0}\left(\widetilde{C}_{S_{L}}-\widetilde{C}_{S_{R}}\right)\frac{\sqrt{Q_{-}}}{m_{b}+m_{c}}m_{+}.
\end{equation}
Similarly, the vector and axial-vector hadronic helicity amplitudes will become
\begin{align}
H_{\pm1/2,t}^{VA} & =\frac{1}{\sqrt{q^{2}}}\left(F_{0}\left(1+\widetilde{C}_{V_{L}}+\widetilde{C}_{V_{R}}\right)\sqrt{Q_{+}}m_{-}\mp G_{0}\left(1+\widetilde{C}_{V_{L}}-\widetilde{C}_{V_{R}}\right)\sqrt{Q_{-}}m_{+}\right),\nonumber \\
H_{\pm1/2,0}^{VA} & =\frac{1}{\sqrt{q^{2}}}\left(F_{+}\left(1+\widetilde{C}_{V_{L}}+\widetilde{C}_{V_{R}}\right)\sqrt{Q_{-}}m_{+}\mp G_{+}\left(1+\widetilde{C}_{V_{L}}-\widetilde{C}_{V_{R}}\right)\sqrt{Q_{+}}m_{-}\right),\nonumber \\
H_{\pm1/2,\pm}^{VA} & =\sqrt{2}\left(F_{\perp}\left(1+\widetilde{C}_{V_{L}}+\widetilde{C}_{V_{R}}\right)\sqrt{Q_{-}}\mp G_{\perp}\left(1+\widetilde{C}_{V_{L}}-\widetilde{C}_{V_{R}}\right)\sqrt{Q_{+}}\right).
\end{align}
Likewise, the non-zero tensor hadronic helicity amplitudes are
\begin{align*}
H_{\pm1/2,t,0}^{\left(T\right)\pm1/2} & =-H_{\pm1/2,+,-}^{\left(T\right)\pm1/2}=\widetilde{C}_{T}\left(h_{+}\sqrt{Q_{-}}\pm\widetilde{h}_{+}\sqrt{Q_{+}}\right),\\
H_{\pm1/2,t,\pm}^{\left(T\right)\mp1/2} & =\mp H_{\pm1/2,0,\pm}^{\left(T\right)\mp1/2}=\widetilde{C}_{T}\frac{\sqrt{2}}{\sqrt{q^{2}}}\left(h_{\perp}\sqrt{Q_{-}}m_{+}\pm\widetilde{h}_{\perp}\sqrt{Q_{+}}m_{-}\right),
\end{align*}
and
\begin{equation}
H_{\lambda_{2},\lambda,\lambda'}^{\left(T\right)\lambda_{1}}=-H_{\lambda_{2},\lambda',\lambda}^{\left(T\right)\lambda_{1}}.
\end{equation}

In the di-leptonic rest frame, the momenta of the final state leptons, and the corresponding spinors are defined in Appendix \ref{hspa-2}. Using them, the non-zero leptonic helicity amplitudes are computed as follows:
\begin{align}
L^{1/2} & =2\sqrt{q^{2}}v,\quad
L_{t}^{1/2}  = 2m_{\tau}v,\quad
L_{0}^{1/2} =-2m_{\tau}v\cos\theta_{\tau},\quad
L_{0}^{-1/2} =2\sqrt{q^{2}}v\sin\theta_{\tau},\nonumber \\
L_{\pm}^{1/2} & =\mp\sqrt{2}m_{\tau}v\sin\theta_{\tau},\quad
L_{\pm}^{-1/2} =\sqrt{2q^{2}}v\left(-1\mp \cos\theta_{\tau}\right),\quad
L_{t,0}^{1/2} =L_{+,-}^{1/2}=-2\sqrt{q^{2}}v\cos\theta_{\tau},\nonumber \\
L_{t,0}^{-1/2} & =L_{+,-}^{-1/2}=2m_{\tau}v\sin\theta_{\tau},\quad
L_{t,\pm}^{1/2} =\mp L_{0,\pm}^{1/2}=\mp\sqrt{2q^{2}}v\sin\theta_{\tau},\quad
L_{t,\pm}^{-1/2}=\mp L_{0,\pm}^{1/2}=\sqrt{2}m_{\tau}v\left(-1\mp \cos\theta_{\tau}\right)
\end{align}
where 
\begin{equation}
L_{\lambda,\lambda'}^{\lambda_{2}}=-L_{\lambda',\lambda}^{\lambda_{2}}.
\end{equation}

\subsection{Observables of $\Lambda_{b}\rightarrow\Lambda_{c}\tau\bar{\nu}_{\tau}$ decay}
In our analysis, we take into account the differential decay rate $d\Gamma/dq^{2}$, lepton forward-backward asymmetry $A_{FB}$, $\Lambda_{c}-$longitudinal polarization fraction $P_{L}^{\Lambda_{c}}$, $\tau-$lepton longitudinal
polarization fraction $P_{L}^{\tau}$, and $\Lambda_{c}-$LFU ratio
$R_{\tau/l}\left(\Lambda_{c}\right)$ that are defined as follows:
\begin{align}
A_{FB}\left(q^{2}\right) & =\frac{\int_{0}^{1}d\cos\theta_{\tau}\left(d^{2}\Gamma/dq^{2}d\cos\theta_{\tau}\right)-\int_{-1}^{0}d\cos\theta_{\tau}\left(d^{2}\Gamma/dq^{2}d\cos\theta_{\tau}\right)}{d\Gamma/dq^{2}},\label{Afb}\\
P_{L}^{\Lambda_{c}}\left(q^{2}\right) & =\frac{\left(d\Gamma/dq^{2}\right)^{\lambda_{2}=1/2}-\left(d\Gamma/dq^{2}\right)^{\lambda_{2}=-1/2}}{d\Gamma/dq^{2}},\label{PLc}\\
P_{L}^{\tau}\left(q^{2}\right) & =\frac{\left(d\Gamma/dq^{2}\right)^{\lambda_{\tau}=1/2}-\left(d\Gamma/dq^{2}\right)^{\lambda_{\tau}=-1/2}}{d\Gamma/dq^{2}},\label{PLt}\\
R_{\tau/\ell}\left(\Lambda_{c}\right) & =\frac{d\Gamma\left(\Lambda_{b}\rightarrow\Lambda_{c}\tau\bar{\nu}_{\tau}\right)/dq^{2}}{d\Gamma\left(\Lambda_{b}\rightarrow\Lambda_{c}l\bar{\nu}_{l}\right)/dq^{2}}.\label{RLc}
\end{align}
The analytical expressions of these angular observables in terms of the helicity amplitudes are given in Appendix \ref{hspa}. As these observables are ratios, consequently, they are largely free from hadronic uncertainties and thus provide excellent tests of the NP effects.

\subsection{Numerical Analysis }
In this section, the prediction for above mentioned physical observables is done in the SM and using the constraints on various NP scenarios.  In Fig. \ref{phen-1d}, we have plotted $d\Gamma/dq^{2}$, $A_{FB}$, $P_{L}^{\Lambda_{c}}$, $P_{L}^{\tau}$, and 
$R_{\tau/\ell}\left(\Lambda_{c}\right)$ with $q^2$ in the SM and by using the parametric space of one-dimension NP scenarios given in Table \ref{1d-table}. The band in each curve shows the theoretical uncertainties from the form factors and other input parameters. The black band shows the SM value. The NP WCs $C_{S_{L}}^{\prime\prime}$, $C_{T}$, $C_{S_{R}}$ and $C_{T}^{\prime}$, $C_{S_{L}}$, $C_{T}^{\prime\prime}$ are represented by pink, blue, and green bands, respectively.
 
From Fig. \ref{phen-1d}(a,b), it is observed that the most significant impact of new physics (NP) on \( \frac{d\Gamma}{dq^{2}} \) occurs in the intermediate \( q^{2} \) region, where the unprimed Wilson coefficient (WC) \( C_{V_{L}} \) exhibits the most substantial effect. Additionally, the tensor operator \( C_{T} \) shows considerable deviations, although its influence is smaller than that of \( C_{V_{L}} \). This indicates that the differential decay rate is sensitive not only to vector-type NP but also to contributions from tensor-type NP.
In comparison, the new left-handed scalar coupling \( C_{S_{L}} \) is outperformed by its right-handed counterpart \( C_{S_{R}} \), which has a slightly greater influence on the decay rate in the low to intermediate \( q^{2} \) region. The primed WC \( C_{V_{L}}^{\prime} \) also significantly impacts the values of \( \frac{d\Gamma}{dq^{2}} \). This is related to \( C_{V_{L}} \) through the Fierz identity, where the interference terms involving \( C_{V_{L}} \) play a crucial role in affecting the differential decay rate.
Similarly, the double-primed right-handed scalar WC \( C_{S_{R}}^{\prime\prime} \) has the largest impact on the values of \( \frac{d\Gamma}{dq^{2}} \) as well, again due to its connection with \( C_{V_{L}} \) established through the Fierz identity. This suggests a strong interplay between scalar and vector currents that influences the decay rate.

In the case of \(A_{FB}\), the deviation from its SM value is observed for the WCs \(C_{S_{R}}\), \(C_{T}\), \(C_{S_{L}}^{\prime\prime}\), and \(C_{T}^{\prime}\), while the effects of \(C_{S_{L}}^{\prime}\) are mild, particularly in the large \(q^{2}\) region. For \(P_{L}^{\Lambda_{c}}\), notable deviations from its SM values are found for \(C_{T}\) and \(C_{S_{L}}^{\prime\prime}\) across all \(q^{2}\) regions, while the effects of \(C_{S_{R}}\), \(C_{T}^{\prime}\), and \(C_{S_{L}}\) are prominent in the low to medium \(q^{2}\) ranges. In the case of $P_{L}^{\tau}$, the effects of $C_{S_{R}}$, $C_{S_{L}}$, $C_{T}^{\prime}$, $C_{S_{L}}^{\prime\prime}$, and $C_{T}^{\prime\prime}$ are prominent, particularly, in the high $q^{2}$ region, whereas the effects of $C_{T}$ are mild. Similar to the case of $P_{L}^{\tau}$, the effects of NP WCs on $R_{\tau/\ell}\left(\Lambda_{c}\right)$  are prominent in the high $q^{2}$ region. Thus, precise measurements of these physical observables across different $q^2$ segments at current and future colliders will be crucial for investigating the status of NP couplings.
\begin{figure}[H]
\centering 
\begin{subfigure}[b]{0.32\textwidth}
\centering 
\includegraphics[width=5.6cm, height=3.5cm]{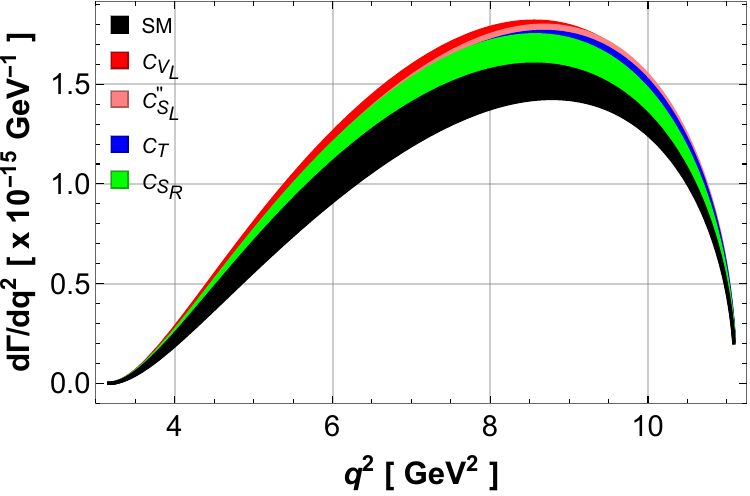}
\caption{}
\end{subfigure}
\begin{subfigure}[b]{0.32\textwidth}
\centering 
\includegraphics[width=5.6cm, height=3.5cm]{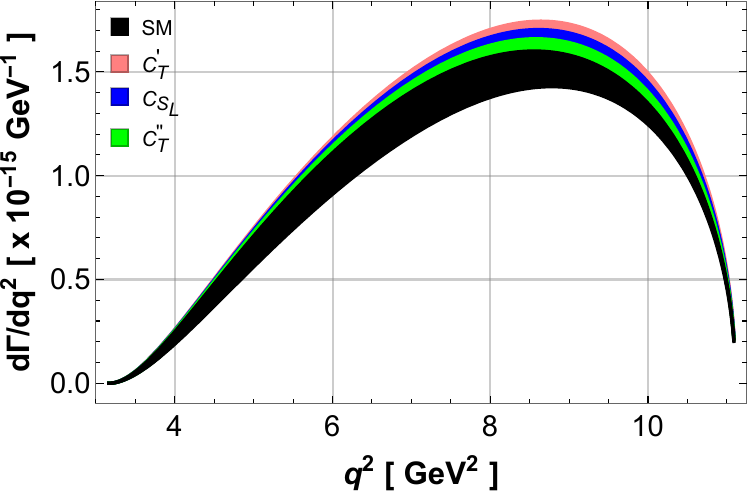} 
\caption{}
\end{subfigure}
\begin{subfigure}[b]{0.32\textwidth}
\centering 
\includegraphics[width=5.6cm, height=3.5cm]{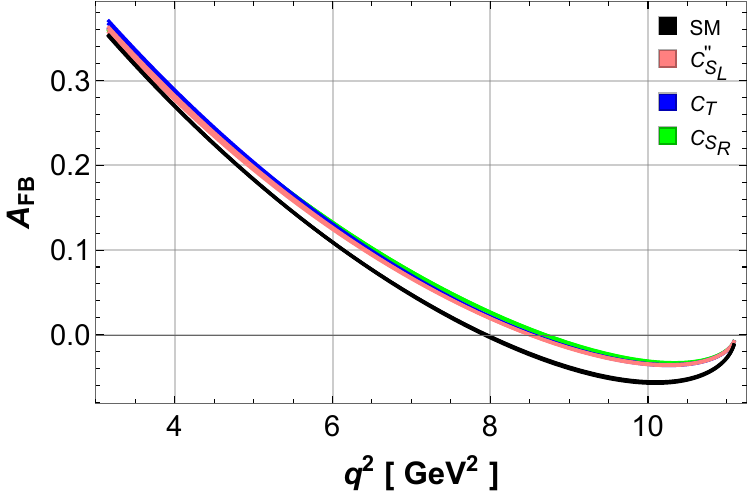}
\caption{}
\end{subfigure}
\begin{subfigure}[b]{0.32\textwidth}
\centering 
\includegraphics[width=5.6cm, height=3.5cm]{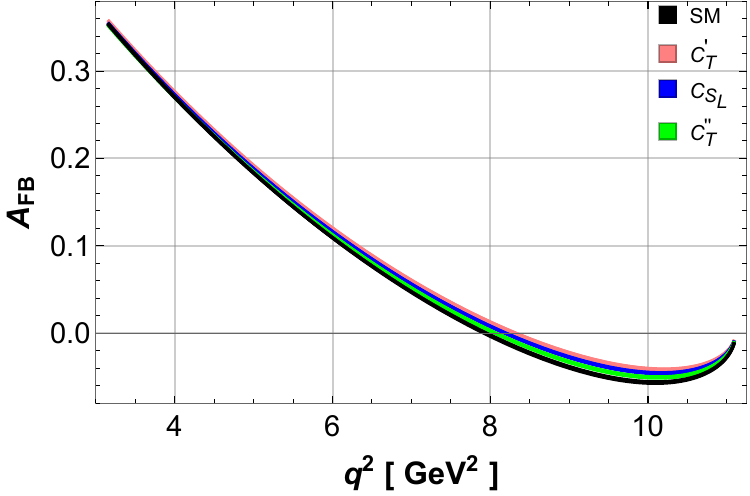}
\caption{}
\end{subfigure}
\begin{subfigure}[b]{0.32\textwidth}
\centering 
\includegraphics[width=5.6cm, height=3.5cm]{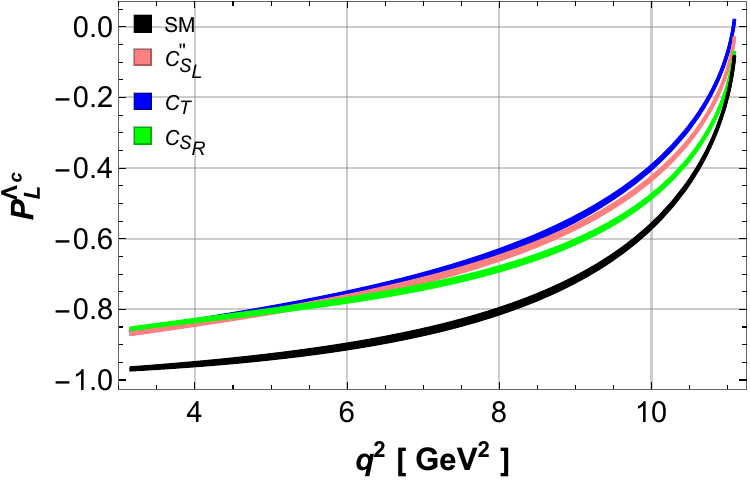} 
\caption{}
\end{subfigure}
\begin{subfigure}[b]{0.32\textwidth}
\centering 
\includegraphics[width=5.6cm, height=3.5cm]{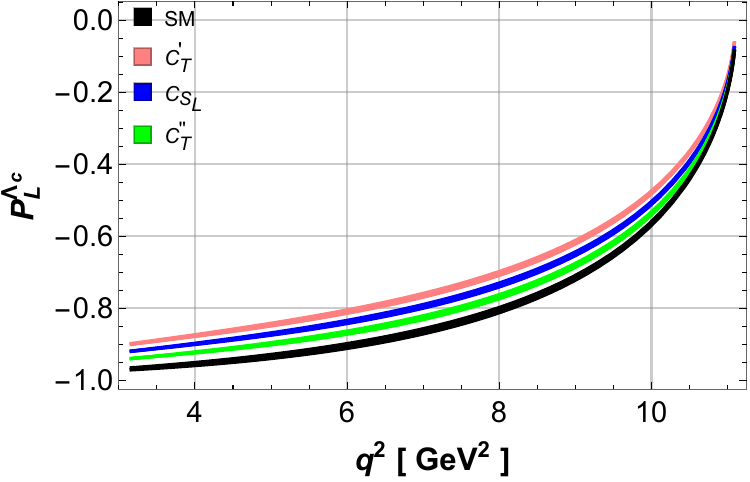}
\caption{}
\end{subfigure}
\begin{subfigure}[b]{0.32\textwidth}
\centering 
\includegraphics[width=5.6cm, height=3.5cm]{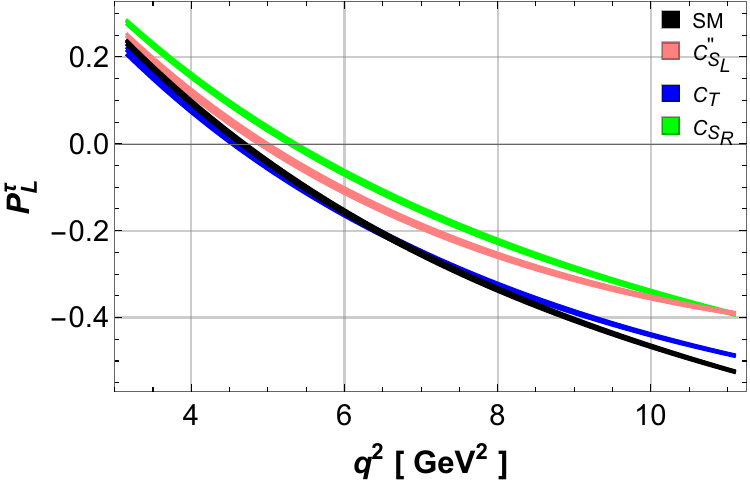}
\caption{}
\end{subfigure}
\begin{subfigure}[b]{0.32\textwidth}
\centering 
\includegraphics[width=5.6cm, height=3.5cm]{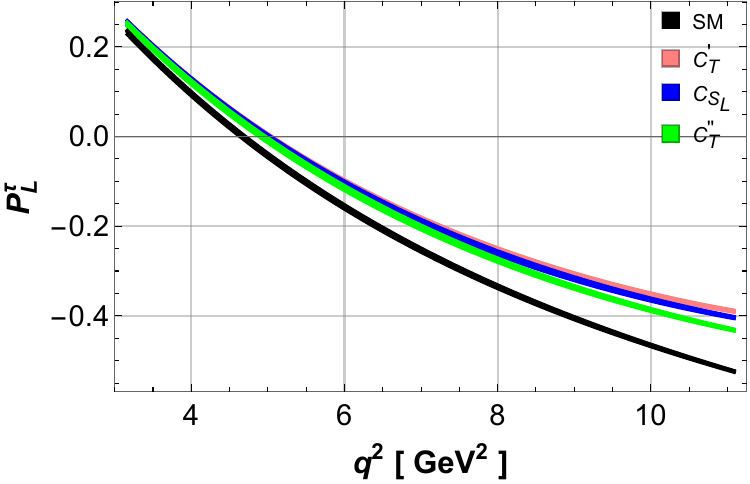} 
\caption{}
\end{subfigure}
\begin{subfigure}[b]{0.32\textwidth}
\centering 
\includegraphics[width=5.6cm, height=3.5cm]{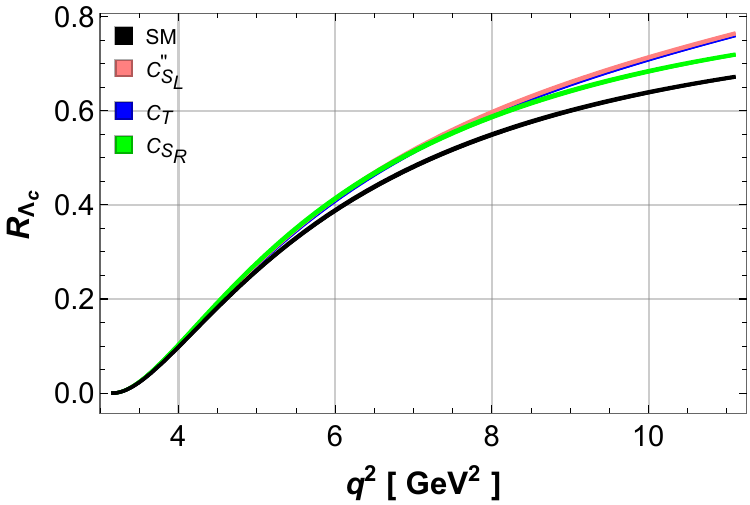}
\caption{}
\end{subfigure}
\begin{subfigure}[b]{0.32\textwidth}
\centering 
\includegraphics[width=5.6cm, height=3.5cm]{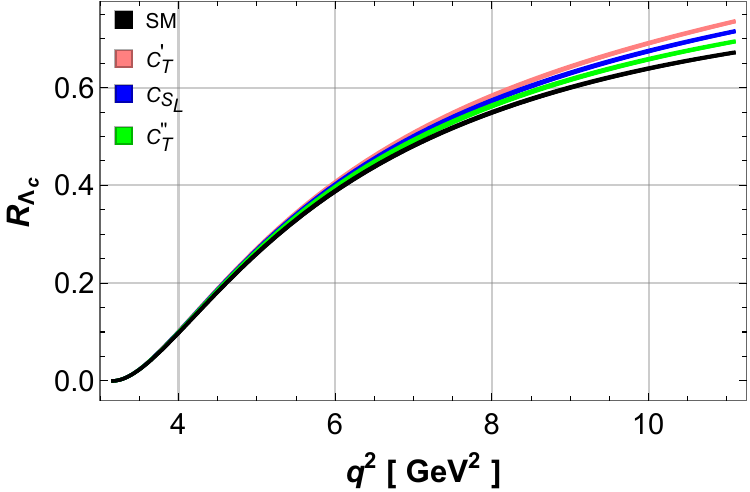}
\caption{}
\end{subfigure}

\caption{\label{phen-1d}The $d\Gamma/dq^{2}$,
 $A_{FB}$, $P_{L}^{\Lambda_{c}}$, $P_{L}^{\tau}$, and 
$R_{\Lambda_c}\equiv R_{\tau/\ell}\left(\Lambda_{c}\right)$ observables exhibited for
various NP coupling as a function of $q^{2}$. The width of each curve
comes from the theoretical uncertainties in hadronic form factors
and quark masses. The SM value is shown in the black band, whereas the NP couplings depicted in color bands.}
\end{figure}

Fig. \ref{phen-2d-plot} illustrates the behavior of above mentioned physical observables as a function $q^2$ for different two-dimensional NP scenarios calculated in Table \ref{2d-table-1}. In the presence of NP, the deviations from the SM predictions are represented by different color bands, each corresponding to specific combinations of WCs: $(C_{S_L}, C_{S_R})$, $(C_{V_L}, C_T)$, $(C_{S_R}, C_T)$, and $(C_{S_L}, C_T)$. The NP effects on the observables are summarized as follows:
\begin{itemize}
    \item $\bm{d\Gamma/dq^{2}:}$ The effects of NP are prominent in the intermediate $q^2$ region for all scenarios for  including unprimed, primed, and double-primed WCs.
    \item $\bm{A_{FB}:}$ The largest effects are observed for $(C_{S_L}, C_{S_R})$, followed by $(C_{S_R}, C_T)$, and then three degenerate scenarios involving $(C_{S_L}, C_T)$. The $(C_{V_L}, C_T)$ shows a minor deviation, lying below the SM prediction in the high $q^2$ region. Notably, the NP scenarios shift the zero value of $A_{FB}$, providing valuable insights into this observable.
    \item $\bm{P_{L}^{\Lambda_{c}}:}$  Significant effects are observed for $(C_{S_L}, C_{S_R})$ in the low to middle $q^2$ region, followed by the combination $(C_{S_R}, C_T)$. The three degenerate scenarios involving $(C_{S_L}, C_T)$ show deviations across the entire $q^2$ region. The $(C_{V_L}, C_T)$ exhibits a smaller deviation, lying below the SM prediction in the low to middle $q^2$ region.
    \item $\bm{P_{L}^{\tau}:}$ The maximum effects occur for $(C_{S_L}, C_{S_R})$ across the entire $q^2$ region. This is followed by deviations in $(C_{S_R}, C_T)$ and the three degenerate $(C_{S_L}, C_T)$ scenarios in the high $q^2$ region. The $(C_{V_L}, C_T)$ exceeds the SM prediction in the low $q^2$ region and falls below it in the high $q^2$ region.
    \item $\bm{R_{\tau/\ell}\left(\Lambda_{c}\right):}$ Significant effects are prominent for four degenerate scenarios involving $(C_{S_R}, C_T)$ and $(C_{S_L}, C_T)$, followed by $(C_{S_L}, C_{S_R})$ in the high $q^2$ region.
\end{itemize}
\begin{figure}[H]
\centering 
\begin{subfigure}[b]{0.32\textwidth}
\centering 
\includegraphics[width=5.6cm, height=3.5cm]{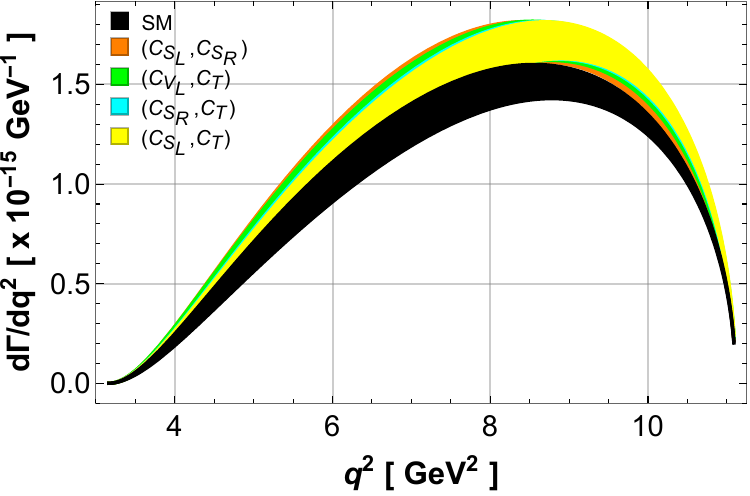}
\caption{}
\end{subfigure}
\begin{subfigure}[b]{0.32\textwidth}
\centering 
\includegraphics[width=5.6cm, height=3.5cm]{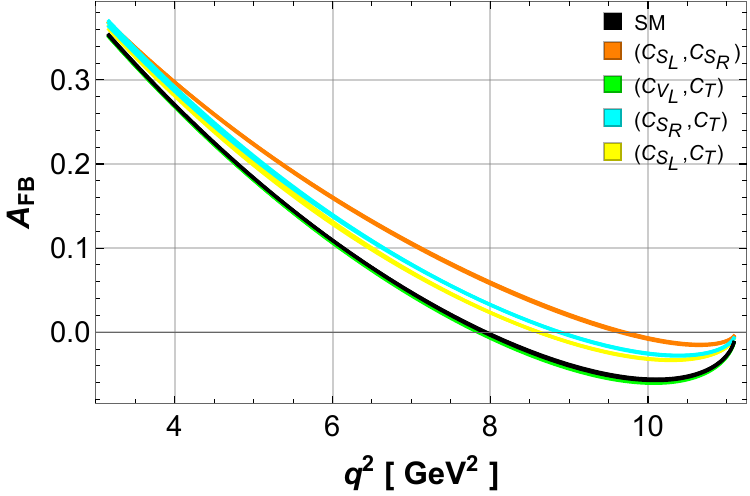} 
\caption{}
\end{subfigure}
\begin{subfigure}[b]{0.32\textwidth}
\centering 
\includegraphics[width=5.6cm, height=3.5cm]{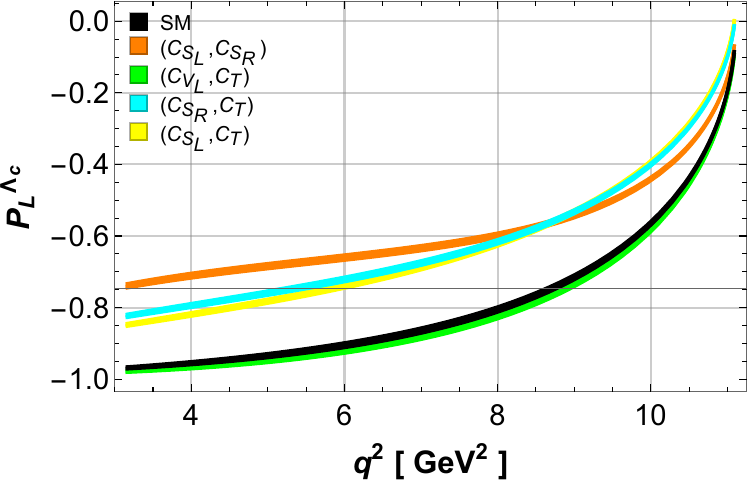}
\caption{}
\end{subfigure}
\begin{subfigure}[b]{0.32\textwidth}
\centering 
\includegraphics[width=5.6cm, height=3.5cm]{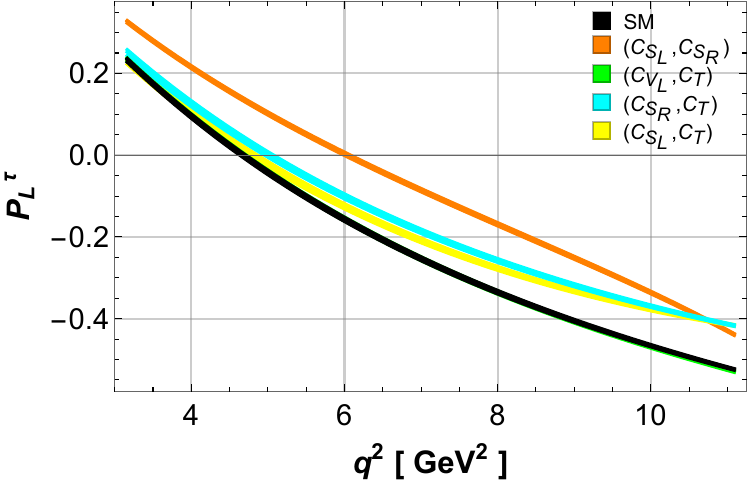}
\caption{}
\end{subfigure}
\begin{subfigure}[b]{0.32\textwidth}
\centering 
\includegraphics[width=5.6cm, height=3.5cm]{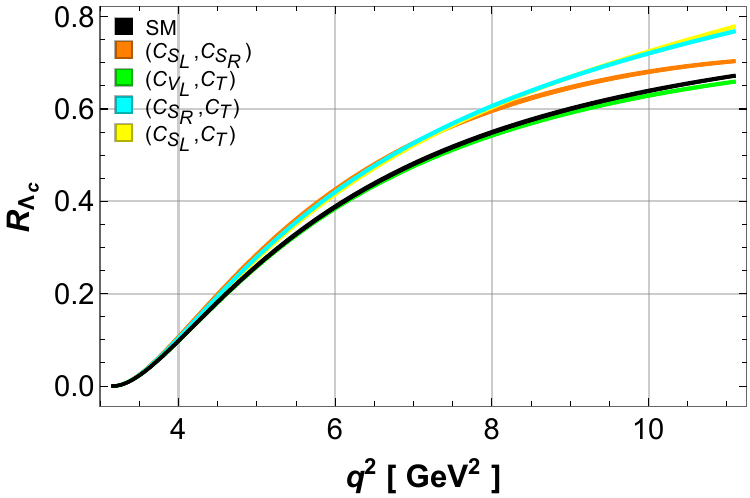} 
\caption{}
\end{subfigure}

\caption{\label{phen-2d-plot}The $d\Gamma/dq^{2}$,
 $A_{FB}$, $P_{L}^{\Lambda_{c}}$, $P_{L}^{\tau}$, and 
$R_{\Lambda_c}\equiv R_{\tau/\ell}\left(\Lambda_{c}\right)$ observables exhibited for
various NP coupling as a function of $q^{2}$. The width of each curve comes from the theoretical uncertainties in hadronic form factors and quark masses. The NP WCs $\left(C_{S_{L}},C_{S_{R}}\right)$, $\left(C_{V_{L}},C_{S_{R}}\right)$, $\left(C_{S_{R}},C_{T}\right)$, $\left(C_{S_{L}},C_{T}\right)$,  for the set of observables $\mathcal{S}_1$
are drawn with orange, green, cyan and yellow colors, respectively.}
\end{figure} 
The average values of these one and two-dimensional WCs scenarios for observable  $d\Gamma/dq^{2}$,
 $A_{FB}$, $P_{L}^{\Lambda_{c}}$, $P_{L}^{\tau}$, and 
$R_{\tau/\ell}\left(\Lambda_{c}\right)$ at BFP are listed in Tables \ref{1d-obs-table} and \ref{2d-obs-table} respectively.
Here, we can observe that except for the decay rate, the errors from the various input parameters do not mimic the NP effects; hence, they are useful probes to establish the NP in these FCCC decays.
\begin{table}[H]
\centering{}%
\renewcommand{\arraystretch}{1.5}
\begin{tabular}{|c|c|c|c|c|c|c|}
\hline 
WC & BFP & $\left\langle d\Gamma/dq^{2}\right\rangle $ & $\left\langle A_{FB}\right\rangle $ & $\left\langle P_{L}^{\Lambda_{c}}\right\rangle $ & $\left\langle P_{L}^{\tau}\right\rangle $ & $\left\langle R_{\tau/\ell}\left(\Lambda_{c}\right)\right\rangle $\tabularnewline
\hline 
\hline 
SM & $C_{i}=0$ & $0.98\pm0.07$ & $0.025\pm0.0018$ & $-0.76\pm0.010$ & $-0.31\pm0.009$ & $0.42\pm0.007$\tabularnewline
\hline 
$C_{V_{L}}$ & $0.31$ & $1.12\pm0.08$ & $0.025\pm0.0018$ & $-0.76\pm0.010$ & $-0.31\pm0.009$ & $0.42\pm0.007$\tabularnewline
\hline 
$C_{S_{L}}^{\prime\prime}$ & $-0.78$ & $1.10\pm0.08$ & $0.042\pm0.0026$ & $-0.62\pm0.009$ & $-0.23\pm0.009$ & $0.46\pm0.008$\tabularnewline
\hline 
$C_{T}$ & $-0.17$ & $1.08\pm0.08$ & $0.044\pm0.0026$ & $-0.60\pm0.010$ & $-0.30\pm0.008$ & $0.46\pm0.009$\tabularnewline
\hline 
$C_{S_{R}}$ & $0.40$ & $1.07\pm0.08$ & $0.049\pm0.0020$ & $-0.65\pm0.008$ & $-0.20\pm0.009$ & $0.45\pm0.007$\tabularnewline
\hline 
$C_{T}^{\prime}$ & $-0.07$ & $1.07\pm0.08$ & $0.037\pm0.0023$ & $-0.66\pm0.008$ & $-0.23\pm0.009$ & $0.45\pm0.007$\tabularnewline
\hline 
$C_{S_{L}}$ & $0.38$ & $1.04\pm0.08$ & $0.033\pm0.0022$ & $-0.69\pm0.008$ & $-0.23\pm0.009$ & $0.44\pm0.007$\tabularnewline
\hline 
$C_{T}^{\prime\prime}$ & $-0.05$ & $1.02\pm0.08$ & $0.029\pm0.0020$ & $-0.72\pm0.009$ & $-0.25\pm0.009$ & $0.43\pm0.006$\tabularnewline
\hline 
$C_{S_{L}}^{\prime}$ & $-0.06$ & $0.99\pm0.07$ & $0.024\pm0.0018$ & $-0.76\pm0.010$ & $-0.30\pm0.009$ & $0.42\pm0.006$\tabularnewline
\hline 
\end{tabular}\caption{\label{1d-obs-table} The BFP of the one-dimensional scenario
for set $\mathcal{S}_1$ with $\mathcal{B}\left(B_{c}^{-}\to\tau^{-}\bar{\nu}_{\tau}\right)<60\%$,
and at BFP, the average value of observables including $\left\langle d\Gamma/dq^{2}\right\rangle $,
$\left\langle A_{FB}\right\rangle $, $\left\langle P_{L}^{\Lambda_{c}}\right\rangle $,
$\left\langle P_{L}^{\tau}\right\rangle $, and $\left\langle R_{\tau/\ell}\left(\Lambda_{c}\right)\right\rangle $.
The uncertainties are due to hadronic form factors and other input parameters.}
\end{table}

\begin{table}[H]
\centering{}%
\renewcommand{\arraystretch}{1.5}
\begin{tabular}{|c|c|c|c|c|c|c|}
\hline 
WCs & BFP & $\left\langle d\Gamma/dq^{2}\right\rangle $ & $\left\langle A_{FB}\right\rangle $ & $\left\langle P_{L}^{\Lambda_{c}}\right\rangle $ & $\left\langle P_{L}^{\tau}\right\rangle $ & $\left\langle R_{\tau/\ell}\left(\Lambda_{c}\right)\right\rangle $\tabularnewline
\hline 
\hline 
SM & $C_{i}=0$ & $0.98\pm0.07$ & $0.025\pm0.0018$ & $-0.76\pm0.010$ & $-0.31\pm0.009$ & $0.42\pm0.007$\tabularnewline
\hline 
$\left(C_{S_{L}},C_{S_{R}}\right)$ & $\left(-1.07,1.28\right)$ & $1.12\pm0.08$ & $0.079\pm0.0015$ & $-0.57\pm0.009$ & $-0.15\pm0.009$ & $0.46\pm0.007$\tabularnewline
\hline 
$\left(C_{V_{L}},C_{S_{R}}\right)$ & $\left(0.29,0.05\right)$, & $1.12\pm0.08$ & $0.028\pm0.0019$ & $-0.75\pm0.010$ & $-0.30\pm0.009$ & $0.43\pm0.007$\tabularnewline
\hline 
$\left(C_{V_{L}},C_{T}\right)$ & $\left(0.35,0.03\right)$ & $1.12\pm0.08$ & $0.022\pm0.0018$ & $-0.78\pm0.010$ & $-0.31\pm0.009$ & $0.42\pm0.006$\tabularnewline
\hline 
$\left(C_{V_{L}}^{\prime},C_{S_{L}}^{\prime}\right)$ & $\left(0.31,-0.04\right)$ & $1.12\pm0.08$ & $0.025\pm0.0018$ & $-0.76\pm0.010$ & $-0.30\pm0.009$ & $0.42\pm0.007$\tabularnewline
\hline 
$\left(C_{S_{R}}^{\prime\prime},C_{T}^{\prime\prime}\right)$ & $\left(0.62,0\right)$ & $1.12\pm0.08$ & $0.025\pm0.0019$ & $-0.76\pm0.010$ & $-0.30\pm0.009$ & $0.42\pm0.007$\tabularnewline
\hline 
$\left(C_{V_{L}},C_{S_{L}}\right)$ & $\left(0.31,0.02\right)$ & $1.12\pm0.08$ & $0.025\pm0.0019$ & $-0.76\pm0.010$ & $-0.30\pm0.009$ & $0.42\pm0.007$\tabularnewline
\hline 
$\left(C_{S_{R}},C_{T}\right)$ & $\left(0.27,-0.13\right)$ & $1.11\pm0.08$ & $0.055\pm0.0025$ & $-0.58\pm0.009$ & $-0.23\pm0.009$ & $0.47\pm0.008$\tabularnewline
\hline 
$\left(C_{V_{L}}^{\prime},C_{T}^{\prime}\right)$ & $\left(0.30,0\right)$ & $1.12\pm0.08$ & $0.025\pm0.0019$ & $-0.75\pm0.010$ & $-0.30\pm0.009$ & $0.43\pm0.007$\tabularnewline
\hline 
$\left(C_{S_{L}}^{\prime\prime},C_{S_{R}}^{\prime\prime}\right)$ & $\left(-0.03,0.61\right)$ & $1.12\pm0.08$ & $0.025\pm0.0019$ & $-0.75\pm0.010$ & $-0.30\pm0.009$ & $0.42\pm0.007$\tabularnewline
\hline 
$\left(C_{S_{L}},C_{T}\right)$ & $\left(0.27,-0.15\right)$ & $1.11\pm0.08$ & $0.046\pm0.0027$ & $-0.58\pm0.009$ & $-0.25\pm0.009$ & $0.47\pm0.008$\tabularnewline
\hline 
\end{tabular}\caption{\label{2d-obs-table} The BFP of the fit for real two-dimensional scenario
for set $\mathcal{S}_1$ with $\mathcal{B}\left(B_{c}^{-}\to\tau^{-}\bar{\nu}_{\tau}\right)<60\%$,
and at BFP, the average value of observables including $\left\langle d\Gamma/dq^{2}\right\rangle $,
$\left\langle A_{FB}\right\rangle $, $\left\langle P_{L}^{\Lambda_{c}}\right\rangle $,
$\left\langle P_{L}^{\tau}\right\rangle $, and $\left\langle R_{\tau/\ell}\left(\Lambda_{c}\right)\right\rangle $.
The uncertainties are due to hadronic form factors and other input parameters.}
\end{table}

Similarly, in Fig. \ref{phen-2dc-plot}, we present the predictions of these observables of $\Lambda_b \to \Lambda_c \tau \nu_{\tau}$ decay by using the parametric space which include $C_{V_R}$, given in Table \ref{global-t2}.
The impact on the values of observables are summarized as follows:
\begin{itemize}
    \item $\bm{d\Gamma/dq^{2}:}$ Significant impact of NP is observed in the intermediate $q^{2}$ region for all NP scenarios given in Table \ref{global-t2}. Notably, the largest effect is found for the $\left(C_{S_{L}}^{\prime},C_{S_{R}}^{\prime}\right)$ scenario. 
    \item $\bm{A_{FB}:}$ The $\left(C_{V_{R}}^{\prime\prime},C_{T}^{\prime\prime}\right)$ scenario shows the maximum effects. Additionally, in the case of $\left(C_{S_{L}}^{\prime},C_{S_{R}}^{\prime}\right)$ scenario, the value is lowered from the SM results in the low $q^{2}$ region.
    \item $\bm{P_{L}^{\Lambda_{c}}:}$ The largest effects are observed for the $\left(C_{V_{R}}^{\prime\prime},C_{T}^{\prime\prime}\right)$ and $\left(C_{V_{R}}^{\prime},C_{S_{L}}^{\prime}\right)$ scenarios in the low to middle $q^{2}$ region.
    \item $\bm{P_{L}^{\tau}:}$ The $\left(C_{V_{R}}^{\prime\prime},C_{T}^{\prime\prime}\right)$ scenario shows the maximum impact across the entire $q^{2}$ range.
    \item $\bm{R_{\tau/\ell}\left(\Lambda_{c}\right):}$ A substantial effect is observed for the $\left(C_{S_{L}}^{\prime},C_{S_{R}}^{\prime}\right)$ scenario, which lowers the SM predictions in the high $q^{2}$ region.
\end{itemize}
The corresponding numerical values of these observables are also calculated in the considered scenarios and are given in Table \ref{2dc-obs-table}. The analysis demonstrates that precise measurements of these observables are also a handy tool to provide valuable insights into NP and for effectively constraining NP models, particularly, the models containing the right-handed operator. 

\begin{figure}[H]
\centering 
\begin{subfigure}[b]{0.32\textwidth}
\centering 
\includegraphics[width=5.6cm, height=3.5cm]{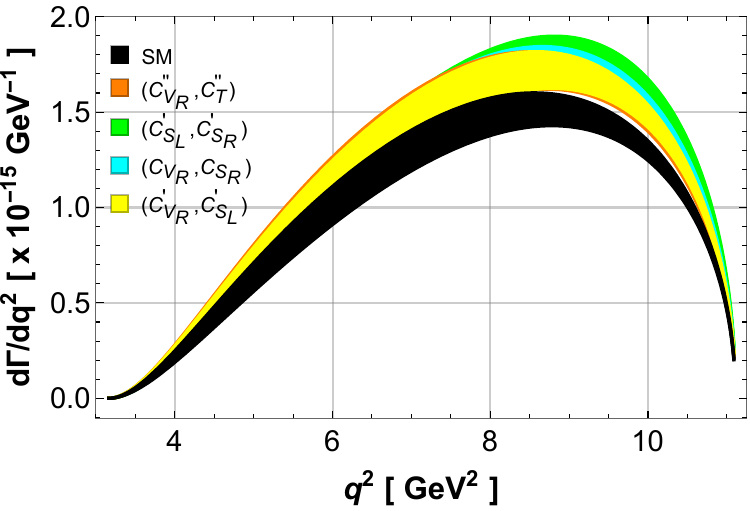}
\caption{}
\end{subfigure}
\begin{subfigure}[b]{0.32\textwidth}
\centering 
\includegraphics[width=5.6cm, height=3.5cm]{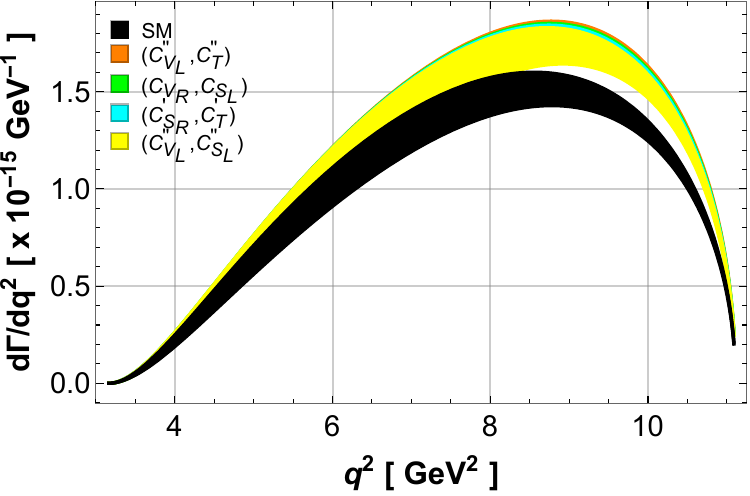} 
\caption{}
\end{subfigure}
\begin{subfigure}[b]{0.32\textwidth}
\centering 
\includegraphics[width=5.6cm, height=3.5cm]{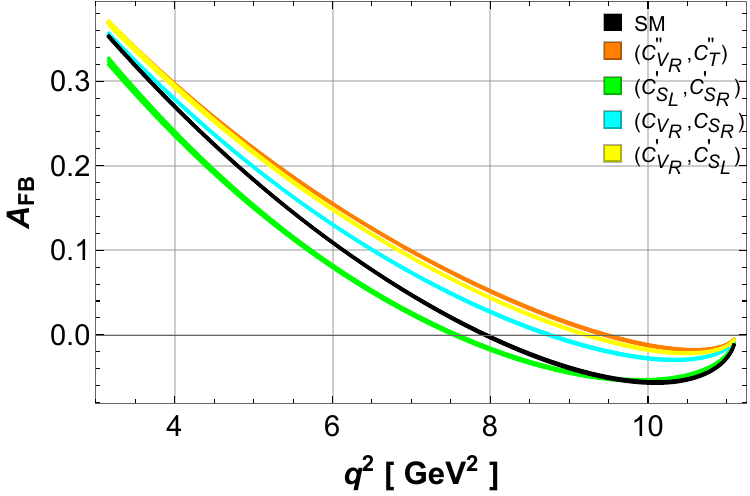}
\caption{}
\end{subfigure}
\begin{subfigure}[b]{0.32\textwidth}
\centering 
\includegraphics[width=5.6cm, height=3.5cm]{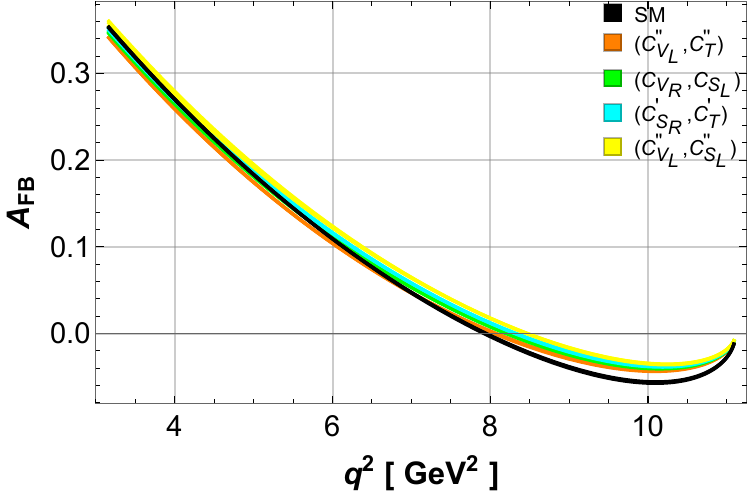}
\caption{}
\end{subfigure}
\begin{subfigure}[b]{0.32\textwidth}
\centering 
\includegraphics[width=5.6cm, height=3.5cm]{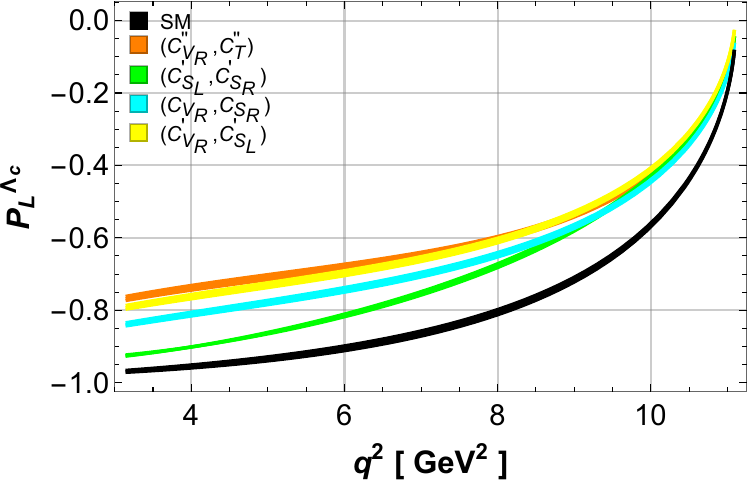} 
\caption{}
\end{subfigure}
\begin{subfigure}[b]{0.32\textwidth}
\centering 
\includegraphics[width=5.6cm, height=3.5cm]{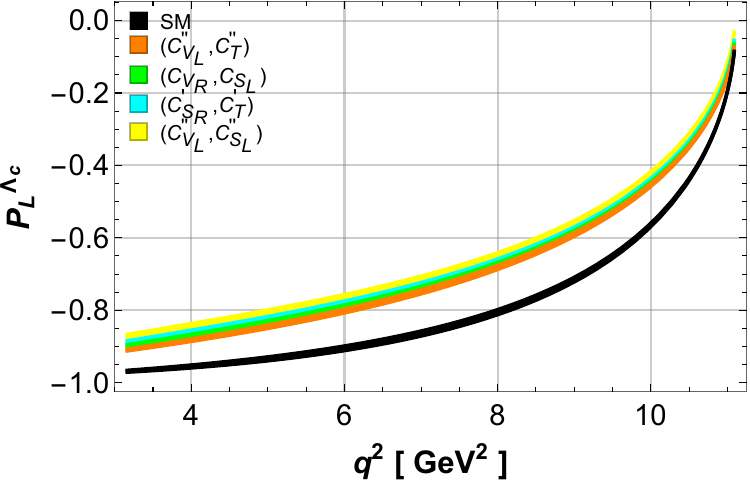}
\caption{}
\end{subfigure}
\begin{subfigure}[b]{0.32\textwidth}
\centering 
\includegraphics[width=5.6cm, height=3.5cm]{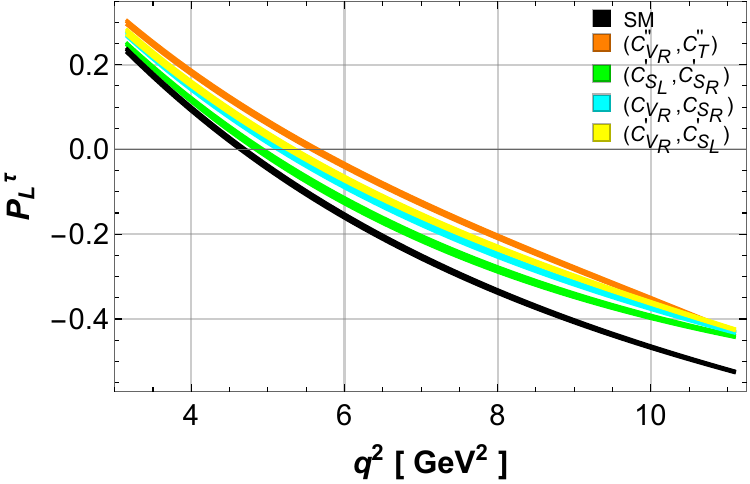} 
\caption{}
\end{subfigure}
\begin{subfigure}[b]{0.32\textwidth}
\centering 
\includegraphics[width=5.6cm, height=3.5cm]{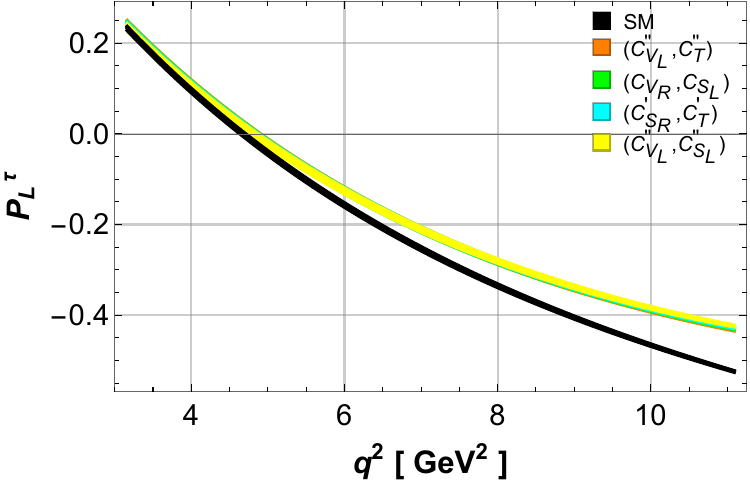}
\caption{}
\end{subfigure}
\begin{subfigure}[b]{0.32\textwidth}
\centering 
\includegraphics[width=5.6cm, height=3.5cm]{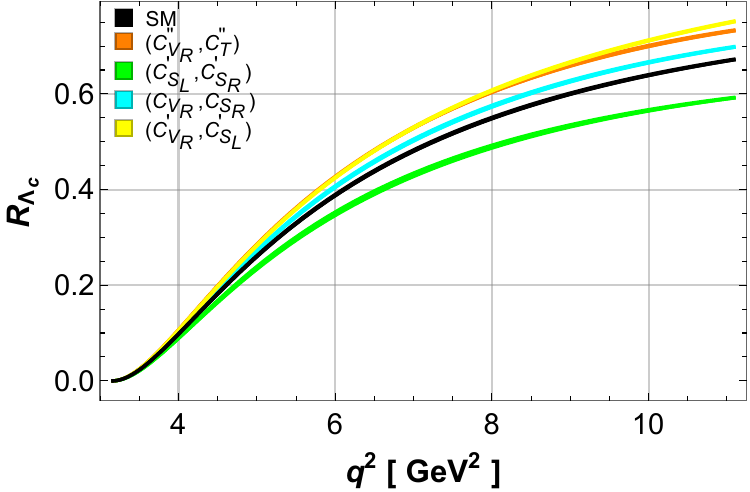}
\caption{}
\end{subfigure}
\begin{subfigure}[b]{0.32\textwidth}
\centering 
\includegraphics[width=5.6cm, height=3.5cm]{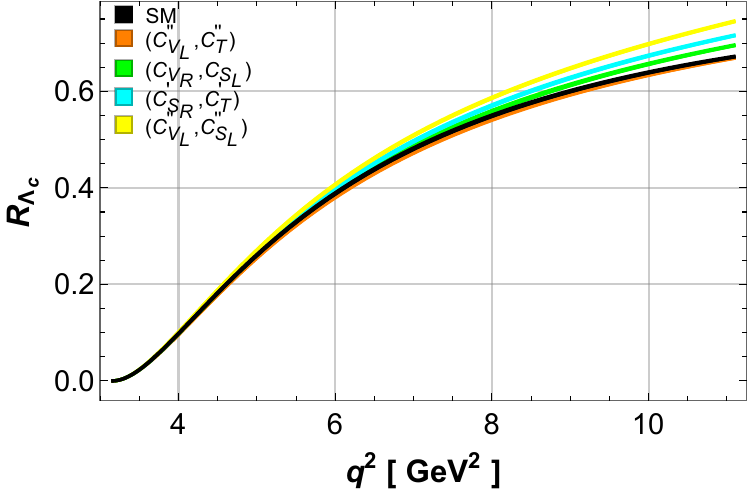} 
\caption{}
\end{subfigure}

\caption{\label{phen-2dc-plot}The $d\Gamma/dq^{2}$,
 $A_{FB}$, $P_{L}^{\Lambda_{c}}$, $P_{L}^{\tau}$, and 
$R_{\Lambda_c}\equiv R_{\tau/\ell}\left(\Lambda_{c}\right)$ observables exhibited for
various NP coupling as a function of $q^{2}$. The width of each curve
comes from the theoretical uncertainties in hadronic form factors
and quark masses. The NP WCs for the set of observables $\mathcal{S}_1$
when we include WC $C_{V_{R}}$.}
\end{figure}

\begin{table}[H]
\centering{}%
\renewcommand{\arraystretch}{1.5}
\begin{tabular}{|c|c|c|c|c|c|c|}
\hline 
WCs & BFP & $\left\langle d\Gamma/dq^{2}\right\rangle $ & $\left\langle A_{FB}\right\rangle $ & $\left\langle P_{L}^{\Lambda_{c}}\right\rangle $ & $\left\langle P_{L}^{\tau}\right\rangle $ & $\left\langle R_{\tau/\ell}\left(\Lambda_{c}\right)\right\rangle $\tabularnewline
\hline 
\hline 
SM & $C_{i}=0$ & $0.98\pm0.07$ & $0.025\pm0.0018$ & $-0.76\pm0.010$ & $-0.31\pm0.009$ & $0.42\pm0.007$\tabularnewline
\hline 
$\left(C_{V_{R}}^{\prime\prime},C_{T}^{\prime\prime}\right)$ & $\left(-0.46,0.12\right)$ & $1.12\pm0.08$ & $0.072\pm0.0019$ & $-0.57\pm0.009$ & $-0.19\pm0.009$ & $0.46\pm0.007$\tabularnewline
\hline 
$\left(C_{S_{L}}^{\prime},C_{S_{R}}^{\prime}\right)$ & $\left(-1.76,1.85\right)$, & $1.15\pm0.08$ & $0.008\pm0.0008$ & $-0.64\pm0.009$ & $-0.26\pm0.010$ & $0.43\pm0.007$\tabularnewline
\hline 
$\left(C_{V_{R}},C_{S_{R}}\right)$ & $\left(-0.30,0.45\right)$ & $1.13\pm0.08$ & $0.049\pm0.0018$ & $-0.61\pm0.009$ & $-0.23\pm0.009$ & $0.42\pm0.006$\tabularnewline
\hline 
$\left(C_{V_{R}}^{\prime},C_{S_{L}}^{\prime}\right)$ & $\left(0.31,0.76\right)$ & $1.11\pm0.08$ & $0.065\pm0.0022$ & $-0.57\pm0.009$ & $-0.21\pm0.009$ & $0.42\pm0.007$\tabularnewline
\hline 
$\left(C_{V_{L}}^{\prime\prime},C_{T}^{\prime\prime}\right)$ & $\left(0.52,-0.10\right)$ & $1.13\pm0.08$ & $0.026\pm0.0016$ & $-0.64\pm0.009$ & $-0.26\pm0.010$ & $0.42\pm0.007$\tabularnewline
\hline 
$\left(C_{V_{R}},C_{S_{L}}\right)$ & $\left(-0.40,0.53\right)$ & $1.13\pm0.08$ & $0.031\pm0.0019$ & $-0.63\pm0.009$ & $-0.26\pm0.009$ & $0.42\pm0.007$\tabularnewline
\hline 
$\left(C_{V_{L}},C_{V_{R}}\right)$ & $\left(0.32,0.01\right)$ & $1.12\pm0.08$ & $0.025\pm0.0019$ & $-0.76\pm0.010$ & $-0.31\pm0.009$ & $0.47\pm0.008$\tabularnewline
\hline 
$\left(C_{S_{R}}^{\prime},C_{T}^{\prime}\right)$ & $\left(0.59,-0.08\right)$ & $1.12\pm0.08$ & $0.035\pm0.0021$ & $-0.62\pm0.009$ & $-0.26\pm0.009$ & $0.43\pm0.007$\tabularnewline
\hline 
$\left(C_{V_{L}}^{\prime\prime},C_{S_{L}}^{\prime\prime}\right)$ & $\left(0.16,-0.74\right)$ & $1.12\pm0.08$ & $0.040\pm0.0024$ & $-0.61\pm0.009$ & $-0.26\pm0.009$ & $0.42\pm0.007$\tabularnewline
\hline 
$\left(C_{V_{R}}^{\prime\prime},C_{S_{L}}^{\prime\prime}\right)$ & $\left(0.16,-1.27\right)$ & $1.10\pm0.08$ & $0.035\pm0.0029$ & $-0.60\pm0.009$ & $-0.26\pm0.009$ & $0.42\pm0.007$\tabularnewline
\hline 
$\left(C_{V_{R}},C_{T}\right)$ & $\left(0.36,-0.28\right)$ & $1.09\pm0.08$ & $0.058\pm0.0034$ & $-0.50\pm0.009$ & $-0.25\pm0.008$ & $0.47\pm0.008$\tabularnewline
\hline 
$\left(C_{V_{R}}^{\prime},C_{T}^{\prime}\right)$ & $\left(0.73,0.20\right)$ & $1.10\pm0.09$ & $0.077\pm0.0011$ & $-0.59\pm0.008$ & $-0.12\pm0.009$ & $0.43\pm0.007$\tabularnewline
\hline 
\end{tabular}\caption{\label{2dc-obs-table} The BFP of the fit for real two WCs scenarios, after inclusion of WC $C_{V_{R}}$
for set $\mathcal{S}_1$ with $\mathcal{B}\left(B_{c}^{-}\to\tau^{-}\bar{\nu}_{\tau}\right)<60\%$,
and at BFP, the average value of observables including $\left\langle d\Gamma/dq^{2}\right\rangle $,
$\left\langle A_{FB}\right\rangle $, $\left\langle P_{L}^{\Lambda_{c}}\right\rangle $,
$\left\langle P_{L}^{\tau}\right\rangle $, and $\left\langle R_{\tau/\ell}\left(\Lambda_{c}\right)\right\rangle $.
The uncertainties are due to hadronic form factors and various input parameters.}
\end{table}

We use $\chi2/dof\sim1$ as the criterion for a good fit. For 1D fits with $dof = 3$ (set $\mathcal{S}_{1}$), we consider scenarios with $\chi2<3$ as allowed; for 2D fits with $dof = 2$ (set $\mathcal{S}_{1}$), we take $\chi2<2$. In the 1D case, three degenerate scenarios involving $C_{V_L}$, $C_{V_L}^{\prime}$ and $C_{V_L}^{\prime\prime}$ and the scenario with $C_{S_L}^{\prime\prime}$ satisfy $\chi2/dof\sim1$ and are considered well-fitted. Due to observed deviations in phenomenological observables (in $\Lambda_{b}\rightarrow\Lambda_{c}\tau\bar{\nu}_{\tau}$),we relax the criterion to $\chi2/dof<1.7$ , allowing two additional 1D scenarios: $C_{S_R}$ and $C_T$. In the 2D case, all scenarios satisfy $\chi2<2$. Among them, $\left(C_{S_{L}},C_{S_{R}}\right)$, shows the most significant deviation across all observables, followed by $\left(C_{S_{R}},C_{T}\right)$, and the three degenerate combinations involving $\left(C_{S_{L}},C_{T}\right)$, $\left(C_{S_{L}}^{\prime},C_{T}^{\prime}\right)$ and $\left(C_{S_{L}}^{\prime\prime},C_{T}^{\prime\prime}\right)$.

\section{Correlating different Physical observables}\label{sumrule}
In Section \ref{sec1}, we have mentioned that the measurements of $R_{\tau/\ell}\left(\Lambda_{c}\right)$, and $R_{\tau/\mu}\left(J/\psi\right)$ are prone to various uncertainties, therefore, it is useful to express them in terms of the observables
with better theoretical control, \textit{i.e.,} $R_{\tau/\mu,e}\left(D\right)$ and $R_{\tau/\mu,e}\left(D^*\right)$. For the first time, these relations are derived in \cite{Fedele:2022iib} and named as the sum rules. In our case, we can write the similar sum rule from the Eqs. (\ref{eqn1}, \ref{eqn2} and \ref{eqn7}) as:
\begin{equation}
\frac{R_{\tau/\ell}\left(\Lambda_{c}\right)}{R_{\tau/\ell}^{\text{SM}}\left(\Lambda_{c}\right)}=0.275\frac{R_{\tau/{\mu,e}}\left(D\right)}{R^{\text{SM}}_{\tau/{\mu,e}}\left(D\right)}+0.725\frac{R_{\tau/{\mu,e}}\left(D^{*}\right)}{R^{\text{SM}}_{\tau/{\mu,e}}\left(D^{*}\right)}+x_{1},\label{sum-n1}
\end{equation}
where small remainder $x_{1}$ can be approximated in terms of WCs at a scale
$m_{b}$ as \cite{Blanke:2018yud}:
\begin{eqnarray}
x_{1}& = &\left.\Re\left[\left(1+C_{V_{L}}\right)\left(0.607C_{V_{R}}^{*}+0.011C_{S_{R}}^{*}+0.341C_{T}^{*}\right)\right]+\Re\left[C_{V_{R}}\left(0.090C_{S_{L}}^{*}+0.080C_{S_{R}}^{*}+0.202C_{T}^{*}\right)\right]\right.\nonumber \\
 &  & \left.+0.013\left(\left|C_{S_{R}}\right|^{2}+\left|C_{S_{L}}\right|^{2}\right)+0.520\Re\left[C_{S_{L}}\left(C_{S_{R}}\right)^{*}\right]-0.044\left|C_{T}\right|^{2}\right.\label{rem1}.
\end{eqnarray}
In Eq. (\ref{sum-n1}), we can see that in $R_{\tau/\ell}\left(\Lambda_c\right)$, the relative weight of the $R_{\tau/{\mu,e}}\left(D^*\right)/R^{\text{SM}}_{\tau/{\mu,e}}\left(D^*\right)$ is $72\%$, and hence with better control over the errors in its measurements and the SM predictions, will help us to predict $R_{\tau/\ell}\left(\Lambda_c\right)$ to good accuracy. Depending on the observation that  $R_{\tau/\mu}\left(J/\psi\right)$ exhibits the same behaviour as $R_{\tau/\ell}\left(\Lambda_c\right)$, therefore, it will be interesting to see if we can establish a similar relation for $R_{\tau/\mu}\left(J/\psi\right)$. The required some rule can be obtained from equations \ref{eqn1}, \ref{eqn2}, and \ref{eqn5} which can be written as follows:
\begin{align}
\frac{R_{\tau/\mu}\left(J/\psi\right)}{R^{\text{SM}}_{\tau/\mu}\left(J/\psi\right)} & =0.006\frac{R_{\tau/{\mu,e}}\left(D\right)}{R^{\text{SM}}_{\tau/{\mu,e}}\left(D\right)}+0.994\frac{R_{\tau/\mu}\left(D^{*}\right)}{R^{\text{SM}}_{\tau/{\mu,e}}\left(D^{*}\right)}+x_{2},\label{sum2}
\end{align}
where the remainder $x_{2}$ can be written as
\begin{eqnarray}
x_{2}& = &\left.-\Re\left[\left(1+C_{V_{L}}\right)\left(0.001C_{V_{R}}^{*}+0.018C_{S_{R}}^{*}+0.259C_{T}^{*}\right)\right]+\Re\left[C_{V_{R}}\left(0.091C_{S_{L}}^{*}-0.109C_{S_{R}}^{*}+0.005C_{T}^{*}\right)\right]\right.\nonumber \\
 &  & \left.-0.006\left(\left|C_{S_{R}}\right|^{2}+\left|C_{S_{L}}\right|^{2}\right)-13.701\left|C_{T}\right|^{2}\right.\label{rem2}.
\end{eqnarray}
In Eq. (\ref{sum2}), the LFU ratio $R_{\tau/\mu}\left(J/\psi\right)$ normalized with the corresponding SM prediction, has negligible dependence on the $R_{\tau/{\mu,e}}\left(D\right)/R^{\text{SM}}_{\tau/{\mu,e}}\left(D\right)$, therefore, the refined measurement of the $R_{\tau/{\mu,e}}\left(D^*\right)$ will help us to get good control over $R_{\tau/\mu}\left(J/\psi\right)$. Also, we can see that if $R_{\tau/{\mu,e}}\left(D\right)$ and $R_{\tau/{\mu,e}}\left(D^{*}\right)$ are enhanced
over their SM values, it follows that $R_{\tau/\mu}\left(J/\psi\right)$ must also experience an enhancement. By incorporating the BFPs of Tables \ref{1d-table} and \ref{2d-table-1}, we find that remainders $x_1$ and $x_2$ in Eqs. (\ref{rem1} and \ref{rem2}), respectively, both are approximated $<10^{-3}$ for all the
NP WCs, which ensure the validity of these sum rules. Being model-independent, these sum rules remain valid in any NP model, indicating that future measurements of $R_{\tau/\ell}\left(\Lambda_{c}\right)$, and $R_{\tau/\ell}\left(J/\psi\right)$ can serve as essential crosschecks for the measurements of $R_{\tau/{\mu,e}}\left(D\right)$ and $R_{\tau/{\mu,e}}\left(D^{*}\right)$. Using the values from Eqs. (\ref{HFLAV-RDDs},\ref{SM-RDDs}) in Eq. (\ref{sum-n1}), we can predict
\begin{align*}
R_{\tau/\ell}\left(\Lambda_{c}\right) & =R^{\text{SM}}_{\tau/\ell}\left(\Lambda_{c}\right)\left(1.135\pm0.045\right) =0.368\pm0.015\pm0.005,
\end{align*}
as given in ref. \cite{Fedele:2022iib}. In this result, the first error comes from the experimental measurements in LFU ratios of $D$ and $D^*$ and the second is due to the form factors uncertainties in the SM predictions of these corresponding ratios. The numerical value is not much different from the previously reported which indicate that the most recent data of $R_{\tau/{\mu,e}}\left(D^{*}\right)$ supports the validity of the above sum rule. Similarly, for the other sum rule (c.f. Eqs. (\ref{sum2})), we have
\begin{align*}
R_{\tau/\mu}\left(J/\psi\right) & =R^{\text{SM}}_{\tau/\mu}\left(J/\psi\right) \left(1.130\pm0.052\right) =0.292\pm0.013\pm0.043.
\end{align*}
It is evident that the SM value of $R_{\tau/\mu}\left(J/\psi\right)$ as well as its updated value derived from the sum rule using the latest data, both fall below the experimental measurements and exhibit a consistent pattern, similar to that of observed in $R_{\tau/{\mu,e}}\left(D^{*}\right)$.  However, in the case of $R_{\tau/\mu}\left(J/\psi\right)$, even though its tensor form factors are not precisely calculated yet, the theoretically predicted values are quite small compared to its experimental value with large uncertainties, $0.71 \pm 0.18\pm0.17$. We expect several planned and current experiments to explore this value further.

We compared our results with some recent literature and the corresponding results are appended in  Table \ref{tab-comp}. This presents a direct comparison between our updated 1D fit results (with real WCs at scale $2\text{TeV}$) and those reported in ref. \cite{Endo:2025cvu} at scale $\mu_{b}$. For each scenario ($C_{S_R}$, $C_{S_L}$, and $C_{T}$), we list the best-fit points (BFPs), pull values, and the deviations in remainders of the sum rules $x_1$ and $x_2$. Since our analysis focuses on real WCs, we note that only the $C_{S_{R}}$ scenario directly match our framework. In our analysis, the $C_{S_{R}}$ scenario yields the BFP of $0.40$ and a pull of $3.17\sigma$, indicating a moderate improvement over the SM. This is consistent with \cite{Endo:2025cvu}, where a pull of $3.9$ was reported. For $C_{S_L}$, we obtain $C_{S_L}=0.38$ ($pull=2.6$), while \cite{Endo:2025cvu} uses a complex WC $-0.57\pm0.86i$ ($pull=4.3$). The comparison is limited due to the real–complex difference, but both fits yield $10^{-3}$ sum rule remainders. In the $C_T$ scenario, our result $C_T=-0.17$ ($pull=3.3$) differs from the reference value $0.02\pm0.13i$ ($pull=3.8$). However, we note that the remainders $x_1$ and $x_2$ between both fits are $1-2\%$ and $0.1\%$ respectively.

\begin{table}[h]
\centering{}
\begin{tabular}{|c|c|c|c|c|}
\hline 
WCs & BFP & Pull & $x_{1}$ & $x_{2}$\tabularnewline
\hline 
\hline 
\multirow{2}{*}{$C_{S_{R}}$} & $0.40\left(2TeV\right)$ & $3.2$ & $0.002$ & $-0.003$\tabularnewline
\cline{2-5} \cline{3-5} \cline{4-5} \cline{5-5} 
 & $0.182\left(\mu_{b}\right)$ & $3.9$ & $0.0007$ & $-0.001$\tabularnewline
\hline 
\multirow{2}{*}{$C_{S_{L}}$} & $0.38\left(2TeV\right)$ & $2.6$ & $0.000$ & $0.000$\tabularnewline
\cline{2-5} \cline{3-5} \cline{4-5} \cline{5-5} 
 & $-0.57\pm0.86i\left(\mu_{b}\right)$ & $4.3$ & $0.001$ & $-0.0008$\tabularnewline
\hline 
\multirow{2}{*}{$C_{T}$} & $-0.17\left(2TeV\right)$ & $3.3$ & $-0.010$ & $0.019$\tabularnewline
\cline{2-5} \cline{3-5} \cline{4-5} \cline{5-5} 
 & $0.02\pm0.13i\left(\mu_{b}\right)$ & $3.8$ & $0.001$ & $0.008$\tabularnewline
\hline 
\end{tabular}\caption{\label{tab-comp}The results of 1-dimensional fit for real WCs scenarios which include
BFPs, pull and remainder of sum rule for $R_{\tau/\ell}\left(\Lambda_{c}\right)$ $x_{1}$ and $R_{\tau/\mu}\left(J/\psi\right)$ $x_{2}$ for
set $\mathcal{S}_1$ at scale $2TeV$ are shown in each first sub-row of each scenario.
Second sub-row for each scenario is results taken from \cite{Endo:2025cvu}
at scale $\mu_{b}$.}

\end{table}

Furthermore, a comparison with \cite{Iguro:2024hyk} is included in Table XIV. Our fitted values for $C_{V_L}=0.31$ ($pull=3.9$), $C_{S_R}=0.40$ ($pull=3.2$), and $C_{T}=-0.17$ ($pull=3.3$) show good agreement with the corresponding values $0.42$, $0.77$, and $0.30$ from \cite{Iguro:2024hyk}, with all scenarios yielding small $x_{1}$ and $x_{2}$ sum rule remainders. This supports the consistency of our results with existing literature at the scale $2\text{TeV}$.

\begin{table}[h]
\centering{}
\begin{tabular}{|c|c|c|c|c|}
\hline 
WCs & BFP & Pull & $x_{1}$ & $x_{2}$\tabularnewline
\hline 
\hline 
$C_{V_{L}}$ & $0.31\left(0.15\right)$ & $3.9$ & $0.000$ & $0.000$\tabularnewline
\hline 
 & $0.42\left(0.12\right)$ & - & $0.000$ & $0.000$\tabularnewline
\hline 
$C_{V_{R}}$ & $-0.26\left(-0.26\right)$ & $1.85$ & $-0.032$ & $0.0001$\tabularnewline
\hline 
 & $0.51\left(0.15\right)$ & - & $0.062$ & $-0.0001$\tabularnewline
\hline 
$C_{S_{L}}$ & $0.38\left(0.27\right)$ & $2.6$ & $0.000$ & $0.000$\tabularnewline
\hline 
 & $0.80\left(0.22\right)$ & - & $0.001$ & $-0.001$\tabularnewline
\hline 
$C_{S_{R}}$ & $0.40\left(0.23\right)$ & $3.2$ & $0.002$ & $-0.003$\tabularnewline
\hline 
 & $0.77\left(0.22\right)$ & - & $0.004$ & $-0.006$\tabularnewline
\hline 
$C_{T}$ & $-0.17\left(0.09\right)$ & $3.3$ & $-0.010$ & $0.019$\tabularnewline
\hline 
 & $0.30\left(0.07\right)$ & - & $0.017$ & $0.021$\tabularnewline
\hline 
\end{tabular}

\caption{The results of 1-dimensional fit for real WCs scenarios which include
BFPs, pull and remainders of sum rule $x_{1}$ and $x_{2}$ for $R_{\tau/\ell}\left(\Lambda_{c}\right)$  and $R_{\tau/\mu}\left(J/\psi\right)$, respectively, for
set $\mathcal{S}_1$ at scale $2\text{TeV}$in the first sub-row of each scenario.
The second sub-row in each scenario shows the corresponding results from \cite{Iguro:2024hyk}, which are
also calculated at $2\text{TeV}$.}

\end{table}
Similar to the sum rule, exploring the correlations among phenomenological observables of $\Lambda_b \to \Lambda_c\tau\nu_{\tau}$ decay,  such as $d\Gamma/dq^{2}$,
 $A_{FB}$, $P_{L}^{\Lambda_{c}}$, $P_{L}^{\tau}$, and
$R_{\tau/l}\left(\Lambda_{c}\right)$ by using $1 \sigma$ parametric space of two-dimensional scenarios would be interesting and insightful. For this purpose, we employed the expressions provided in Appendix \ref{obsRLc}, and the corresponding plots are presented in Fig. \ref{corr-2d}. We observed that the differential decay rate \( \frac{d\Gamma}{dq^{2}} \), when correlated with \( A_{FB} \), \( P_{L}^{\Lambda_{c}} \), \( P_{L}^{\tau} \), and \( R_{\tau/l}(\Lambda_{c}) \), demonstrates a high degree of positive correlation for the WCs combinations \( (C_{S_{L}}, C_{S_{R}}) \), \( (C_{S_{R}}, C_{T}) \), and \( (C_{S_{L}}, C_{T}) \). Among these, the \( (C_{S_{L}}, C_{S_{R}}) \) shows direct correlations were found between $d\Gamma/dq^{2}$ and $P_{L}^{\tau}$ for WC $\left(C_{S_{L}},C_{S_{R}}\right)$, and has the largest $p-$value, indicating a stronger impact on the fit to the experimental data. A moderate negative correlation was found for \( (C_{V_{L}}, C_{T}) \) across all observables except for \( P_{L}^{\tau} \), which shows a positive behavior. Additionally, \( P_{L}^{\tau} \) did not vary significantly with changes in \( \frac{d\Gamma}{dq^{2}} \) for \( (C_{V_{L}}, C_{T}) \).

The correlation plots of $A_{FB}$ with other observables demonstrate a strong correlation for the combinations \( (C_{S_{L}}, C_{S_{R}}) \), \( (C_{S_{R}}, C_{T}) \), and \( (C_{S_{L}}, C_{T}) \). We found direct correlation between $A_{FB}$ and $P_{L}^{\Lambda_{c}}$ for three degenerate scenarios involving WCs $\left(C_{S_{L}},C_{T}\right)$. Moreover, for $\left(C_{V_{L}},C_{T}\right)$, we observe a direct correlation passing through the SM value, shown by a red star. A similar behavior of $\left(C_{V_{L}},C_{T}\right)$ is observed when $P_{L}^{\Lambda_{c}}$ is correlated with $P_{L}^{\tau}$ and $R_{\tau/\ell}\left(\Lambda_{c}\right)$ as well as when $P_{L}^{\tau}$ and $R_{\tau/\ell}\left(\Lambda_{c}\right)$ are correlated. It is worth mentioning that the correlations among the observables are illustrated using the parametric space of those WC combinations that exhibit significant deviations from their measured values, as discussed in the phenomenology section. Fig. \ref{corr-2dc-1} and \ref{corr-2dc-2} show the correlation plots when $C_{V_{R}}$ is included in the analysis.

\begin{figure}[H]
\centering 
\begin{subfigure}[b]{0.32\textwidth}
\centering 
\includegraphics[width=5.6cm, height=3.5cm]{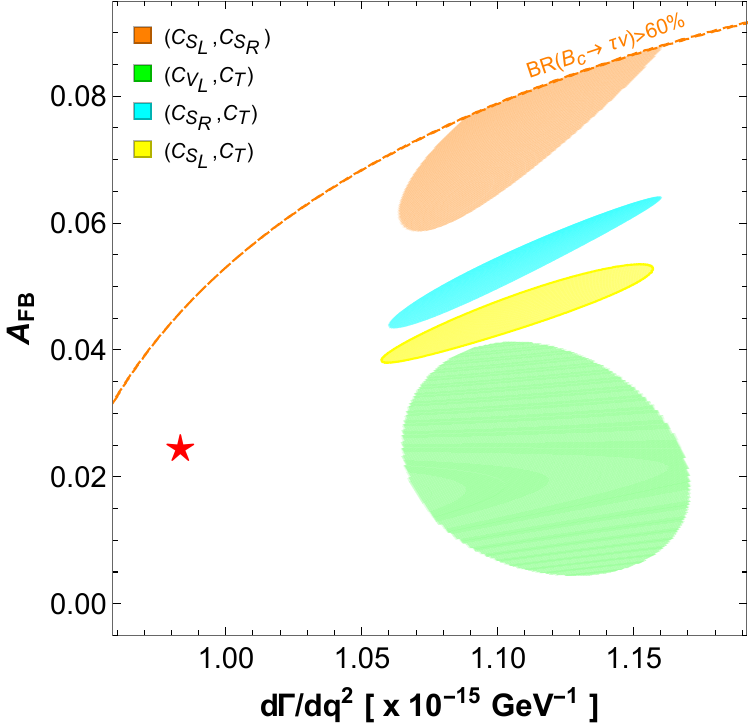}
\caption{}
\end{subfigure}
\begin{subfigure}[b]{0.32\textwidth}
\centering 
\includegraphics[width=5.6cm, height=3.5cm]{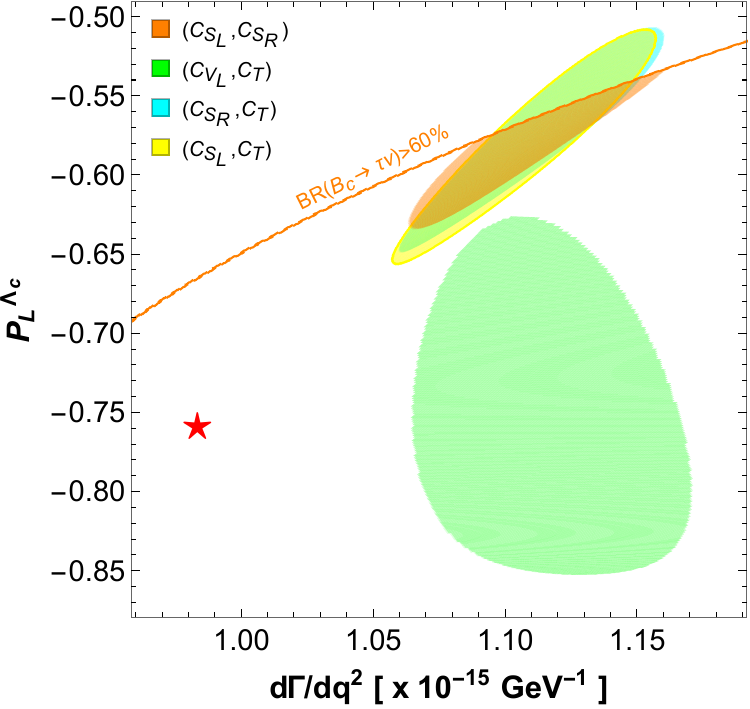} 
\caption{}
\end{subfigure}
\begin{subfigure}[b]{0.32\textwidth}
\centering 
\includegraphics[width=5.6cm, height=3.5cm]{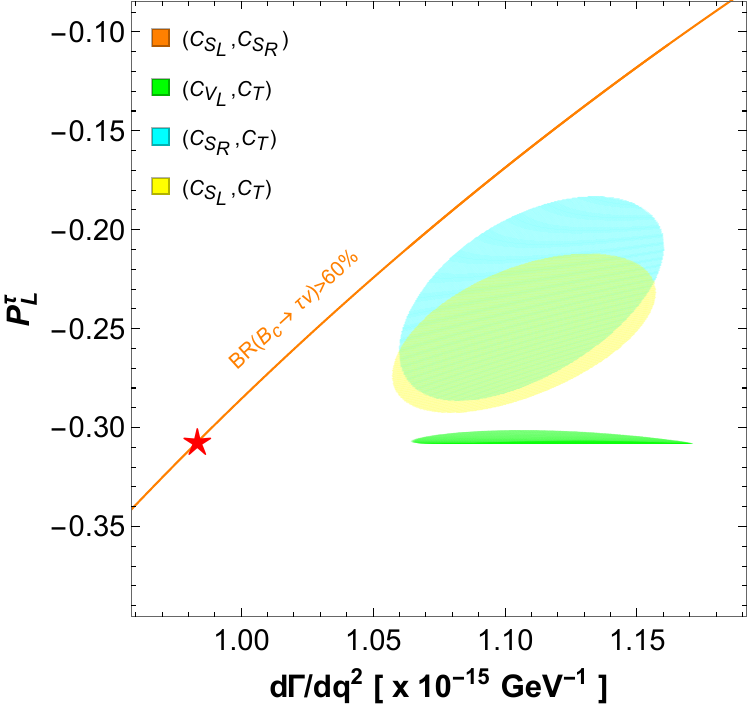}
\caption{}
\end{subfigure}
\begin{subfigure}[b]{0.32\textwidth}
\centering 
\includegraphics[width=5.6cm, height=3.5cm]{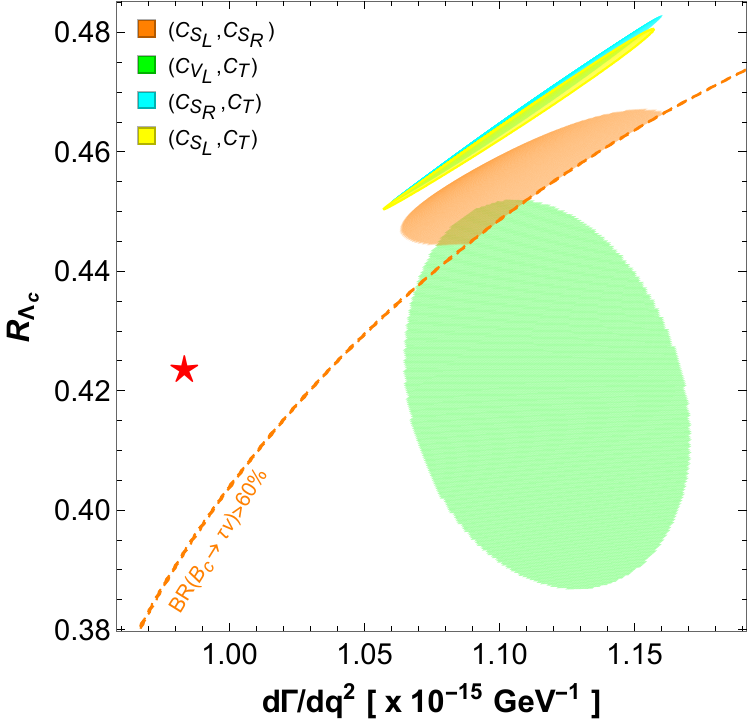}
\caption{}
\end{subfigure}
\begin{subfigure}[b]{0.32\textwidth}
\centering 
\includegraphics[width=5.6cm, height=3.5cm]{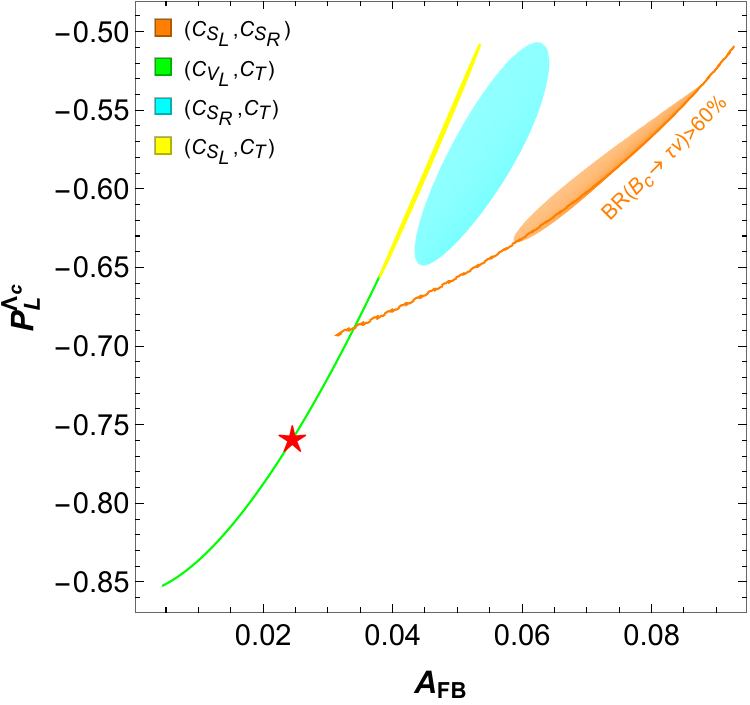} 
\caption{}
\end{subfigure}
\begin{subfigure}[b]{0.32\textwidth}
\centering 
\includegraphics[width=5.6cm, height=3.5cm]{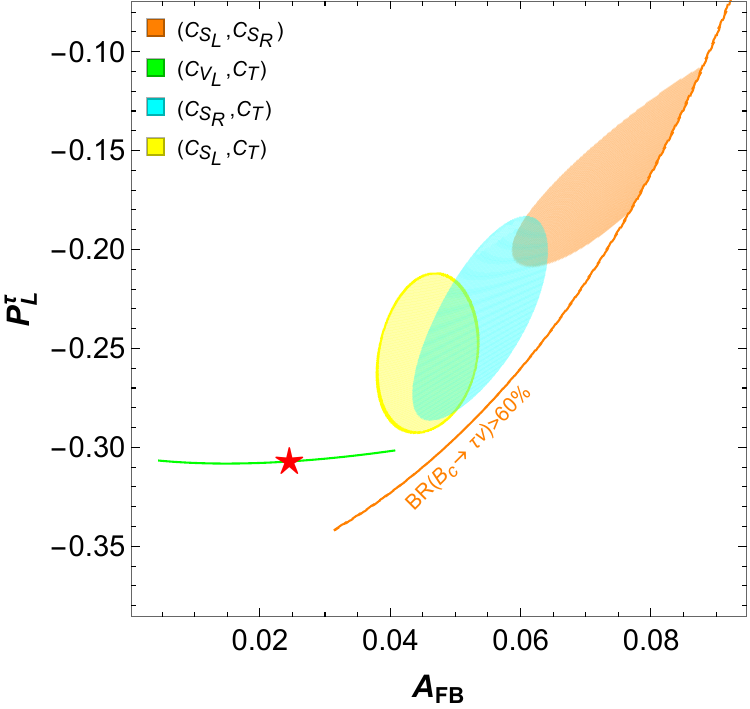}
\caption{}
\end{subfigure}
\begin{subfigure}[b]{0.32\textwidth}
\centering 
\includegraphics[width=5.6cm, height=3.5cm]{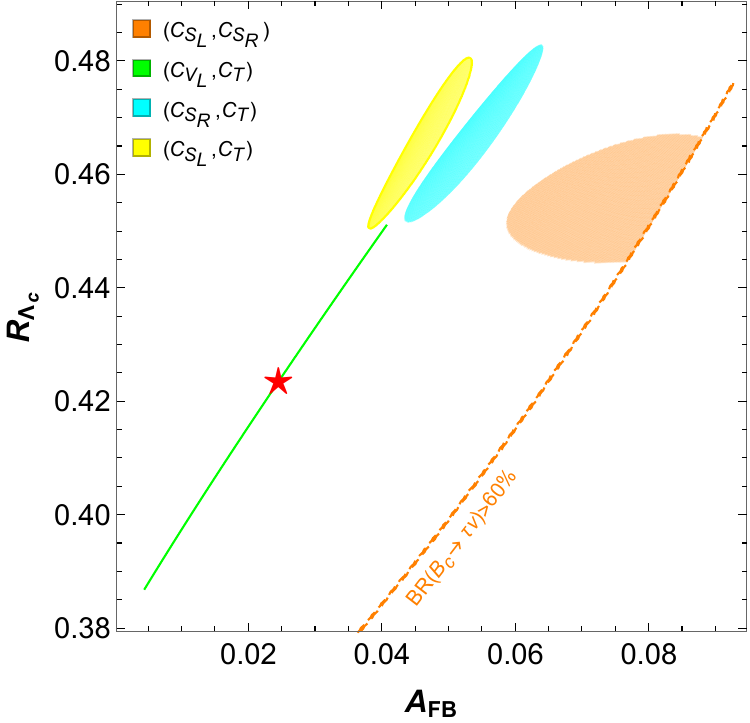}
\caption{}
\end{subfigure}
\begin{subfigure}[b]{0.32\textwidth}
\centering 
\includegraphics[width=5.6cm, height=3.5cm]{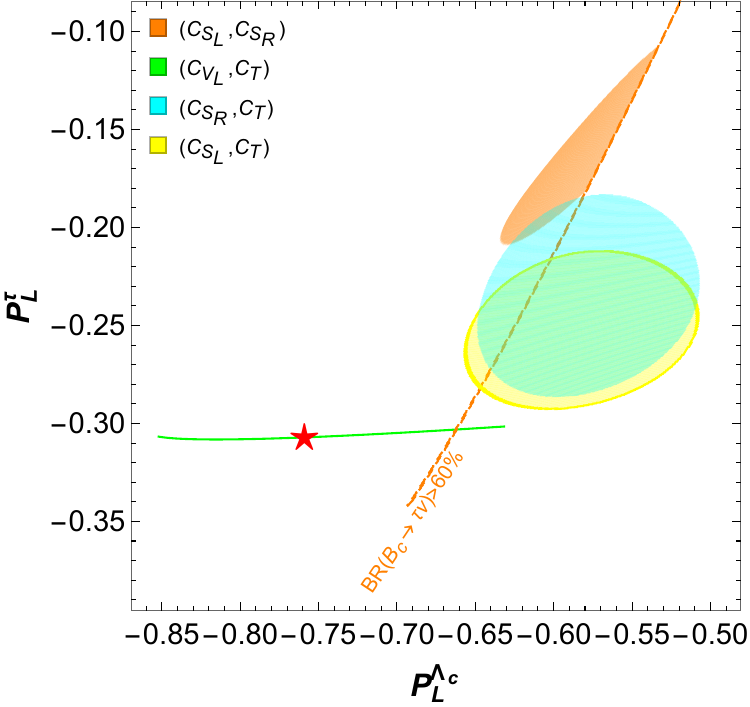} 
\caption{}
\end{subfigure}
\begin{subfigure}[b]{0.32\textwidth}
\centering 
\includegraphics[width=5.6cm, height=3.5cm]{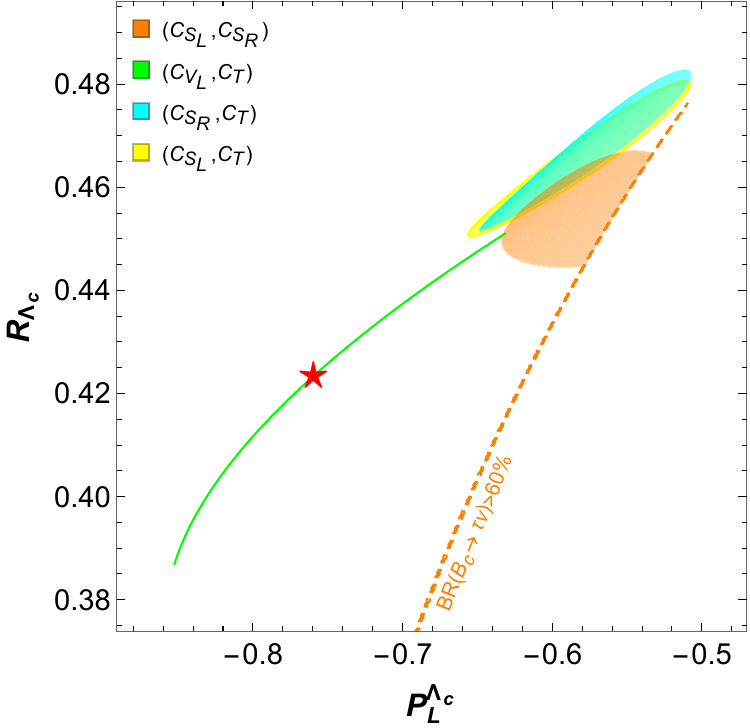}
\caption{}
\end{subfigure}
\begin{subfigure}[b]{0.32\textwidth}
\centering 
\includegraphics[width=5.6cm, height=3.5cm]{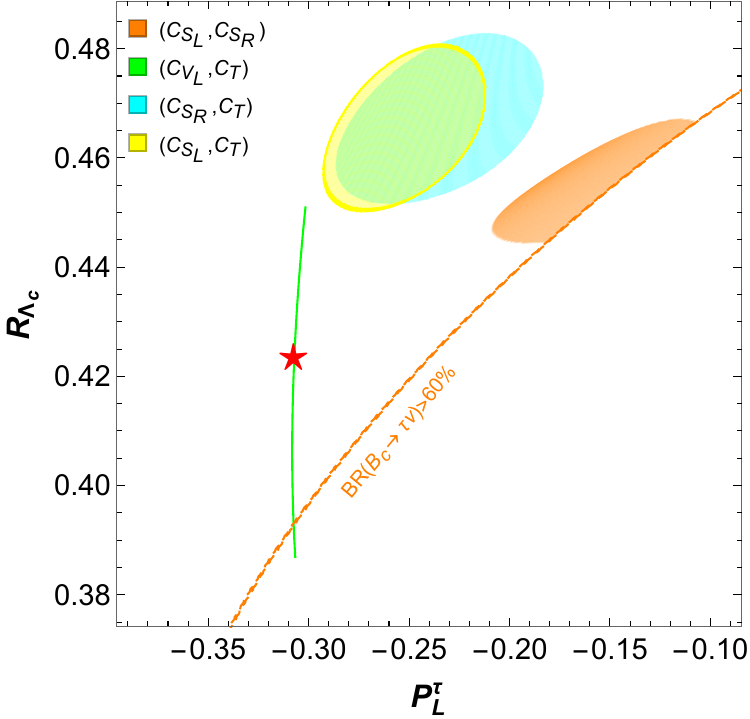}
\caption{}
\end{subfigure}

\caption{\label{corr-2d}Preferred $1\sigma$ regions for the four two-WCs  scenarios for set $\mathcal{S}_1$ in the $d\Gamma/dq^{2}-A_{FB}$ plane (first row), $d\Gamma/dq^{2}-P_{L}^{\Lambda_{c}}$ plane (second row), $d\Gamma/dq^{2}-P_{L}^{\tau}$ plane (third row), $d\Gamma/dq^{2}-R_{\tau/\ell}\left(\Lambda_{c}\right)$ plane (fourth row), and $A_{FB}-P_{L}^{\Lambda_{c}}$ plane (last row), for the $BR\left(B_{c}\to\tau\bar{\nu}_\tau\right)<60\%$. The regions
of the plot in the left panel correspond to the unprimmed WCs scenarios, the middle corresponds to primmed and the right corresponds to double primmed WCs scenarios. The solid lines refer to a constraint on $\mathcal{B}\left(B^-_{c}\to\tau^-\nu_\tau\right)<60\%$. The red stars represent SM predictions. Legends are same as depicted in \ref{phen-2d-plot}}
\end{figure}

\begin{figure}[H]
\centering 
\begin{subfigure}[b]{0.32\textwidth}
\centering 
\includegraphics[width=5.6cm, height=3.5cm]{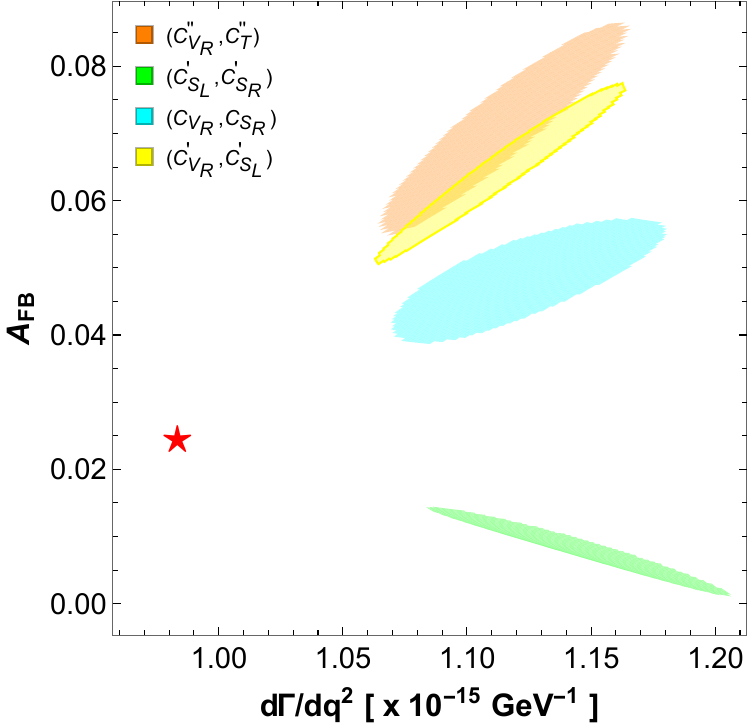}
\caption{}
\end{subfigure}
\begin{subfigure}[b]{0.32\textwidth}
\centering 
\includegraphics[width=5.6cm, height=3.5cm]{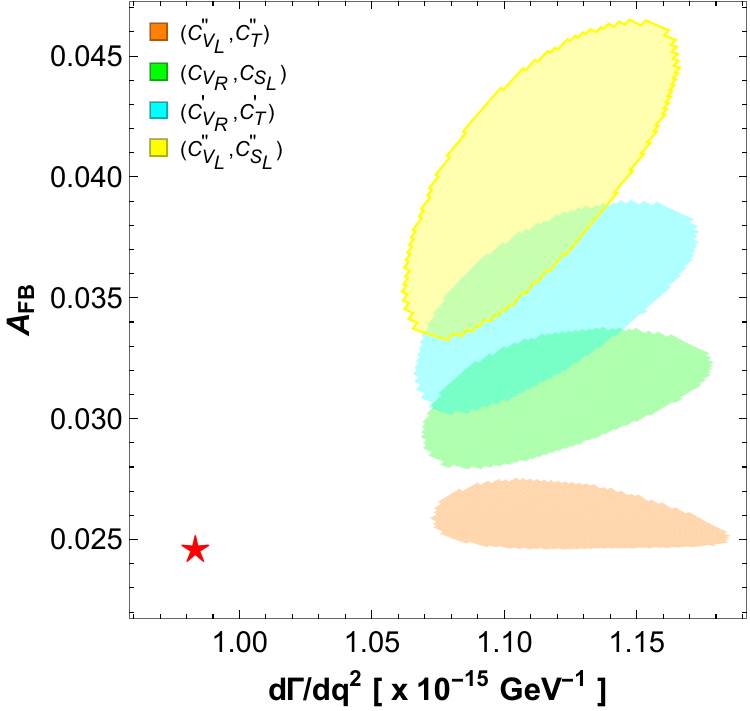} 
\caption{}
\end{subfigure}
\begin{subfigure}[b]{0.32\textwidth}
\centering 
\includegraphics[width=5.6cm, height=3.5cm]{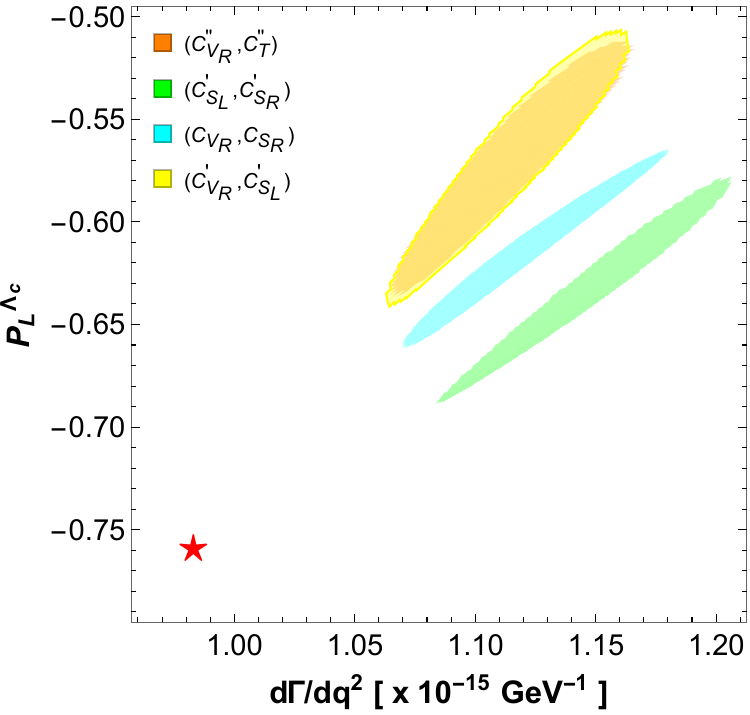}
\caption{}
\end{subfigure}
\begin{subfigure}[b]{0.32\textwidth}
\centering 
\includegraphics[width=5.6cm, height=3.5cm]{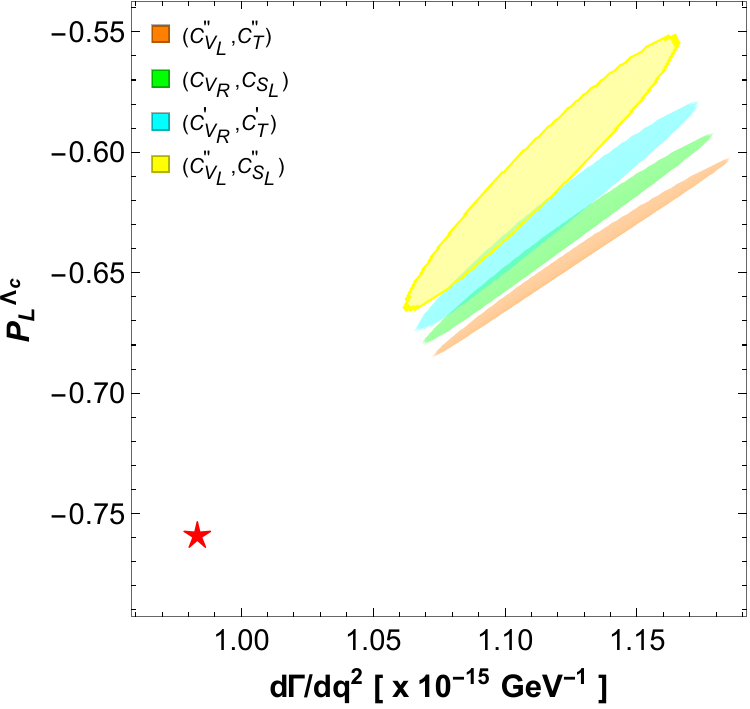}
\caption{}
\end{subfigure}
\begin{subfigure}[b]{0.32\textwidth}
\centering 
\includegraphics[width=5.6cm, height=3.5cm]{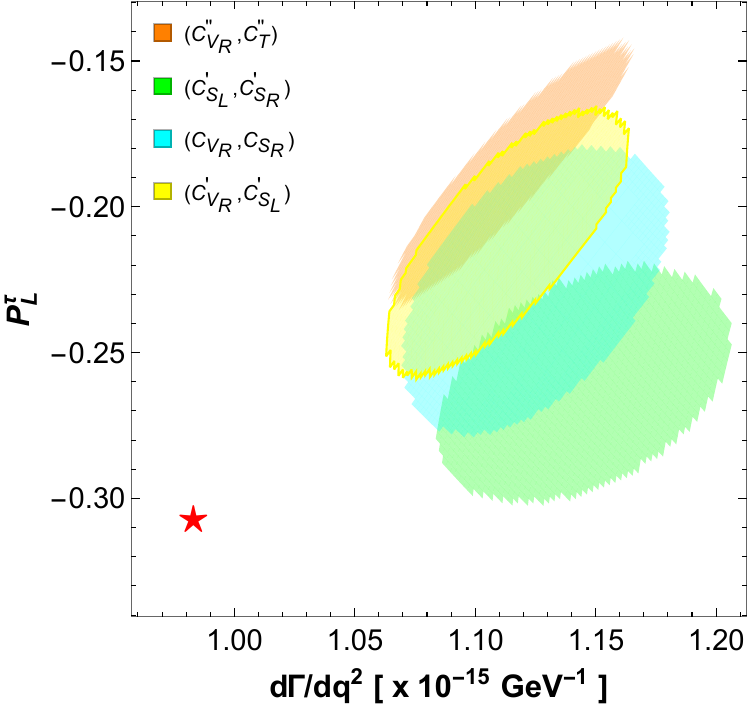} 
\caption{}
\end{subfigure}
\begin{subfigure}[b]{0.32\textwidth}
\centering 
\includegraphics[width=5.6cm, height=3.5cm]{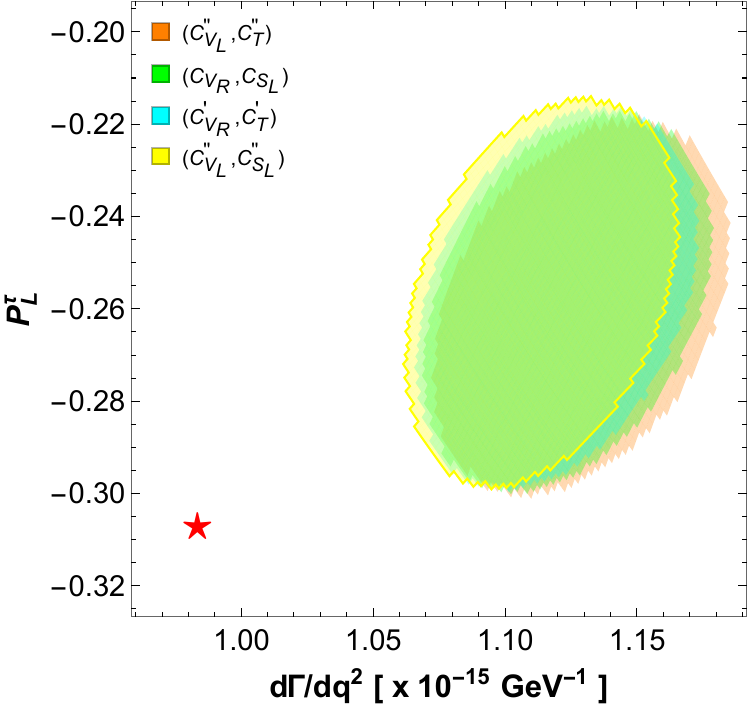}
\caption{}
\end{subfigure}
\begin{subfigure}[b]{0.32\textwidth}
\centering 
\includegraphics[width=5.6cm, height=3.5cm]{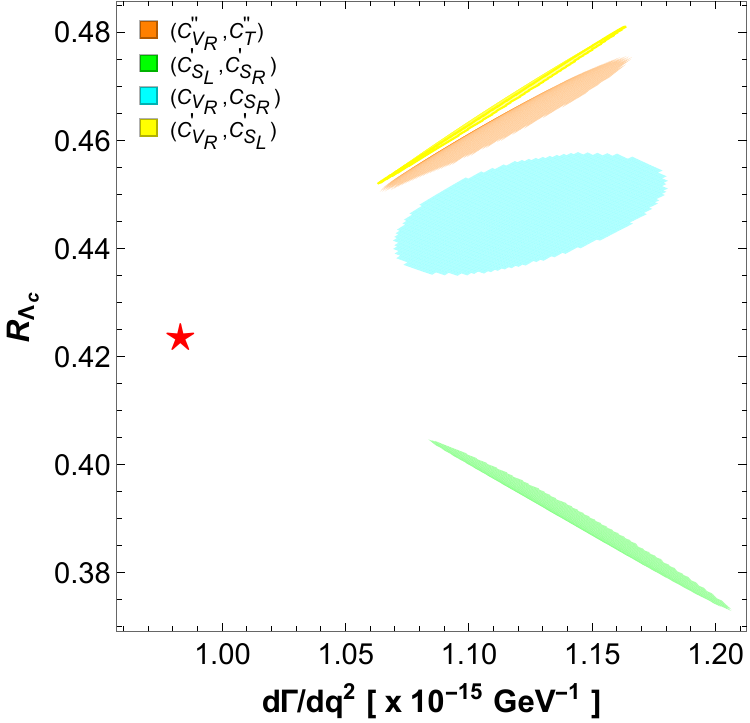}
\caption{}
\end{subfigure}
\begin{subfigure}[b]{0.32\textwidth}
\centering 
\includegraphics[width=5.6cm, height=3.5cm]{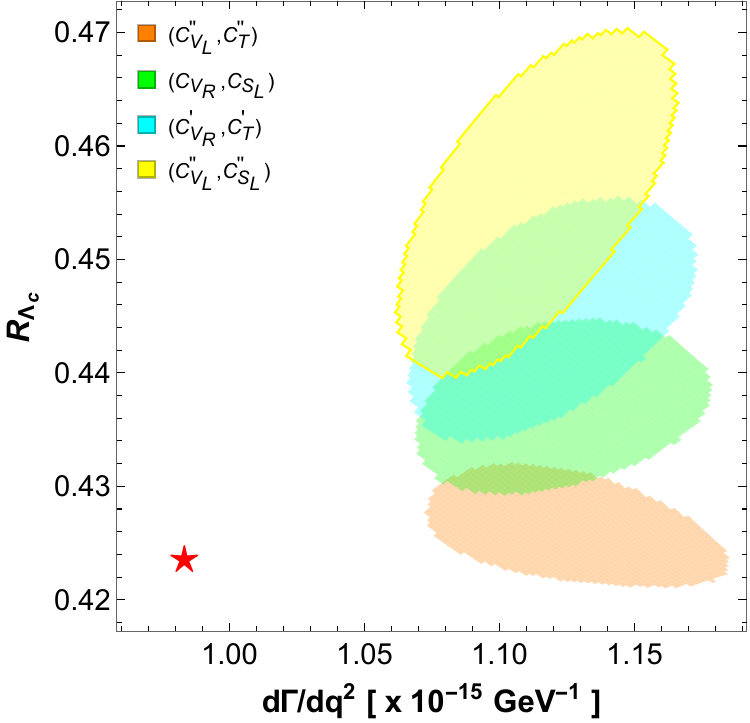} 
\caption{}
\end{subfigure}
\begin{subfigure}[b]{0.32\textwidth}
\centering 
\includegraphics[width=5.6cm, height=3.5cm]{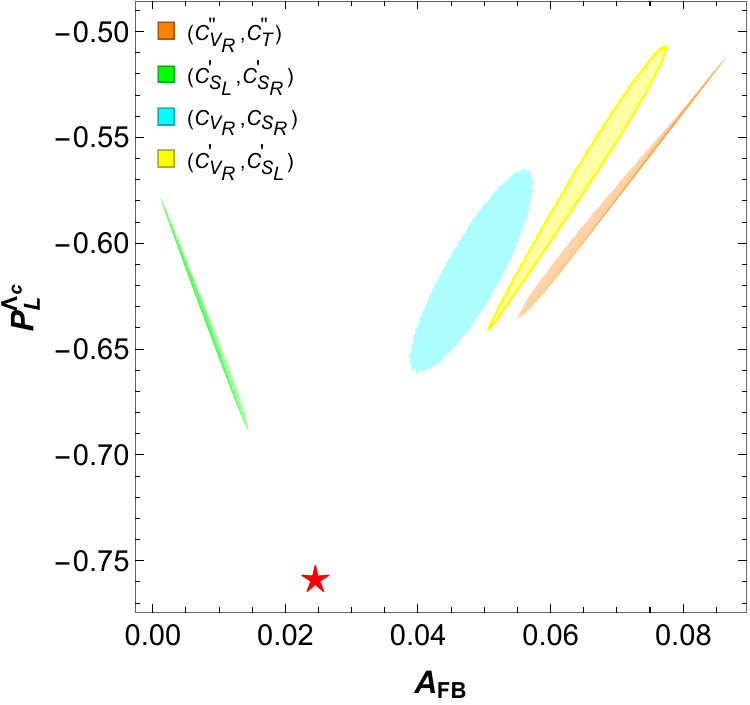}
\caption{}
\end{subfigure}
\begin{subfigure}[b]{0.32\textwidth}
\centering 
\includegraphics[width=5.6cm, height=3.5cm]{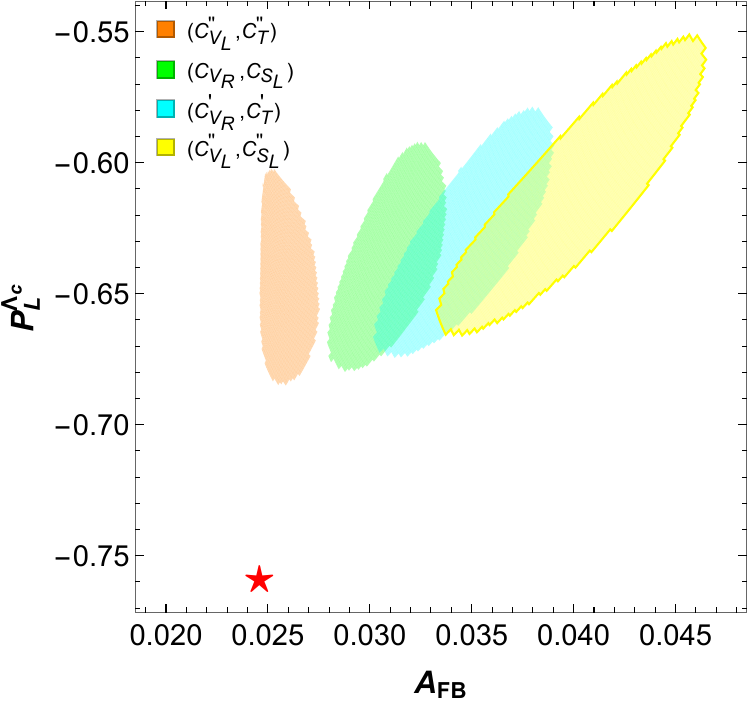}
\caption{}
\end{subfigure}
\begin{subfigure}[b]{0.32\textwidth}
\centering 
\includegraphics[width=5.6cm, height=3.5cm]{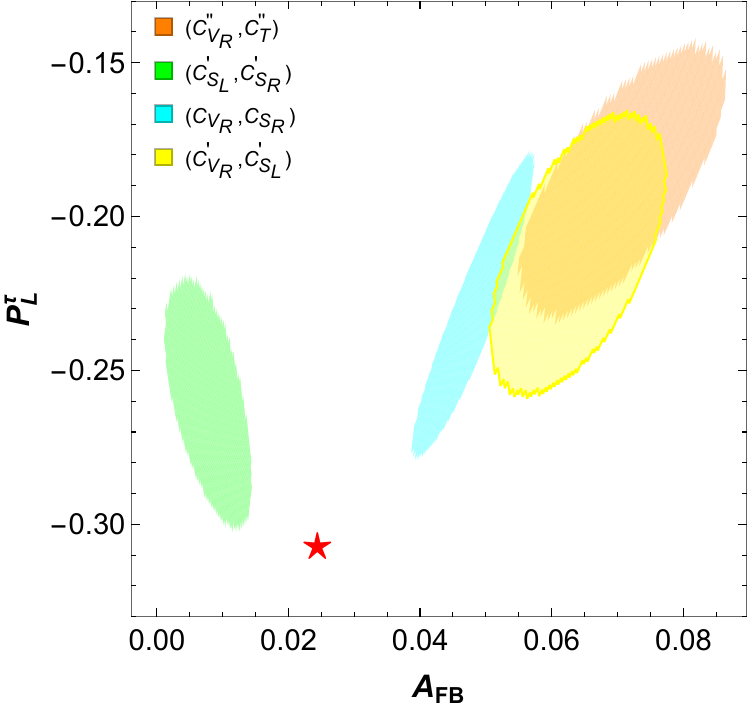}
\caption{}
\end{subfigure}
\begin{subfigure}[b]{0.32\textwidth}
\centering 
\includegraphics[width=5.6cm, height=3.5cm]{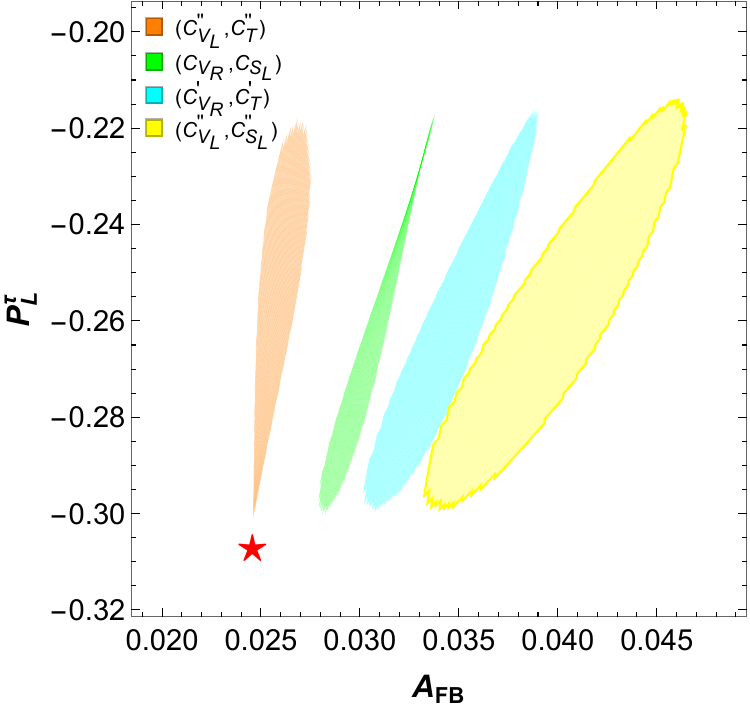}
\caption{}
\end{subfigure}
\begin{subfigure}[b]{0.32\textwidth}
\centering 
\includegraphics[width=5.6cm, height=3.5cm]{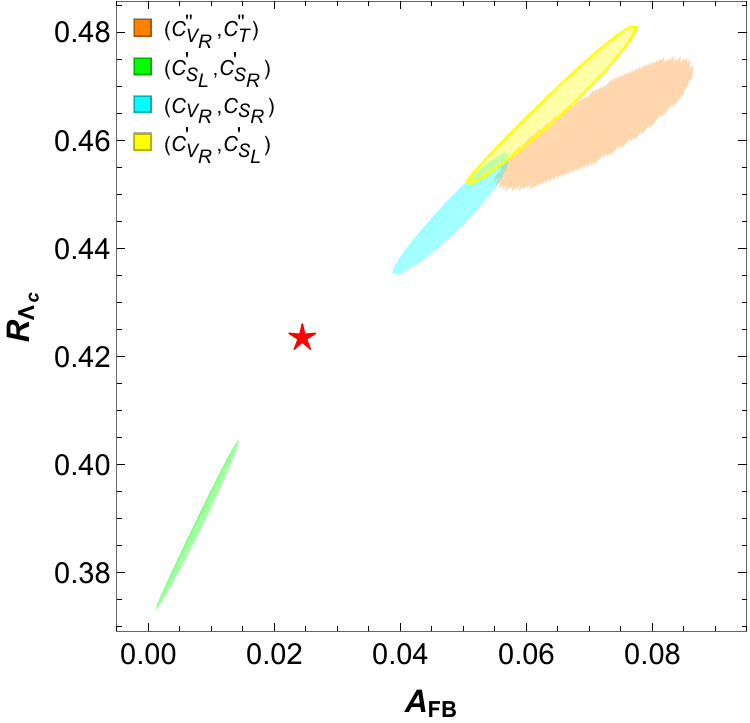}
\caption{}
\end{subfigure}
\begin{subfigure}[b]{0.32\textwidth}
\centering 
\includegraphics[width=5.6cm, height=3.5cm]{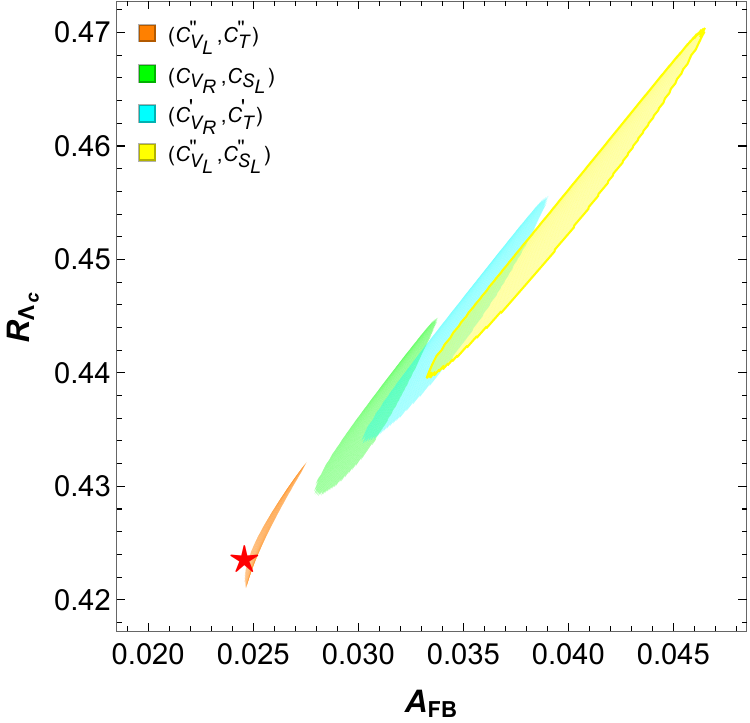}
\caption{}
\end{subfigure}

\caption{\label{corr-2dc-1}Preferred $1\sigma$ regions for the four two-WCs  scenarios for set $\mathcal{S}_1$ when we include WC $C_{V_{R}}$ in the $d\Gamma/dq^{2}-A_{FB}$ plane (a) and (b), $d\Gamma/dq^{2}-P_{L}^{\Lambda_{c}}$ plane (c) and (d), $d\Gamma/dq^{2}-P_{L}^{\tau}$ plane (e) and (f), $d\Gamma/dq^{2}-R_{\tau/\ell}\left(\Lambda_{c}\right)$ plane (g) and (h), $A_{FB}-P_{L}^{\Lambda_{c}}$ plane (i) and (j), $A_{FB}-P_{L}^{\tau}$ plane (k) and (l), and $A_{FB}-R_{\tau/\ell}\left(\Lambda_{c}\right)$ plane (m) and (n)  for the $BR\left(B_{c}\to\tau\bar{\nu}_\tau\right)<60\%$. The red stars represent SM predictions. Legends are depicted in plot.}
\end{figure}

\begin{figure}[H]
\centering 
\begin{subfigure}[b]{0.32\textwidth}
\centering 
\includegraphics[width=5.6cm, height=3.5cm]{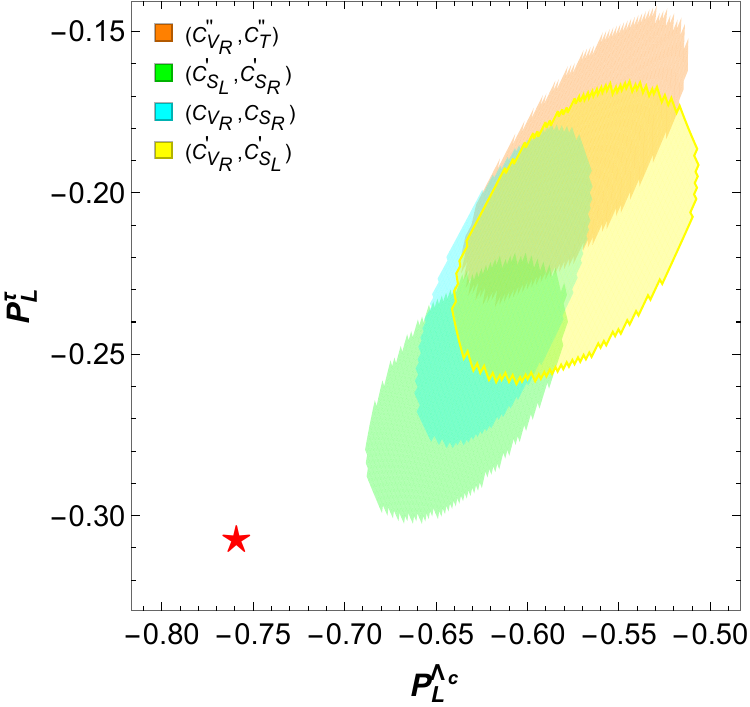}
\caption{}
\end{subfigure}
\begin{subfigure}[b]{0.32\textwidth}
\centering 
\includegraphics[width=5.6cm, height=3.5cm]{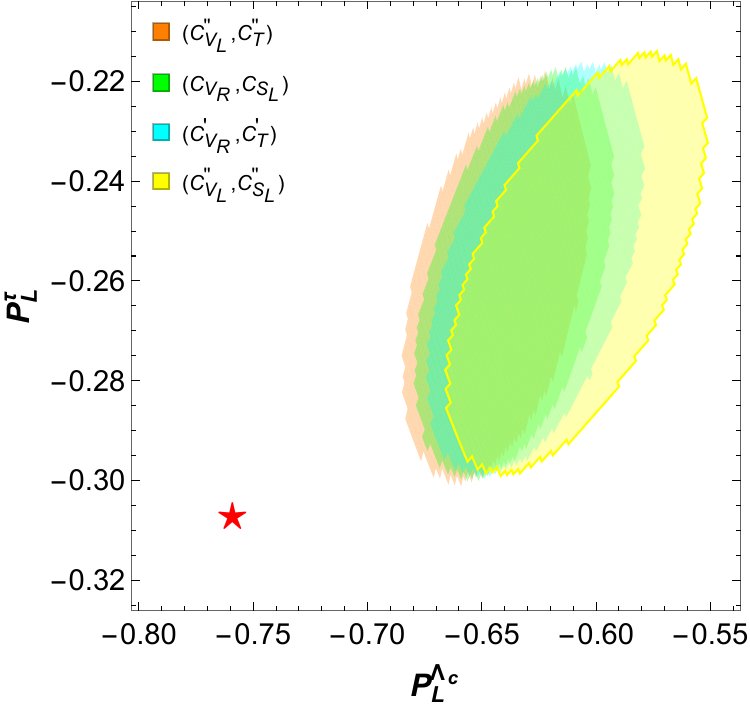} 
\caption{}
\end{subfigure}
\begin{subfigure}[b]{0.32\textwidth}
\centering 
\includegraphics[width=5.6cm, height=3.5cm]{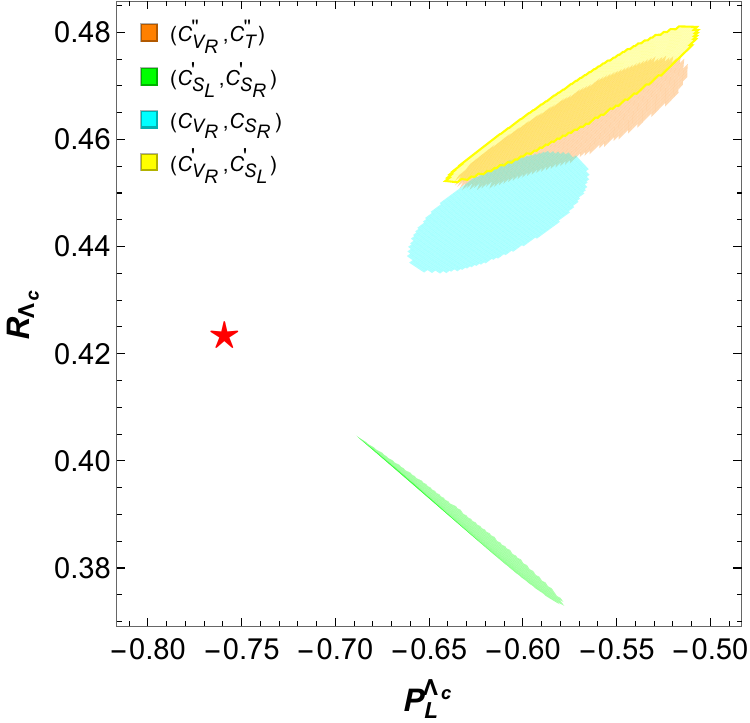}
\caption{}
\end{subfigure}
\begin{subfigure}[b]{0.32\textwidth}
\centering 
\includegraphics[width=5.6cm, height=3.5cm]{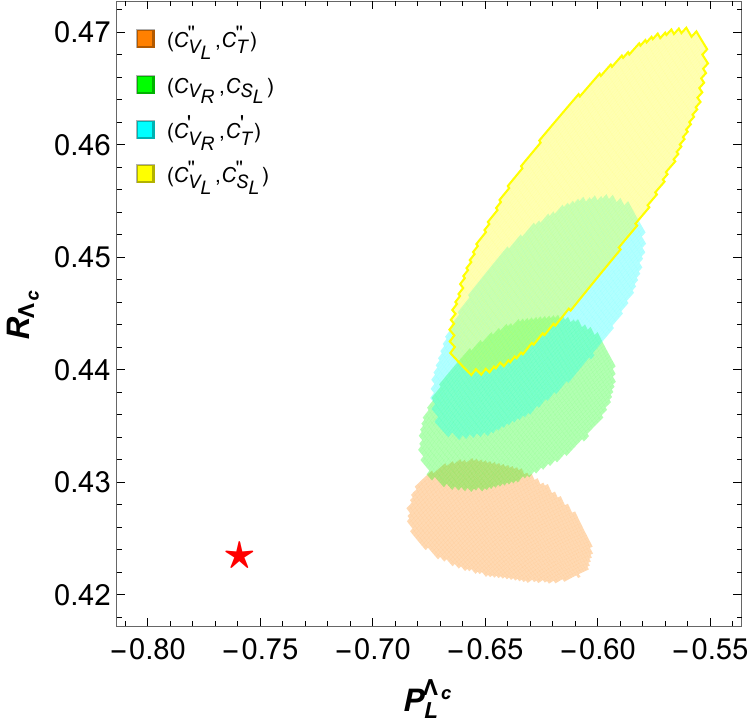}
\caption{}
\end{subfigure}
\begin{subfigure}[b]{0.32\textwidth}
\centering 
\includegraphics[width=5.6cm, height=3.5cm]{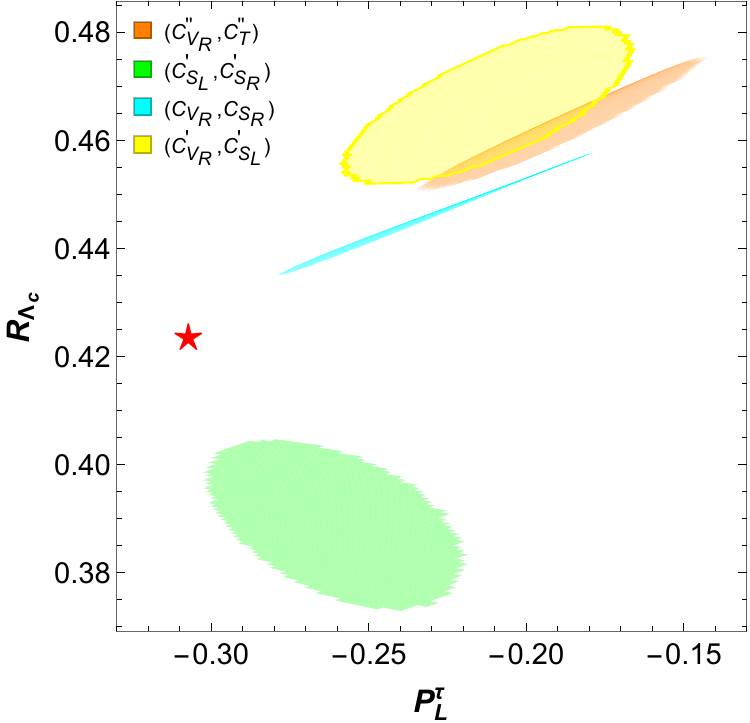} 
\caption{}
\end{subfigure}
\begin{subfigure}[b]{0.32\textwidth}
\centering 
\includegraphics[width=5.6cm, height=3.5cm]{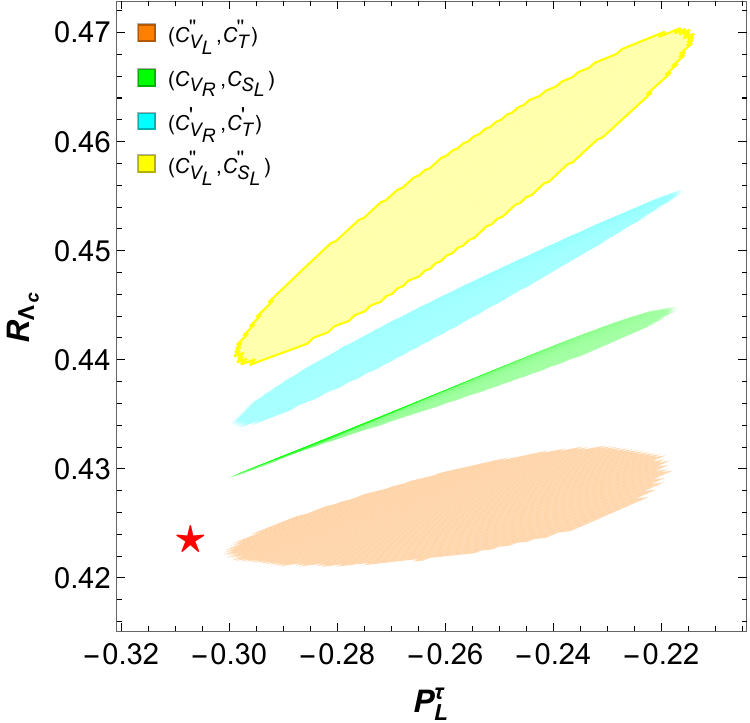}
\caption{}
\end{subfigure}

\caption{\label{corr-2dc-2}Preferred $1\sigma$ regions for the four two-WCs  scenarios for set $\mathcal{S}_1$ when we include WC $C_{V_{R}}$ in the $P_{L}^{\Lambda_{c}}-P_{L}^{\tau}$ plane (a) and (b), $P_{L}^{\Lambda_{c}}-R_{\tau/\ell}\left(\Lambda_{c}\right)$ plane (c) and (d), $P_{L}^{\tau}-R_{\tau/\ell}\left(\Lambda_{c}\right)$ plane (e) and (f) for the $BR\left(B_{c}\to\tau\bar{\nu}_\tau\right)<60\%$. The red stars represent SM predictions. Legends are depicted in plots.}
\end{figure}

\section{Conclusion}\label{sec6}
The experimental results of the  $R_{\tau/{\mu,e}}\left(D^{(*)}\right)$ from BaBar and Belle and the LHCb experiment show $3.31\sigma$ deviations from their SM predictions. By using the latest HFLAV results and considering the branching ratio constraints $60\%$, $30\%$ and $10\%$ from the lifetime of the $B_c$ meson, we determine the values of the Wilson coefficients (WCs) for each New Physics (NP) four-fermion operator with specific Lorentz structures. Our $\chi^2$ analysis shows that the WC $\left(C_{S_{L}},C_{S_{R}}\right)$ scenario has the largest $p-value$, $92\%$, and maximum pull from the SM, $3.87$, indicates that this scenario significantly improves the fit to the data over the SM alone. Furthermore, this scenario is sensitive to $B_c\to \tau\nu$ branching fraction constraints, highlighting the importance of precise branching fraction measurements. Additionally, the three one-dimensional NP scenarios: $C_{V_{L}}$, $C_{V_{L}}^{\prime}$, and $C_{S_{R}}^{\prime\prime}$ have the second-largest $p-value$, $92\%$, showing multiple ways to introduce NP.

The baryonic decay $\Lambda_{b}\rightarrow\Lambda_{c}\tau\bar{\nu}_{\tau}$ provides a valuable avenue for further exploring the $R_{\tau/{\mu,e}}\left(D^{(*)}\right)$ anomaly. This anomaly arises from observed discrepancies between experimental measurements and SM predictions in the ratios of branching fractions for semileptonic B- meson decays involving $\tau$ leptons compared to those involving lighter leptons. In this study, we investigate the impact of NP couplings with various Lorentz structures, including scalar, vector, and tensor interactions. The different observables in $\Lambda_{b}\rightarrow\Lambda_{c}\tau\bar{\nu}_{\tau}$ decays, such as differential decay rate $d\Gamma/dq^{2}$,
 lepton forward-backward asymmetry $A_{FB}$, $\Lambda_{c}-$longitudinal
polarization fraction $P_{L}^{\Lambda_{c}}$, $\tau-$lepton longitudinal polarization fraction $P_{L}^{\tau}$, and $\Lambda_{c}-$LFU ratio
$R_{\tau/\ell}\left(\Lambda_{c}\right)$, are used to distinguish between various NP operators. Unlike mesonic decays, where form factors have been extensively studied and determined from experimental data, the form factors for the baryonic decay $\Lambda_{b}\rightarrow\Lambda_{c}$ remain undetermined experimentally. This lack of experimental data makes it crucial to rely on theoretical methods, particularly lattice QCD, to derive these form factors. 
After constraining the parametric space of NP WCs, we investigated their impacts on various observables in the decays of $\Lambda_{b} \rightarrow \Lambda_{c} \tau \bar{\nu}_{\tau}$. Our findings indicate that these observables are sensitive to the presence of NP. Measurements of these observables will play a crucial role in distinguishing between different NP operators. In the one-dimensional case, we found a significant deviation for the WC $C_{S_{L}}^{\prime\prime}$ across all observables, followed by $C_{S_{R}}$ and $C_{T}$ (except for $P_{L}^{\tau}$). Additionally, $C_{V_{L}}, C_{V_{L}}^{\prime}$, and $C_{S_{R}}^{\prime\prime}$ exhibited the largest deviations for the differential decay rate, with the second-largest degenerate p-value of $92\%$. Notably, there was a minor deviation below the SM value for $C_{S_{L}}^{\prime}$ in $A_{FB}$. In the two-dimensional case, the maximum deviation was observed for the pair $(C_{S_{L}}, C_{S_{R}})$ for all observables, except for a notable second-largest deviation in $R_{\tau/l}\left(\Lambda_{c}\right)$. This scenario also presented the largest p-value of $92\%$. A maximum deviation was found in the differential decay rate $d\Gamma/dq^{2}$ across all WC scenarios. Furthermore, we discovered three degenerate solutions involving the WCs $(C_{S_{L}}, C_{T})$, which displayed the second-largest deviations in nearly all observables. Conversely, the deviation for WCs $(C_{V_{L}}, C_{T})$ was lower than that of the SM for $A_{FB}$, $P_{L}^{\Lambda_{c}}$, and $R_{\tau/l}\left(\Lambda_{c}\right)$.

In the last part, we validated and slightly updated the sum rule for $R_{\tau/\ell}\left(\Lambda_{c}\right)$ and derived the sum rule for $R_{\tau/\mu}(J/\psi)$ by linking it with $R_{\tau/{\mu,e}}\left(D^{\left(*\right)}\right)$. Notably, $R_{\tau/\mu}(J/\psi)$ is primarily influenced by $R_{\tau/{\mu,e}}\left(D^*\right)$. Using the BFP and recent $R_{\tau/{\mu,e}}\left(D^{\left(*\right)}\right)$ measurements, we determined the central value of  $R_{\tau/\ell}\left(\Lambda_{c}\right)=0.368$ and $R_{\tau/\mu}(J/\psi)=0.292$. Furthermore, the correlation analysis highlights significant relationships between the phenomenological observables in $\Lambda_{b}\rightarrow\Lambda_{c}\tau\bar{\nu}_{\tau}$ decay. Specifically, the direct correlations were found between $d\Gamma/dq^{2}$ and $P_{L}^{\tau}$ for WC $\left(C_{S_{L}},C_{S_{R}}\right)$; and between $A_{FB}$ and $P_{L}^{\Lambda_{c}}$ for three degenerate scenarios involving WCs $\left(C_{S_{L}},C_{T}\right)$. We hope our findings can be tested at the LHCb and future high-energy experiments dedicated to $b-$ decays.

\section*{Data Availability Statement}

No Data associated in the manuscript.

\appendix
\numberwithin{equation}{section}

\section{Expressions of Physical observables in terms of NP WCs}\label{AppendixA}
The expressions of the physical observables used to fit the data are given below \cite{Arslan:2023wgk}:
\begin{eqnarray}
R_{\tau/{\mu,e}}(D) & = & R_{\tau/{\mu,e}}^{\text{SM}}(D)\left\{ \left|1+\widetilde{C}_{V_{L}}+\widetilde{C}_{V_{R}}\right|^{2}+1.01\left|\widetilde{C}_{S_{L}}+\widetilde{C}_{S_{R}}\right|^{2}+0.84\left|\widetilde{C}_{T}\right|^{2}\right.\nonumber \\
 &  & \left.+1.49\Re\left[\left(1+\widetilde{C}_{V_{L}}+\widetilde{C}_{V_{R}}\right)\left(\widetilde{C}_{S_{L}}+\widetilde{C}_{S_{R}}\right)^{*}\right]+1.08\Re\left[\left(1+\widetilde{C}_{V_{L}}+\widetilde{C}_{V_{R}}\right)\left(\widetilde{C}_{T}\right)^{*}\right]\right\} ,\label{eqn1}\\
R_{\tau/{\mu,e}}(D^{*}) & = & R_{\tau/{\mu,e}}^{\text{SM}}(D^{*})\left\{ \left|1+\widetilde{C}_{V_{L}}\right|^{2}+\left|\widetilde{C}_{V_{R}}\right|^{2}+0.04\left|\widetilde{C}_{S_{L}}-\widetilde{C}_{S_{R}}\right|^{2}+16\left|\widetilde{C}_{T}\right|\right.\nonumber \\
 &  & \left.-1.83\Re\left[\left(1+\widetilde{C}_{V_{L}}\right)\left(\widetilde{C}_{V_{R}}\right)^{*}\right]-0.11\Re\left[\left(1+\widetilde{C}_{V_{L}}-\widetilde{C}_{V_{R}}\right)\left(\widetilde{C}_{S_{L}}-\widetilde{C}_{S_{R}}\right)^{*}\right]\right.\nonumber \\
 &  & \left.-5.17\Re\left[\left(1+\widetilde{C}_{V_{L}}\right)\left(\widetilde{C}_{T}\right)^{*}\right]+6.60\Re\left[\widetilde{C}_{V_{R}}\left(\widetilde{C}_{T}\right)^{*}\right]\right\} ,\label{eqn2}\\
 P_{\tau}\left(D\right) & = & P_{\tau}^{SM}\left(D\right)\left(\frac{R_{\tau/{\mu,e}}(D)}{R_{\tau/{\mu,e}}^{\text{SM}}(D)}\right)^{-1}\left\{ \left|1+\widetilde{C}_{V_{L}}\widetilde{C}_{V_{R}}\right|^{2}+3.04\left|\widetilde{C}_{S_{L}}+\widetilde{C}_{S_{R}}\right|^{2}+0.17\left|\widetilde{C}_{T}\right|^{2}\right.\nonumber \\
 &  & \left.+4.50\Re\left[\left(1+\widetilde{C}_{V_{L}}+\widetilde{C}_{V_{R}}\right)\left(\widetilde{C}_{S_{L}}+\widetilde{C}_{S_{R}}\right)^{*}\right]-1.09\Re\left[\left(1+\widetilde{C}_{V_{L}}+\widetilde{C}_{V_{R}}\right)\left(\widetilde{C}_{T}\right)^{*}\right]\right.\label{eqn8}\\
P_{\tau}\left(D^{*}\right) & = & P_{\tau}^{SM}\left(D^{*}\right)\left(\frac{R_{\tau/{\mu,e}}(D^{*})}{R_{\tau/{\mu,e}}^{\text{SM}}(D^{*})}\right)^{-1}\left\{ \left|1+\widetilde{C}_{V_{L}}\right|^{2}+\left|\widetilde{C}_{V_{R}}\right|^{2}-0.07\left|\widetilde{C}_{S_{L}}-\widetilde{C}_{S_{R}}\right|^{2}-1.85\left|\widetilde{C}_{T}\right|^{2}\right.\nonumber \\
 &  & \left.-1.79\Re\left[\left(1+\widetilde{C}_{V_{L}}\right)\left(\widetilde{C}_{V_{R}}\right)^{*}\right]+0.23\Re\left[\left(1+\widetilde{C}_{V_{L}}-\widetilde{C}_{V_{R}}\right)\left(\widetilde{C}_{S_{L}}-\widetilde{C}_{S_{R}}\right)^{*}\right]\right.\nonumber \\
 &  & \left.\left.-3.47\Re\left[\left(1+\widetilde{C}_{V_{L}}\right)\left(\widetilde{C}_{T}\right)^{*}\right]+4.41\Re\left[\widetilde{C}_{V_{R}}\left(\widetilde{C}_{T}\right)^{*}\right]\right\} \right.\label{eqn3}\\
F_{L}\left(D^{*}\right) & = & F_{L}^{SM}\left(D^{*}\right)\left(\frac{R_{\tau/{\mu,e}}(D^{*})}{R_{\tau/{\mu,e}}^{\text{SM}}(D^{*})}\right)^{-1}\left\{ \left|1+\widetilde{C}_{V_{L}}-\widetilde{C}_{V_{R}}\right|^{2}+0.08\left|\widetilde{C}_{S_{L}}-\widetilde{C}_{S_{R}}\right|^{2}+6.9\left|\widetilde{C}_{T}\right|^{2}\right.\nonumber \\
 &  & \left.\left.-0.25\Re\left[\left(1+\widetilde{C}_{V_{L}}-\widetilde{C}_{V_{R}}\right)\left(\widetilde{C}_{S_{L}}-\widetilde{C}_{S_{R}}\right)^{*}\right]-4.3\Re\left[\left(1+\widetilde{C}_{V_{L}}-\widetilde{C}_{V_{R}}\right)\left(\widetilde{C}_{T}\right)^{*}\right]\right\} \right.\label{eqn4}\\
R_{\tau/{\mu}}(J/\psi) & = & R_{\tau/{\mu}}^{\text{SM}}(J/\psi)\left\{ \left|1+\widetilde{C}_{V_{L}}\right|^{2}+\left|\widetilde{C}_{V_{R}}\right|^{2}+0.04\left|\widetilde{C}_{S_{L}}-\widetilde{C}_{S_{R}}\right|^{2}+14.7\left|\widetilde{C}_{T}\right|^{2}\right.\nonumber \\
 &  & \left.-1.82\Re\left[\left(1+\widetilde{C}_{V_{L}}\right)\left(\widetilde{C}_{V_{R}}\right)^{*}\right]+0.1\Re\left[\left(1+\widetilde{C}_{V_{L}}-\widetilde{C}_{V_{R}}\right)\left(\widetilde{C}_{S_{L}}-\widetilde{C}_{S_{R}}\right)^{*}\right]\right.\nonumber \\
 &  & \left.-5.39\Re\left[\left(1+\widetilde{C}_{V_{L}}\right)\left(\widetilde{C}_{T}\right)^{*}\right]+6.57\Re\left[\widetilde{C}_{V_{R}}\left(\widetilde{C}_{T}\right)^{*}\right]\right\} ,\label{eqn5}\\
R_{\tau/{\ell}}\left(\Lambda_{c}\right) & = & R_{\tau/{\ell}}^{\text{SM}}\left(\Lambda_{c}\right)\left\{ \left|1+\widetilde{C}_{V_{L}}\right|^{2}+\left|\widetilde{C}_{V_{R}}\right|^{2}+0.32\left(\left|\widetilde{C}_{S_{L}}\right|^{2}+\left|\widetilde{C}_{S_{R}}\right|^{2}\right)+10.4\left|\widetilde{C}_{T}\right|^{2}\right.\nonumber \\
 &  & \left.-0.72\Re\left[\left(1+\widetilde{C}_{V_{L}}\right)\left(\widetilde{C}_{V_{R}}\right)^{*}\right]+0.5\Re\left[\left(1+\widetilde{C}_{V_{L}}\right)\left(\widetilde{C}_{S_{R}}\right)^{*}+\widetilde{C}_{V_{R}}\left(\widetilde{C}_{S_{L}}\right)^{*}\right]\right.\nonumber \\
 &  & \left.+0.33\Re\left[\left(1+\widetilde{C}_{V_{L}}\right)\left(\widetilde{C}_{S_{L}}\right)^{*}+\widetilde{C}_{V_{R}}\left(\widetilde{C}_{S_{R}}\right)^{*}\right]+0.52\Re\left[\widetilde{C}_{S_{R}}\left(\widetilde{C}_{S_{L}}\right)^{*}\right]\right.\nonumber \\
 &  & \left.-3.11\Re\left[\left(1+\widetilde{C}_{V_{L}}\right)\left(\widetilde{C}_{T}\right)^{*}\right]+4.88\Re\left[\widetilde{C}_{V_{R}}\left(\widetilde{C}_{T}\right)^{*}\right]\right\} \label{eqn7}
\end{eqnarray}
Similarly, the expression of branching ratio of the $B_{c}\to\tau\bar{\nu}_{\tau}$
decay read as 
\begin{eqnarray*}
\mathcal{B}\left(B_{c}^{-}\rightarrow\tau^{-}\bar{\nu}_{\tau}\right) & = & \mathcal{B}\left(B_{c}^{-}\rightarrow\tau^{-}\bar{\nu}_{\tau}\right)^{\text{SM}}\left\{ \left|1+\widetilde{C}_{V_{L}}-\widetilde{C}_{V_{R}}-4.35\left(\widetilde{C}_{S_{L}}-\widetilde{C}_{S_{R}}\right)\right|^{2}\right\} ,
\end{eqnarray*}
where in the SM $\mathcal{B}\left(B_{c}^{-}\rightarrow\tau^{-}\bar{\nu}_{\tau}\right)^{\text{SM}}\approx 0.022$  \cite{Iguro:2022yzr}.  

\numberwithin{equation}{section}
\section{Goodness of fit}\label{GoF}

We accomplish $\chi^2$ to test the hypothesis about the distribution
of observables in distinct effective operators. This helps us to quantify the discrepancy between the
theoretical and experimental data used to fit. The $\chi^{2}$ expression is coded
as \cite{ParticleDataGroup:2020ssz,Workman:2022ynf}:
\begin{equation}
\chi^{2}\left(C_{X_{M}}\right)=\sum_{i,j}^{N_{obs}}\left[O_{i}^{\text{exp.}}-O_{i}^{\text{th}}\left(C_{X_{M}}\right)\right]C_{ij}^{-1}\left[O_{j}^{\text{exp.}}-O_{j}^{\text{th}}\left(C_{X_{M}}\right)\right]\label{B.1},
\end{equation}
where $N_{obs}$ is the number of observables, $O_{i}^{exp.}$ are
the data from experiments and $O_{i}^{th}$ are the observables theoretical
parameters, which in our case are real functions of scalar, vector and tensor WCs $C_{X_{M}}$
$(X=S,V,T)$ and $(M=L,R)$ and covariance matrix $C_{ij}=\rho_{ij}$, where $\rho_{ij}$ is correlation between observables $O_{i}$ and $O_{j}$. In our case, the only correlation
is between the observables $R_{\tau/{\mu,e}}\left(D\right)$ and $R_{\tau/{\mu,e}}\left(D^{*}\right)$, as provided by HFLAV Moriond 2024. For all other observables in the fit, the correlations
are assumed to be negligible or unavailable, and hence those contributions
are treated as uncorrelated. Now to implement the correlation we have to write
the full $\chi^2-$function in Eq. (\ref{B.1}) using the corresponding
$C_{ij}$ matrix for $R_{\tau/{\mu,e}}\left(D\right)$ and $R_{\tau/{\mu,e}}\left(D^{*}\right)$. The correlation between $R_{\tau/{\mu,e}}(D)$ and $R_{\tau/{\mu,e}}(D^{*})$ in terms of $pull$ is calculated using
\begin{equation}
\chi_{R_{\tau/{\mu,e}}\left(D\right)-R_{\tau/{\mu,e}}\left(D^{*}\right)}^{2}=\frac{\chi_{R_{\tau/{\mu,e}}\left(D\right)}^{2}+\chi_{R_{\tau/{\mu,e}}\left(D^{*}\right)}^{2}-2*\rho*\text{pull}_{R_{\tau/{\mu,e}}\left(D\right)}*\text{pull}_{R_{\tau/{\mu,e}}\left(D^{*}\right)}}{1-\rho^{2}},
\end{equation}
where, the correlation value $\rho=-0.39$ is taken from the HFLAV. Firstly, we work out how many degrees of freedom
($dof$) we have, which is equal to $N_{dof}=N_{obs}-N_{par}$, where $N_{obs}$ is the number of independent observables used to fit,
$N_{par}$ is the number of free parameters to be fitted for each parameter. For the real WCs, we have  $N_{par}=1\left(2\right)$  for $1\left(2\right)-$ dimension(s) and
$N_{obs}=4$ and $5$ for the sets of observables $\mathcal{S}_1$ and $\mathcal{S}_2$. As a second step, we obtain the minimum value of $\chi^{2}$ for each parameter
to acquire Best fit points (BFP). Third, we used the value of
$\chi^{2}$ to obtain $p-\text{value}$. The $p-\text{value}$ for the hypothesis can
be calculated as \cite{ParticleDataGroup:2020ssz,Workman:2022ynf}
\begin{equation}
p=\intop_{\chi^{2}}^{\infty}f\left(z;n_{d}\right)dz,
\end{equation}
where $f(z;n_{d})$ is the $\chi^{2}$ probability distribution function
and $n_{d}$ is the number of degrees of freedom. The $p-\text{value}$ quantify
the consistency between data and the hypothesis of the NP scenario. Finally,
we estimate the value of $\text{pull}$ from the SM in units of standard deviation
($\sigma$) determined by 
\begin{equation}
\text{pull}_{SM}=\sqrt{\chi_{\text{SM}}^{2}-\chi_{\text{min}}^{2}},
\end{equation}
where $\chi_{\text{\text{SM}}}^{2}=\chi^{2}(0)$.

\section{Helicity spinors and polarization vectors}\label{hspa}

In this appendix, the spinors and polarization vectors used to calculate
the helicity amplitudes for the decay $\Lambda_{b}\rightarrow\Lambda_{c}\tau\bar{\nu}_{\tau}$
are presented.

\subsection{$\Lambda_{b}$ rest frame}

To calculate the hadronic helicity amplitudes, we have consider
\begin{equation}
p_{1}^{\mu}=\left(m_{1},0,0,0\right),\quad p_{2}^{\mu}=\left(E_{2},0,0,\left|q\right|\right),\quad p_{1}^{\mu}=\left(q_{0},0,0,-\left|q\right|\right),
\end{equation}
where $q^{\mu}$is the four-momentum of the virtual vector boson in
the $\Lambda_{b}$ rest frame, and 
\begin{align}
q_{0} & =\frac{m_{1}^{2}-m_{2}^{2}+q^{2}}{2m_{1}},\quad E_{2}=\frac{m_{1}^{2}+m_{2}^{2}-q^{2}}{2m_{1}},\quad\left|q\right|=\frac{1}{2m_{1}}\sqrt{Q_{+}Q_{-}},
\end{align}
and $Q_{\pm}=\left(m_{1}\pm m_{2}\right)^{2}-q^{2}.$ 

The baryon spinors are then given by \cite{Auvil:1966eao}
\begin{align}
\overline{u}_{2}\left(\pm\frac{1}{2},p_{2}\right) & =\sqrt{E_{2}+m_{2}}\left(\chi_{\pm}^{\dagger},\frac{\mp\left|p_{2}\right|}{E_{2}+m_{2}}\chi_{\pm}^{\dagger}\right),\nonumber \\
u_{1}\left(\pm\frac{1}{2},p_{1}\right) & =\sqrt{2m_{1}}\left(\begin{array}{c}
\chi_{\pm}\\
0
\end{array}\right),
\end{align}
where $\chi_{+}=\left(\begin{array}{c}
1\\
0
\end{array}\right)$ and $\chi_{-}=\left(\begin{array}{c}
0\\
1
\end{array}\right)$ are usual Pauli two-spinors.

The polarization vectors of the virtual vector boson are \cite{Auvil:1966eao}
\begin{align}
\epsilon^{\mu}\left(t\right) & =\frac{1}{\sqrt{q^{2}}}\left(q_{0},0,0,-\left|q\right|\right),\quad\epsilon^{\mu}\left(0\right)=\frac{1}{\sqrt{q^{2}}}\left(\left|q\right|,0,0,-q_{0}\right),\quad\epsilon^{\mu}\left(\pm1\right)=\frac{1}{\sqrt{2}}\left(0,\mp1,i,0\right).\label{hpol}
\end{align}
and the orthonormality and completeness relation
\begin{equation}
\sum_{\mu}\epsilon^{*\mu}\left(m\right)\epsilon_{\mu}^{*}\left(n\right)=g_{mn},\quad\sum_{m,n}\epsilon_{\mu}^{*}\left(m\right)\epsilon_{\nu}^{*}\left(n\right)=g_{\mu\nu},\quad m,n\in\left\{ t,0,\pm1\right\} \label{ortho}
\end{equation}
where $g_{mn}=diag\left(1,-1,-1,-1\right).$ 

\subsection{HQET form factors}
The $\Lambda_{b}\rightarrow\Lambda_{c}$ hadronic matrix elements can be written in terms of ten helicity form factors $\left\{ F_{0,+,\perp},G_{0,+,\perp},h_{+,\perp},\widetilde{h}_{+,\perp}\right\}$ \cite{Detmold:2015aaa,Datta:2017aue,Yan:2019hpm}.
Following Ref. \cite{Detmold:2015aaa}, the lattice calculations are fitted by two
(Bour-rely-Caprini-Lellouch) BCL $z-$parameterization. In the so-called ``nominal'' fit, a form factor $f$ reduces to the form
\[
f\left(q^{2}\right)=\frac{1}{1-q^{2}/\left(m_{pole}^{f}\right)^{2}}\left\{ a_{0}^{f}+a_{1}^{f}z^{f}\left(q^{2}\right)\right\},
\]
while a form factor $f$ in the higher-order fit is given by 
\[
f_{HO}\left(q^{2}\right)=\frac{1}{1-q^{2}/\left(m_{pole}^{f}\right)^{2}}\left\{ a_{0,HO}^{f}+a_{1,HO}^{f}z^{f}\left(q^{2}\right)+a_{2,HO}^{f}\left(z^{f}\left(q^{2}\right)\right)^{2}\right\} ,
\]
where
\[
t_{0}=\left(m_{1}-m_{2}\right)^{2},\quad t_{+}^{f}=\left(m_{pole}^{f}\right)^{2},\quad z^{f}\left(q^{2}\right)=\frac{\sqrt{t_{+}^{f}-q^{2}}-\sqrt{t_{+}^{f}-t_{0}}}{\sqrt{t_{+}^{f}-q^{2}}+\sqrt{t_{+}^{f}-t_{0}}}.
\]
The values of the fit parameters and all the pole masses are taken from \cite{Datta:2017aue}.
\subsection{Dilepton rest frame}\label{hspa-2}

To calculate the leptonic helicity amplitudes,we work in the rest
frame of the virtual vector boson, which is equal to the rest frame
of the $\tau\bar{\nu}_{\tau}$ dilepton ysytem. we have
\begin{equation}
p_{\tau}^{\mu}=\left(E_{\tau},0,0,\left|p_{\tau}\right|\right),\quad p_{\bar{\nu}_{\tau}}^{\mu}=\left(E_{\nu},0,0,-\left|p_{\tau}\right|\right),
\end{equation}
where 
\begin{equation}
\left|p_{\tau}\right|=\frac{1}{2}\sqrt{q^{2}}v^{2},\quad E_{\tau}=\left|p_{\tau}\right|+\frac{m_{\tau}^{2}}{\sqrt{q^{2}}},\quad v=\sqrt{1-\frac{m_{\tau}^{2}}{q^{2}}},
\end{equation}
and $\theta_{\tau}$ is the angle between the three-momentum of the
$\tau$ and $\Lambda_{c}$ in this frame.

The lepton spinors are then given by 
\begin{align}
\overline{u}_{\tau}\left(\pm\frac{1}{2},p_{\tau}\right) & =\sqrt{E_{\tau}+m_{\tau}}\left(\chi_{\pm}^{\dagger},\frac{\mp\left|p_{\tau}\right|}{E_{\tau}+m_{\tau}}\chi_{\pm}^{\dagger}\right),\nonumber \\
u_{\bar{\nu}_{\tau}}\left(\pm\frac{1}{2},p_{1}\right) & =\sqrt{E_{\nu}}\left(\begin{array}{c}
\chi_{+}\\
-\chi_{+}
\end{array}\right).
\end{align}

We then rotate about the $y$ axis by the angle $\theta_{\tau}$ so
that after the rotation, the three-momentum of the $\Lambda_{c}$
points in the $+z$ direction. The two-spinors transform as
\begin{equation}
\chi'_{\pm}=e^{-i\theta_{\tau}\sigma_{2}/2}\chi_{\pm}=\left(\begin{array}{cc}
\cos\left(\theta_{\tau}/2\right) & -\sin\left(\theta_{\tau}/2\right)\\
\sin\left(\theta_{\tau}/2\right) & \cos\left(\theta_{\tau}/2\right)
\end{array}\right)\chi_{\pm},
\end{equation}
and
\[
\chi_{\pm}^{\dagger}=\chi_{\pm}^{\dagger}\left(\begin{array}{cc}
\cos\left(\theta_{\tau}/2\right) & \sin\left(\theta_{\tau}/2\right)\\
-\sin\left(\theta_{\tau}/2\right) & \cos\left(\theta_{\tau}/2\right)
\end{array}\right),
\]
and the full lepton spinors after the rotation are 
\begin{align}
\overline{u}_{\tau}\left(\frac{1}{2},p_{\tau}\right) & =\sqrt{E_{\tau}+m_{\tau}}\left(\cos\left(\theta_{\tau}/2\right),\sin\left(\theta_{\tau}/2\right),\frac{-\left|p_{\tau}\right|}{E_{\tau}+m_{\tau}}\cos\left(\theta_{\tau}/2\right),\frac{-\left|p_{\tau}\right|}{E_{\tau}+m_{\tau}}\sin\left(\theta_{\tau}/2\right)\right),\nonumber \\
\overline{u}_{\tau}\left(-\frac{1}{2},p_{\tau}\right) & =\sqrt{E_{\tau}+m_{\tau}}\left(-\sin\left(\theta_{\tau}/2\right),\cos\left(\theta_{\tau}/2\right),\frac{-\left|p_{\tau}\right|}{E_{\tau}+m_{\tau}}\sin\left(\theta_{\tau}/2\right),\frac{\left|p_{\tau}\right|}{E_{\tau}+m_{\tau}}\cos\left(\theta_{\tau}/2\right)\right),\\
u_{\bar{\nu}_{\tau}}\left(\pm\frac{1}{2},p_{1}\right) & =\sqrt{E_{\nu}}\left(\begin{array}{c}
\cos\left(\theta_{\tau}/2\right)\\
\sin\left(\theta_{\tau}/2\right)\\
-\cos\left(\theta_{\tau}/2\right)\\
-\sin\left(\theta_{\tau}/2\right)
\end{array}\right).
\end{align}

The polarization vectors of the virtual vector boson are
\begin{align}
\epsilon^{\mu}\left(t\right) & =\left(1,0,0,0\right),\quad\epsilon^{\mu}\left(0\right)=\left(0,0,0,-1\right),\quad\epsilon^{\mu}\left(\pm1\right)=\frac{1}{\sqrt{2}}\left(0,\mp1,i,0\right).
\end{align}
which can be obtained from Eq. \ref{hpol} by Lorentz transformation
and satisfy the orthonormality and completeness relation in Eq. \ref{ortho}. 

\section{Observables in $\Lambda_{b}\rightarrow\Lambda_{c}\tau\bar{\nu}_{\tau}$
decay}\label{obsRLc}

\subsection{Differential decay rate}

We obtain the analytical expression of Differential decay rate by
integrating two-fold decay rate in Eq. (\ref{DecayRate}) w.r.t $\cos\theta_{\tau},$
as
\begin{equation}
\frac{d\Gamma}{dq^{2}}=N_{\varLambda_{c}}\left\{ A_{1}^{VA}+\frac{m_{\tau}^{2}}{2q^{2}}A_{2}^{VA}+\frac{3}{2}A_{3}^{SP}+8\left(1+\frac{2m_{\tau}^{2}}{q^{2}}\right)A_{4}^{T}+\frac{3m_{\tau}^{2}}{\sqrt{q^{2}}}\left(A_{5}^{VA-SP}+4A_{6}^{VA-T}\right)\right\} ,
\end{equation}
with
\begin{equation}
N_{\varLambda_{c}}=\frac{G_{F}^{2}\left|V_{cb}\right|^{2}}{384\pi^{3}}\left(1-\frac{m_{\tau}^{2}}{q^{2}}\right)^{2}\frac{q^{2}\sqrt{Q_{+}Q_{-}}}{m_{1}^{3}},
\end{equation}
and
\begin{align}
A_{1}^{VA} & =\sum_{s=\pm 1/2}\left|H_{s,0}^{VA}\right|^{2}+\left|H_{1/2,+}^{VA}\right|^{2}+\left|H_{-1/2,-}^{VA}\right|^{2},\nonumber \\
A_{2}^{VA} & =A_{1}^{VA}+3\sum_{s=\pm 1/2}\left|H_{s,t}^{VA}\right|^{2},\nonumber \\
A_{3}^{SP} & =\sum_{s=\pm 1/2}\left|H_{s}^{SP}\right|^{2}\nonumber \\
A_{4}^{T} & =\sum_{s=\pm 1/2}\left|H_{s,t,0}^{T,s}\right|^{2}+\left|H_{1/2,t,+}^{T,-1/2}\right|^{2}+\left|H_{-1/2,t,-}^{T,1/2}\right|^{2},\nonumber \\
A_{5}^{VA-SP} & =\sum_{s=\pm 1/2}\Re\left[H_{s}^{SP*}H_{s,t}^{VA}\right],\nonumber \\
A_{6}^{VA-T} & =\sum_{s=\pm 1/2}\Re\left[H_{s,0}^{VA*}H_{s,t,0}^{T,s}\right]+\Re\left[H_{1/2,+}^{VA*}H_{1/2,t,+}^{T,-1/2}\right]+\Re\left[H_{-1/2,-}^{VA*}H_{-1/2,t,-}^{T,1/2}\right].
\end{align}

\subsection{Forward-backward asymmetry}

For the forward-backward asymmetry in Eq.(\ref{Afb}), we obtain
\begin{equation}
\frac{dA_{FB}}{dq^{2}}=\frac{N_{\varLambda_{c}}}{d\Gamma/dq^{2}}\frac{3}{4}\left\{ B_{1}^{VA}+\frac{2m_{\tau}^{2}}{q^{2}}\left(B_{2}^{VA}+8B_{3}^{T}\right)+\frac{2m_{\tau}^{2}}{\sqrt{q^{2}}}\left(B_{4}^{VA-SP}+4A_{5}^{VA-T}\right)+8B_{6}^{SP-T}\right\} ,
\end{equation}
with
\begin{align}
B_{1}^{VA} & =\left|H_{1/2,+}^{VA}\right|^{2}-\left|H_{-1/2,-}^{VA}\right|^{2},\nonumber \\
B_{2}^{VA} & =\sum_{s=\pm 1/2}\Re\left[H_{s,t}^{VA*}H_{s,0}^{VA}\right],\nonumber \\
B_{3}^{T} & =\left|H_{1/2,t,+}^{T,-1/2}\right|^{2}-\left|H_{-1/2,t,-}^{T,1/2}\right|^{2},\nonumber \\
B_{4}^{VA-SP} & =\sum_{s=\pm 1/2}\Re\left[H_{s}^{SP*}H_{s,0}^{VA}\right],\nonumber \\
B_{5}^{VA-T} & =\sum_{s=\pm 1/2}\Re\left[H_{s,t}^{VA*}H_{s,t,0}^{T,s}\right]+\Re\left[H_{1/2,+}^{VA*}H_{1/2,t,+}^{T,-1/2}\right]+\Re\left[H_{-1/2,-}^{VA*}H_{-1/2,t,-}^{T,1/2}\right],\nonumber \\
B_{6}^{SP-T} & =\sum_{s=\pm 1/2}\Re\left[H_{s}^{SP*}H_{s,t,0}^{T,s}\right].
\end{align}

\subsection{$\Lambda_{c}-$longitudinal polarization fraction}

For the $\Lambda_{c}-$longitudinal polarization fraction in Eq.(\ref{PLc}),
we obtain
\begin{equation}
\frac{dP_{L}^{\Lambda_{c}}}{dq^{2}}=\frac{N_{\varLambda_{c}}}{d\Gamma/dq^{2}}\frac{1}{2}\left\{ 2C_{1}^{VA}+\frac{m_{\tau}^{2}}{q^{2}}C_{2}^{VA}+3C_{3}^{SP}+16\left(1+\frac{2m_{\tau}^{2}}{q^{2}}\right)C_{4}^{T}+\frac{6m_{\tau}}{\sqrt{q^{2}}}\left(C_{5}^{VA-SP}+4C_{6}^{VA-T}\right)\right\} ,
\end{equation}
with 
\begin{align}
C_{1}^{VA} & =\left|H_{1/2,0}^{VA}\right|^{2}-\left|H_{-1/2,0}^{VA}\right|^{2}+\left|H_{1/2,+}^{VA}\right|^{2}-\left|H_{-1/2,-}^{VA}\right|^{2},\nonumber \\
C_{2}^{VA} & =C_{1}^{VA}+3\left(\left|H_{1/2,t}^{VA}\right|^{2}-\left|H_{1/2,t}^{VA}\right|^{2}\right),\nonumber \\
C_{3}^{SP} & =\left|H_{1/2}^{SP}\right|^{2}-\left|H_{-1/2}^{SP}\right|^{2},\nonumber \\
C_{4}^{T} & =\sum_{s=\pm 1/2} 2s\left|H_{s,t,0}^{T,s}\right|^{2}+\left|H_{1/2,t,+}^{T,-1/2}\right|^{2}-\left|H_{-1/2,t,-}^{T,1/2}\right|^{2},\nonumber \\
C_{5}^{VA-SP} & =\sum_{s=\pm 1/2} 2s\Re\left[H_{s}^{SP*}H_{s,t}^{VA}\right],\nonumber \\
C_{6}^{VA-T} & =\sum_{s=\pm 1/2} 2s\Re\left[H_{s,0}^{VA*}H_{s,t,0}^{T,s}\right]+\Re\left[H_{1/2,+}^{VA*}H_{1/2,t,+}^{T,-1/2}\right]-\Re\left[H_{-1/2,-}^{VA*}H_{-1/2,t,-}^{T,1/2}\right].
\end{align}

\subsection{$\tau-$lepton longitudinal polarization fraction}

For the $\tau-$lepton longitudinal polarization fraction in Eq.(\ref{PLt}),
we obtain
\begin{equation}
\frac{dP_{L}^{\tau}}{dq^{2}}=\frac{N_{\varLambda_{c}}}{d\Gamma/dq^{2}}\frac{1}{2}\left\{ -2D_{1}^{VA}+\frac{m_{\tau}^{2}}{q^{2}}D_{2}^{VA}+3D_{3}^{SP}+16\left(1-\frac{2m_{\tau}^{2}}{q^{2}}\right)D_{4}^{T}+\frac{m_{\tau}}{\sqrt{q^{2}}}\left(6D_{5}^{VA-SP}-8D_{6}^{VA-T}\right)\right\} ,
\end{equation}
with 
\begin{align}
D_{1}^{VA} & =\sum_{s=\pm 1/2}\left|H_{s,0}^{VA}\right|^{2}+\left|H_{1/2,+}^{VA}\right|^{2}-\left|H_{-1/2,-}^{VA}\right|^{2},\nonumber \\
D_{2}^{VA} & =D_{1}^{VA}+3\sum\left|H_{s,t}^{VA}\right|^{2},\nonumber \\
D_{3}^{SP} & =\sum\left|H_{s}^{SP}\right|^{2},\nonumber \\
D_{4}^{T} & =\sum_{s=\pm 1/2}\left|H_{s,t,0}^{T,s}\right|^{2}+\left|H_{1/2,t,+}^{T,-1/2}\right|^{2}-\left|H_{-1/2,t,-}^{T,1/2}\right|^{2},\nonumber \\
D_{5}^{VA-SP} & =\sum_{s=\pm 1/2}\Re\left[H_{s}^{SP*}H_{s,t}^{VA}\right],\nonumber \\
D_{6}^{VA-T} & =\sum_{s=\pm 1/2}\Re\left[H_{s,0}^{VA*}H_{s,t,0}^{T,s}\right]+\Re\left[H_{1/2,+}^{VA*}H_{1/2,t,+}^{T,-1/2}\right]-\Re\left[H_{-1/2,-}^{VA*}H_{-1/2,t,-}^{T,1/2}\right].
\end{align}

\end{document}